\documentclass[useAMS,usenatbib,usegraphicx]{mn2e}
\title[A spectroscopic analysis of NLS1s]{A spectroscopic analysis of a sample of narrow-line Seyfert 1 galaxies selected from the Sloan Digital Sky Survey}
\author[V. Cracco et al.]{V. Cracco,$^{1}$\thanks{E-mail:valentina.cracco@unipd.it} S. Ciroi,$^{1}$ M. Berton,$^{1}$  F. Di Mille,$^{2}$  L. Foschini,$^{3}$ 
G. La Mura,$^{1}$ 
\newauthor P. Rafanelli$^{1}$\\
$^{1}$Department of Physics and Astronomy, Padova University, Vicolo dell'Osservatorio 3, 35122 Padova, Italy\\
$^{2}$Las Campanas Observatory, Casilla 601, La Serena, Chile \\
$^{3}$INAF/Osservatorio Astronomico di Brera, via E. Bianchi 46, 23807 Merate (LC), Italy }

\begin{document}

\date{Accepted *** Received ***; in original form ***}

\pagerange{\pageref{firstpage}--\pageref{lastpage}} \pubyear{2002}

\maketitle

\label{firstpage}

\begin{abstract}
We revisited the spectroscopic characteristics of narrow-line Seyfert 1 galaxies (NLS1s) by analysing a homogeneous sample of 296 NLS1s at redshift between 0.028 and 0.345, extracted from the Sloan Digital Sky Survey (SDSS-DR7) public archive. We confirm that NLS1s are mostly characterized by Balmer lines with Lorentzian profiles, lower black hole masses and higher Eddington ratios than classic broad-line Seyfert 1 (BLS1s), but they also appear to be active galactic nuclei (AGNs) contiguous with BLS1s and sharing with them common properties. Strong Fe\,{\sc ii} emission does not seem to be a distinctive property of NLS1s, as low values of Fe\,{\sc ii}/H$\beta$ are equally observed in these AGNs. 
Our data indicate that Fe\,{\sc ii} and Ca\,{\sc ii} kinematics are consistent with the one of H$\beta$. On the contrary, O\,{\sc i}\,$\lambda$8446 seems to be systematically narrower and it is likely emitted by gas of the broad-line region more distant from the ionizing source and showing different physical properties. 
Finally, almost all NLS1s of our sample show radial motions of the narrow-line region highly-ionised gas. The mechanism responsible for this effect is not yet clear, but there are hints that very fast outflows require high continuum luminosities ($>10^{44}$ erg s$^{-1}$) or high Eddington ratios ($\log(L_{\rm bol}/L_{\rm Edd})>-0.1$).
\end{abstract}


\section{Introduction}
Known for thirty years \citep{1985ApJ...297..166O}, narrow-line Seyfert 1 galaxies (NLS1s) were conventionally classified as those Seyfert 1 galaxies having full width at half maximum (FWHM) of H$\beta$ lower than 2000 km s$^{-1}$ \citep{1989ApJ...342..224G} and showing the flux ratio between [O\,{\sc iii}]$\lambda$5007 and the total (broad plus narrow) H$\beta$ smaller than 3 \citep{1985ApJ...297..166O}. Most of them are also characterized by emission of strong Fe\,{\sc ii} multiplets, [Fe\,{\sc vii}]$\lambda$6087 and [Fe\,{\sc x}]$\lambda$6374, and in some cases also by [Fe\,{\sc xi}]$\lambda$7892 and [Fe\,{\sc xiv}]$\lambda$5303 \citep{2008MNRAS.385...53M, 2011nlsg.confE...2P}. In addition, NLS1s show a steep soft X-ray slope \citep{1996A&A...305...53B}, rapid and large soft X-ray variability \citep{2004MNRAS.347..269G}, a weak big blue bump in the optical/UV range \citep[e.g.][]{1994A&A...283L...9M, 1995MNRAS.276.1281P}, bright IR emission \citep{1996ApJS..106..341M}, and nuclear super-solar metallicity \citep[and references therein]{2000MNRAS.314L..17M, 2001A&A...374..914K}. Some NLS1s show also radio and gamma-ray emission, signs of the presence of a relativistic jet \citep{2009ApJ...707L.142A, 2015A&A...575A..13F}. 
NLS1s are generally believed to be powered by less massive black holes (BHs) than classic broad-line Seyfert 1 galaxies (BLS1s) \citep{1996A&A...305...53B, 2000NewAR..44..491P} and accreting up to the Eddington limit \citep{1992ApJS...80..109B}. 
Therefore, it is likely that NLS1s are a young phase of BLS1s \citep{2000MNRAS.314L..17M}.
Notwithstanding all these established properties, some questions still remain unsolved. 

Models by \citet{2012MNRAS.426.3086G} suggest that NLS1s are expected to have broad-line region (BLR) lines with Lorentzian profiles caused by a macroscopic turbulent motion of gaseous clouds, which, especially at large radii, contributes significantly to the line wings. At the same time, \citet{2011Natur.470..366K, 2013A&A...549A.100K} showed that the profiles of permitted lines in active galactic nuclei (AGNs) consist mainly of rotationally broadened Lorentzian functions. In the past, other authors favored a fitting with a single Lorentzian function \citep[see e.g.][]{1996ApJS..106..341M, 2001A&A...372..730V, 2002ApJ...566L..71S}. However, there is no consensus on this topic, also because it is still debated whether a Lorentzian function is the most appropriate representation of the line profile, since a double Gaussian, or even a multi-Gaussian approach seemed to provide a more statistically robust fitting \citep[see e.g.][]{2005ApJ...623..700D, 2008MNRAS.385...53M}.

Strong multiplets of Fe\,{\sc ii} are often detected in NLS1s spectra.
The excitation mechanism of these spectral lines is still uncertain. Photoionization models cannot easily account for strong Fe\,{\sc ii} emission \citep{2000NewAR..44..531C}, since the region emitting these multiplets is a weakly ionized part of the BLR. Assuming a high over-abundance of iron also does not reproduce the intensity of the multiplets \citep{2000NewAR..44..531C}. Shocks induced by winds and/or outflows have been invoked to heat the gas and cause the collisional excitation of the upper energy levels of the Fe\,{\sc ii} transitions \citep{2000NewAR..44..531C, 2006A&A...451..851V, 2010ApJ...721.1835S}. 
Fe\,{\sc ii} emission could depend on the physical conditions of the low-ionization clouds of the BLR which emits also O\,{\sc i}\,$\lambda$8446, and Ca\,{\sc ii}\,$\lambda$8498,8542,8662 (hereafter CaT). Indeed, the ionization potentials of O\,{\sc i} and Ca\,{\sc ii} (13.6 and 11.9 eV, respectively) are close to that of Fe\,{\sc ii} (16.2 eV).
These lines are not yet extensively studied likely because of their proximity to the NIR domain, which requires spectroscopic infrared observations when the redshift is larger than 0.1. CaT is a transition onto a metastable triplet produced by collisional excitation and likely emitted by a high-density, optically-thick neutral gas \citep{1988ApJ...330..751P, 2007ApJ...663..781M, 2013AstRv...8d...4M, 2015ApJS..217....3M}. O\,{\sc i}\,$\lambda$8446 can be emitted in less extreme conditions  \citep{2013AstRv...8d...4M, 2015ApJS..217....3M}, but its origin is not yet completely understood. The O\,{\sc i}\,$\lambda$8446 line is generally believed to be produced by Ly-$\beta$ fluorescence, that excites electrons from level 2p$^4$ 3P to the higher level 3d 3D$^0$. The other possible physical processes are collisional excitation and recombination. Their contribution to O\,{\sc i}\,$\lambda$8446 can be evaluated on the basis of the presence and intensity of O\,{\sc i}\,$\lambda$7774. 
The ratio 7774/8446 is expected to be $\sim$0.3 in case of collision \citep{1980ApJ...238...10G}, while it should be $\sim$1.1 in case of recombination \citep{2008ApJS..174..282L}.
A comparison among the kinematics of Fe\,{\sc ii}, O\,{\sc i} and CaT in UV, optical and NIR has been recently used to discuss the site of the Fe\,{\sc ii} emitting region, which seems to be the same of O\,{\sc i} and CaT \citep{2016arXiv160205159M} and located in an outer part of the BLR with respect to hydrogen \citep{2015ApJS..221...35K}. On the contrary, reverberation-mapping analyses of a small number of AGNs suggest that both Fe\,{\sc ii} and hydrogen lines are produced by the same ionized gas \citep{2015ApJ...804..138H}.

Seyfert 1 galaxies, and NLS1 in particular, are known to show often blue asymmetric profiles in their [O\,{\sc iii}]$\lambda\lambda4959,5007$ emission lines \citep{1981ApJ...247..403H,1985ApJ...294..106V,1985MNRAS.213....1W,1991ApJS...75..383V}. These profiles consist mainly of two components: 1) a narrow \textit{core} component (FWHM$\sim$200--500 km\,s$^{-1}$), believed to be emitted by gas located far from the active nucleus and whose kinematics is dominated by the gravitational potential of the host galaxy, as shown by \citet{1996ApJ...465...96N}; 2) a broad component (FWHM$\sim$500--1000 km\,s$^{-1}$), called \textit{wing}, believed to be emitted by gas located closer to the active nucleus and whose width is dominated by turbulent outflows in a direction perpendicular to the accretion disk   \citep{2001A&A...372..730V, 2005MNRAS.364..187B, 2008ApJ...680..926K, 2011ApJ...737...71Z}.
Furthermore, in some AGNs the high ionization emission lines show a systematic difference in velocity when compared to the low ionization lines. First detected by \citet{1976ApJ...208...37P} in I\,Zw\,1, this effect was studied more recently by several authors \citep[see e.g.][]{2002ApJ...576L...9Z,2003MNRAS.345.1133M,2005ApJ...618..601A,2005AJ....130..381B,2005MNRAS.364..187B,2008ApJ...680..926K,2011ApJ...737...71Z,2015arXiv151107138M,2015arXiv151202642Z}. AGNs with [O\,{\sc iii}] showing a velocity shift of $\Delta v<-250$ km s$^{-1}$ with respect to H$\beta$  were called \textit{blue outliers} by \citet{2002ApJ...576L...9Z}. More recently, \citet{2008ApJ...680..926K} applied a less-restrictive upper limit of $-150$ km s$^{-1}$. This blueshift could be due to an outflow associated with a disk wind \citep{2002ApJ...576L...9Z}, or with a hot decelerating wind \citep{2008ApJ...680..926K} combined with a source of opacity. \citet{1982ApJ...263...79G} found that also the high-ionization lines of BLR, and in particular C\,{\sc iv} line, are blueshifted by about 600 km\,s$^{-1}$ with respect to the rest-frame. 
A recent work by \citet{2013ApJ...769...30G} points out how much these questions about radial motions of ionized gas in AGNs are still open. On the basis of velocity--resolved reverberation mapping, these authors suggested that blueshifting in BLR could be due to Rayleigh scattering associated to inflowing gas towards the super-massive BH (SMBH). 

We addressed these open questions by selecting a sample of NLS1s and we focused our work on the following topics: 1) the analysis of the shape of the BLR H$\beta$ emission line to quantify how much a Lorentzian profile is typical in these AGNs and to investigate if the width of H$\beta$ is connected to other physical parameters of the active nucleus like the Eddington ratio; 2) the analysis of the BLR emission lines from Fe\,{\sc ii}, O\,{\sc i} and Ca\,{\sc ii}  to clarify if a strong Fe\,{\sc ii} emission is a distinctive property of NLS1s and to explore the physical properties of the low ionization gas; 
3) the analysis of the radial motions of ionized gas observed in the narrow-line region (NLR) to understand which are the main drivers and whether the blue wings are related to the shift of the core component. In Sections 2 and 3 we describe the selection of the sample of NLS1--candidate galaxies with its main properties, and the details of the spectroscopic analysis. The results of the analysis are shown in Section 4 and discussed in Section 5. In Section 6, we summarize the results of this work.

We adopt a cosmology with $H_0=70$ km s$^{-1}$ Mpc$^{-1}$, $\Omega_M=0.3$ and $\Omega_{\Lambda}=0.7$ to compare our results with the most recent papers.

\section{Sample}\label{sample}

We extracted our NLS1s sample from the archive of the Sloan Digital Sky Survey \citep[SDSS--DR7,][]{2009ApJS..182..543A}. We selected emission-line objects, classified as ``galaxies'' or ``QSO'' by SDSS, with redshift between 0.02 and 0.35 in order to have a spectral range including [O\,{\sc ii}]$\lambda$3727 and [S\,{\sc ii}]$\lambda\lambda$6717,6731. The FWHM of H$\beta$ was forced in the range 800--3000 km s$^{-1}$, the lower limit was decided on the basis of the measurements of Seyfert 2 and intermediate-type Seyfert galaxies by \citet{2012MNRAS.427.1266V}, and the upper limit was arbitrarily fixed larger than 2000  km s$^{-1}$ to avoid possible losses caused by inaccurate measurements.
Finally, we chose objects showing H$\beta$ and oxygen lines, which are prominent features in the optical spectra of AGNs, and with signal-to-noise ratio (S/N) of [O\,{\sc i}]$\lambda$6300 larger than 3. This last criterion was used because [O\,{\sc i}] is weak, therefore its detection ensures the presence of the other brighter emission lines with a good S/N. Furthermore, this emission line is stronger in AGNs than in H\,{\sc ii} regions because of the shape of the ionizing spectrum. In Section \ref{NLR_fitting} we discussed whether or not this criterion could introduce a bias.  

The query result is a sample of 2372 objects contaminated by LINERs, Seyfert 2 and intermediate-type Seyfert galaxies. We checked each spectrum by visual inspection, doing three iterations. At first, we set a flag to mark LINERs and Seyfert 2 galaxies, and another flag for intermediate-type Seyfert and NLS1--candidates. Then, we performed a second iteration by checking again each object of these sub-samples to be sure that our classification was correct. After a third final check, we got a list of 360 NLS1--candidates out of 2372 sources. 

\begin{figure}
\includegraphics[width=\columnwidth]{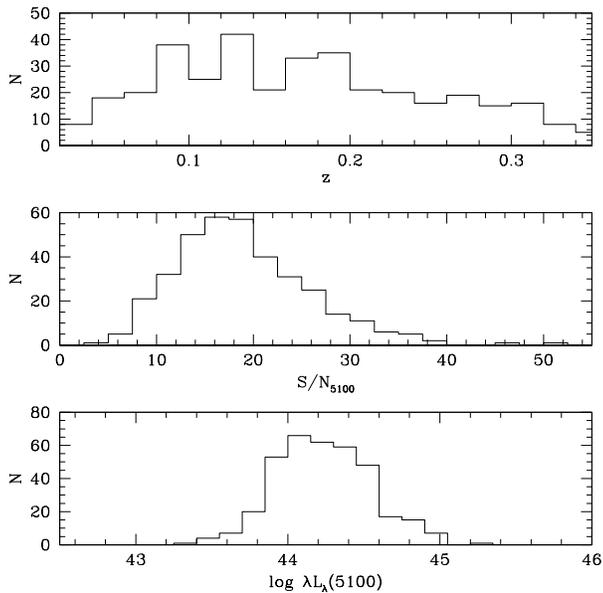} 
\caption{The histograms refer to the distributions of values of the redshift (top), S/N of the continuum at rest-frame 5100 \AA\ (middle) and the luminosity of the continuum at rest-frame 5100 \AA\ ($\lambda$L$_{\lambda}(5100)$ (bottom) for our sample of 360 NLS1--candidates.}
\label{z_hist}
\end{figure}

\begin{figure}
\includegraphics[width=\columnwidth]{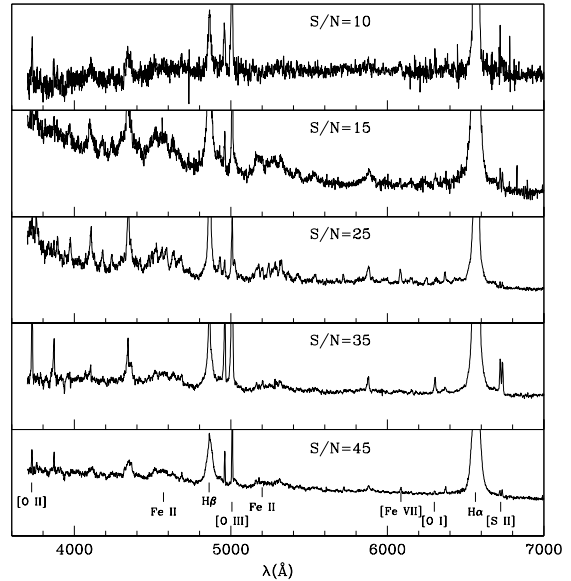} 
\caption{ Examples of spectra with different S/N ratio. From top to bottom: SDSS J091037.28+112523.8, SDSS J145840.41+195218.6, SDSS J120719.81+241155.8, SDSS J094842.67+502931.4, SDSS J133435.38+575015.6.}
\label{SN_spettri}
\end{figure}

Our data are homogeneously distributed in the range of redshift 0.028--0.345 (Fig.~\ref{z_hist}, top panel). 
The S/N of the continuum at rest-frame 5100 \AA\ varies between 4 and 51 (Fig.~\ref{z_hist}, middle panel) with median value of 19 and interquartile range (IQR) of 9. Finally, the logarithm of the luminosity of the continuum at rest-frame 5100 \AA\ ($\lambda$L$_{\lambda}(5100)$), in unit of erg s$^{-1}$, shows a range of values of 43.2--45.3 (Fig.~\ref{z_hist}, bottom panel) with a median value of 44.2 and an IQR of 0.43.
Fig.~\ref{SN_spettri} shows some examples of spectra with different values of S/N = 10, 15, 25, 35 and 45, from top to bottom. 

\section{Data Analysis}

The spectra were corrected for Galactic reddening using the {\sc iraf} task {\sc deredden} and the values of $V$--band extinction found in the NASA/IPAC Extragalactic Database (NED), then they were shifted to rest--frame  with the {\sc iraf} task {\sc newredshift} and their dispersion was modified to a value of 1 \AA\ px$^{-1}$ with the {\sc iraf} task {\sc dispcor}. After these preliminary procedures, we fitted narrow emission lines by means of {\sc pan} (Peak ANalysis)\footnote{http://ifs.wikidot.com/pan}, a public {\sc idl} software with a graphical interface based on {\sc mpfit} \citep{2009ASPC..411..251M}. 
After subtracting the underlying continuum of each line, we found that a single Gaussian function was suitable for fitting the [O\,{\sc ii}], [Fe\,{\sc vii}], [O\,{\sc i}] and [S\,{\sc ii}] doublets emission lines.
We did not use two Gaussians for the [O\,{\sc ii}] doublet because, although their separation of 3 \AA\ is at the resolution limit of our data, the two lines are not resolved and their summed profiles give rise to a single smoothed spectral line. 
To properly model the H$\beta$ and [O\,{\sc iii}] emission lines the Fe\,{\sc ii} multiplets lines had to be subtracted. This was accomplished by using the software\footnote{http://servo.aob.rs/FeII\_AGN/} developed by \citet{2010ApJS..189...15K} and \citet{2012ApJS..202...10S}. After the Fe\,{\sc ii} subtraction, we fitted the [O\,{\sc iii}] doublet by using first a single Gaussian function for each line and then by adding more components in order to fit asymmetric profiles. 
The H$\beta$ profile was fitted by means of a single Gaussian function for the narrow component (H$\beta_{\rm n}$) with FWHM, central wavelength and [O\,{\sc iii}]/H$\beta$ ratio fixed only when necessary, and a single Lorentzian function or 2 Gaussian functions for the broad component (H$\beta_{\rm b}$).
For consistency, O\,{\sc i} $\lambda 8446$ and CaT were fitted with the same function used for the corresponding H$\beta$ broad profile. 

Throughout this paper we consider a correlation strong when either the Pearson or the Spearman rank coefficients ($r$ and $r_{\rm s}$) are larger than 0.7 and weak when they are between 0.3 and 0.5. We consider a correlation significant when the probability that the correlation is not present, the  p-value, is lower than 0.001. We report the significance only for correlation coefficients lower than 0.7. The same threshold of 0.001 was used for the Kolmogorov-Smirnov (K--S) test.
To describe the statistical distributions, we use the median as typical value, and, as dispersion, the standard deviation in case of normal distributions and the interquartile range (that is, the difference between the upper and lower quartiles) in generic distributions.

\subsection{Error analysis}\label{errors}
The flux uncertainties were estimated assuming that the determination of the continuum level is the main source of error. Accordingly, we calculated the relative errors as the ratio between the intensity of the peak of each emission line over the continuum and the standard deviation of the continuum measured in a spectral region near the emission line. The median values are around 10--20 per cent for stronger lines, such as [S\,{\sc ii}] and [O\,{\sc iii}], up to 60--65 per cent for the weak [O\,{\sc i}].

The uncertainties of FWHM, density, velocity shift and continuum slope were estimated with the following method. 
For each quantity we reconstructed one or more model spectra obtained from our fitting and we added them to 100 randomly-selected spectra with different S/N, whose lines and continuum were previously subtracted.
The same fitting procedure was applied and the distributions of the measurements  were analysed. 
Almost normal distributions centred in the expected values were found.

For the FWHM of [S\,{\sc ii}] and of [O\,{\sc iii}] core and wing components we found standard deviations of the distributions corresponding to errors of 5 up to 10 per cent. 
For the density we obtained a percentage error of about 25 per cent. 
The measurements of the [O\,{\sc iii}] velocity shift have a normal distribution with $\sigma=10$ km\,s$^{-1}$. 

For the continuum slope, errors range between 10 and 50 per cent. The fitting of the slope can be affected by the host galaxy contribution and by the internal extinction. The estimate of the contribution of the former is controversial because of the lack of the absorption features. Many of them are suppressed by the non-thermal continuum or confused in the noise of the spectra. We measured the equivalent width (EW) of Ca\,K\,$\lambda3934.8$ in 157 out of our 360 spectra, and in 113 the values are smaller than 1.5 \AA. Only 44 objects show strong absorption features. We tested the effects of the galaxy spectrum using a stellar template with solar metallicity and age of 11 Gyr, following \citet{2015ApJ...804..138H}, and three AGN spectra whose continuum was fitted with power-law indices $\alpha=$ $-0.5$, $-1.0$ and $-1.5$. The stellar template was normalized and multiplied by scaling factors from 0.1 to 0.9 and the resulting spectra were subtracted from the three normalized AGN spectra. In this way we obtained spectra with a different contribution of the host galaxy with respect to the AGN continuum. We measured the slope and analysed the values of $\alpha$ as a function of the EW of Ca\,K. The average change of $\alpha$ values for low-EW objects is about 50 per cent, while it can be 2.5 times larger for high-EW objects. The extinction effects were estimated by using the typical A$(V)$ values found in \citet{2002ApJ...572...94R}. We calculated a median extinction A$(V)=0.53$ with a standard deviation of 0.31. By assuming A$(V)=0.5$ and the extinction law by \citet{1989ApJ...345..245C}, the slope becomes steeper of about $-0.5$, regardless of the input value of $\alpha$. 

For the FWHM of the Fe\,{\sc ii} multiplets the errors are about 5--12 per cent, for values between $\sim1000$ and $\sim2300$ km\,s$^{-1}$.
This can be justified by the fact that the multiplets are made by numerous transitions which are close in wavelengths, therefore the more the velocity dispersion increases, the more the global profile becomes smoothed and difficult to be precisely fitted. This effect can be even stronger when the EW of Fe\,{\sc ii} is small. An additional source of error could be the presence of the Fe\,{\sc ii} narrow emissions, which is not considered by the software. To evaluate their contribution it is definitely not an easy task. We made some tests using the list of transitions and relative intensities for the broad and narrow component of Fe\,{\sc ii} published by \citet{2004A&A...417..515V} (see their tables A.1 and A.2). We reconstructed four templates for the broad component, using Lorentzian functions having FWHM = 1000, 1300, 2000, 2700 km s$^{-1}$, and two templates for the narrow component, using Gaussian functions having FWHM = 300 and 600 km s$^{-1}$. These input values have been chosen on the basis of our measurements of H$\beta_{\rm b}$ and NLR emission lines. Each broad template was summed to 100 narrow templates having fixed FWHM and different intensities. To apply realistic scale factors, we used the data of 4178 objects from \citet[][their table 1]{2011ApJ...736...86D}. First, we calculated the intensity ratios between their narrow and broad Fe\,{\sc ii} fluxes, then we generated 100 random numbers within the interval of observed values (0.005--0.5), making sure that they had a distribution similar to the real one. Finally, we combined the 100 broad plus narrow templates with 100 Fe\,{\sc ii}-subtracted spectra and we applied the fitting software. 
The spectra obtained from the broad template alone were also fitted as a comparison. 
We calculated the ratios between the FWHMs obtained for spectra with only the broad component and those with broad plus narrow components. For each value of the broad component, the ratios do not depend on the value of the narrow component, whose width, indeed, does not affect the result of the fitting. By comparing different values of broad FWHM, we found that when the FWHM is approximately lower than $1300$\,km\,s$^{-1}$, the ratio is $\sim 1$ with a $\sigma$ of about 3--4 per cent, while for larger FWHMs (2000--2700\,km\,s$^{-1}$) the ratio increases systematically toward 1.1--1.2 with a $\sigma$ of about 9--16 per cent. In conclusion, for small values of FWHM the possible presence of a narrow component does not affect our measurements of the FWHM of Fe\,{\sc ii} multiplets. For larger values of FWHM, the multiples appear blurred, making the narrow component more visible and giving, as a result, smaller values of FWHM.

\subsection{Broad-line region}

\subsubsection{H$\beta$ fitting}\label{BLRem}
In order to better analyse the H$\beta$ line, both the [O\,{\sc iii}] and Fe\,{\sc ii} contributions must be subtracted. The analysis of the H$\alpha$ line was omitted because our spectra do not have both S/N and resolution high enough to fit its profile without introducing many constraints that would imply significant uncertainties. We decided to model the H$\beta$ broad component (H$\beta_{\rm b}$) with a Lorentzian function following \citet{2012MNRAS.426.3086G} and \citet{2011Natur.470..366K, 2013A&A...549A.100K}. This approach was successful for 335 sources, only 25 spectra showed strong asymmetries with red and/or blue wings, which could not be reproduced by a Lorentzian function. In these cases two Gaussians were used to fit the broad component. The contribution of the narrow component of H$\beta$ (H$\beta_{\rm n}$) was taken into account by using a constrained Gaussian function, whose FWHM was fixed according to the values measured for [O\,{\sc ii}]$\lambda$3727, [O\,{\sc i}]$\lambda$6300 and [S\,{\sc ii}]$\lambda\lambda$6717,6731. The central wavelength was let free to vary in 30 per cent of the spectra. In the remaining cases, we had to fix it according to the values of [O\,{\sc ii}] in order to improve the fitting. We applied an additional constraint, forcing the ratio [O\,{\sc iii}]$\lambda$5007/H$\beta_{\rm n}$ to vary in the range 2--20, a reasonable interval of values for the NLR of Seyfert galaxies \citep{2012MNRAS.427.1266V}. The flux ratio between the narrow and the broad component of H$\beta$ emission line has a distribution ranging from  0.01 to 0.34, with a median value of 0.05 and an IQR of 0.04. 323 out of 360 objects have H$\beta_{\rm n}$/H$\beta_{\rm b}$ lower than 0.1, therefore in most cases H$\beta_{\rm n}$ contributes typically with less than 10 per cent to the total flux of H$\beta$.

For H$\beta_{\rm b}$ fitted by two Gaussians, we obtained the FWHM after the correction for instrumental resolution. In case of a Lorentzian fitting, it was not possible to obtain the intrinsic FWHM by deconvolving the instrumental width from the measured one. Therefore, we decided to convolve Lorentzian functions having FWHM from 5 to 40 \AA\ with a Gaussian function with FWHM = 2.7 \AA\ (the instrumental resolution at $\lambda=4861$ \AA). We measured the width of these profiles and we calculated the relation to convert from the observed FWHM to the intrinsic one ($ \rm FWHM_{int}=1.015 \times FWHM_{obs} - 0.673$). We applied this relation to correct for the instrumental resolution and we converted the values into velocities.

The sample of NLS1s was defined using the criterion of FWHM(H$\beta_{\rm b}$) $< 2000$ km s$^{-1}+ 10$ per cent. The tolerance was chosen on the basis of the instrumental width (167 km s$^{-1}$).
In support of this choice, we checked how the FWHM measurement changes with decreasing S/N. We created a Lorentzian profile with FWHM=2000 km s$^{-1}$ and we added noise, in order to obtain spectra having H$\beta_{\rm b}$ with S/N from 5 to 45 with step of 5. 
These spectra have been fitted with the same method used for our data and the measured values of FWHM are 2300 km s$^{-1}$ for S/N = 5,  2100  km s$^{-1}$ for S/N = 10, up to 2000 km s$^{-1}$  for S/N = 45. The S/N of H$\beta_{\rm b}$ for the sample of spectra fitted with a Lorentzian function ranges between 6 and 87 and its distribution has a median value of approximately 24 with IQR of 16. 
Therefore, we can be confident that our upper limit of 2200 km s$^{-1}$ is acceptable and that the measured values of FWHM do not strongly depend on the fitting and on the quality of the spectra.

By applying our criterion, we selected 296 NLS1s out of 360 candidates. We classified the remaining 64 objects as BLS1s, but we did not consider them as a real control sample. We stress that this sample is not representative of the whole class of BLS1s, having FWHM values too close to those of NLS1s (FWHM(H$\beta_{\rm b}$)$\sim$2200--3700\,km\,s$^{-1}$). 

The black hole masses was estimated by assuming that the BLR clouds are virialized \citep{1999ApJ...526..579W} and by means of the usual formula:
\begin{equation}
  M_{\rm BH} = \frac{R_{\rm BLR}~ v^2}{G} .
\end{equation}
The radius of the BLR ($R_{\rm BLR}$) was obtained through the H$\beta_{\rm b}$ luminosity ($L({\rm H\beta_b}$))\citep{2010ApJ...723..409G}:
\begin{equation}
\log \frac{R_{\rm BLR}}{10~ {\rm lt-days}} = {\rm A_2} \times \log \frac{{L \rm(H\beta_{\rm b})}}{10^{43}~ {\rm erg~ s^{-1}}} + {\rm B_2},
\end{equation}
where A$_2=0.53\pm0.04$, B$_2=0.85\pm0.05$. The velocity $v$ was obtained from the FWHM(H$\beta_{\rm b}$) assuming an isotropic distribution of the BLR clouds ($v=\frac{\sqrt{3}}{2}\,{\rm FWHM}$). This choice was made to be consistent with other authors like \citet{2006ApJS..166..128Z} and \citet{2012AJ....143...83X}.
In order to recover the rotational component of the FWHM, it would be necessary to remove the turbulent one. Since this procedure is not trivial \citep{2013A&A...549A.100K} and we needed to compare our results with those of other authors, we preferred not to apply any model-dependent correction to our FWHM measurements.

It is not simple to estimate the uncertainties of the measurements of the black hole mass which is calculated as a function of the gas velocity dispersion and of the BLR radius. We can assume an uncertainty of 10 per cent for the FWHM, depending on the fitting procedure and the S/N. The radius was obtained from a relation depending on the $L({\rm H\beta_b})$, which is affected by the flux calibration errors and by the intrinsic variability of the sources. On the basis of the light curves of 17 AGNs published by \citet{2000ApJ...533..631K}, we inferred a median L(H$\beta_{\rm b}$) variation of 20 per cent and we assumed this value as a typical error due to the use of a single epoch measurement. Then we calculated the $R_{\rm BLR}$ error and we obtained a median value of 30 per cent. By adding the velocity and radius errors, the BH mass error is 50 per cent, corresponding to about 0.2 dex. Taking into account the highest luminosity variations observed in  \citet{2000ApJ...533..631K} (40 per cent) and the highest BLR radius errors (50 per cent), the uncertainty of the BH mass grows to 0.3 dex. This is however a lower limit, because it does not account for the photometric calibration error. We mention that other authors \citep{2001A&A...377...52W,2004ApJ...606L..41G,2007ApJ...667L..33K} preferred to assume a conservative upper limit of 0.5 dex.

From the BH mass of each object, we calculated the Eddington luminosity, $L_{\rm Edd} = 1.3 \times 10^{38} M_{\rm BH}/{\rm M_{\odot}}$ erg s$^{-1}$, and then the Eddington ratio between the bolometric luminosity ($L_{\rm bol}$) and the Eddington luminosity, assuming  $L_{\rm bol} = 9\times \lambda L_{\lambda}$(5100) \citep{2000ApJ...533..631K}. The uncertainty of the Eddington ratio depends on the continuum luminosity variations, the BH mass errors, the bolometric correction chosen to calculate the bolometric luminosity and the flux calibration error. As done for BH mass, we estimated the errors of continuum luminosity from \citet{2000ApJ...533..631K} and we obtained a median value of 27 per cent. By assuming an error of 50 per cent for the BH masses, the resulting lower limit for the uncertainty of the Eddington ratio is about 0.33 dex.

\subsubsection{Fe\,{\sc ii} multiplets fitting}
A precise fitting and subtraction of the Fe\,{\sc ii} spectrum is mandatory to analyse the [O\,{\sc iii}]+H$\beta$ spectral region. The code we used creates a template with 65 Fe\,{\sc ii} emission lines with Gaussian profile. The lines are identified as the strongest within 4000--5500 \AA\ spectral range and are sorted in 5 line groups. The input has to be a continuum-subtracted spectrum in the range 4000--5500 \AA. Furthermore, it needs guesses for 8 free parameters: the width and shift of the Gaussian functions, the intensity of the 5 groups of multiplets, and the excitation temperature. Intensities and widths were chosen according to the values of each spectrum. In particular, by assuming that Fe\,{\sc ii} multiplets are emitted mostly by the BLR, we used  rough estimates of the H$\beta$ width as guessing values of the Fe\,{\sc ii} one. 

\begin{figure}
\centering
\includegraphics[width=\columnwidth]{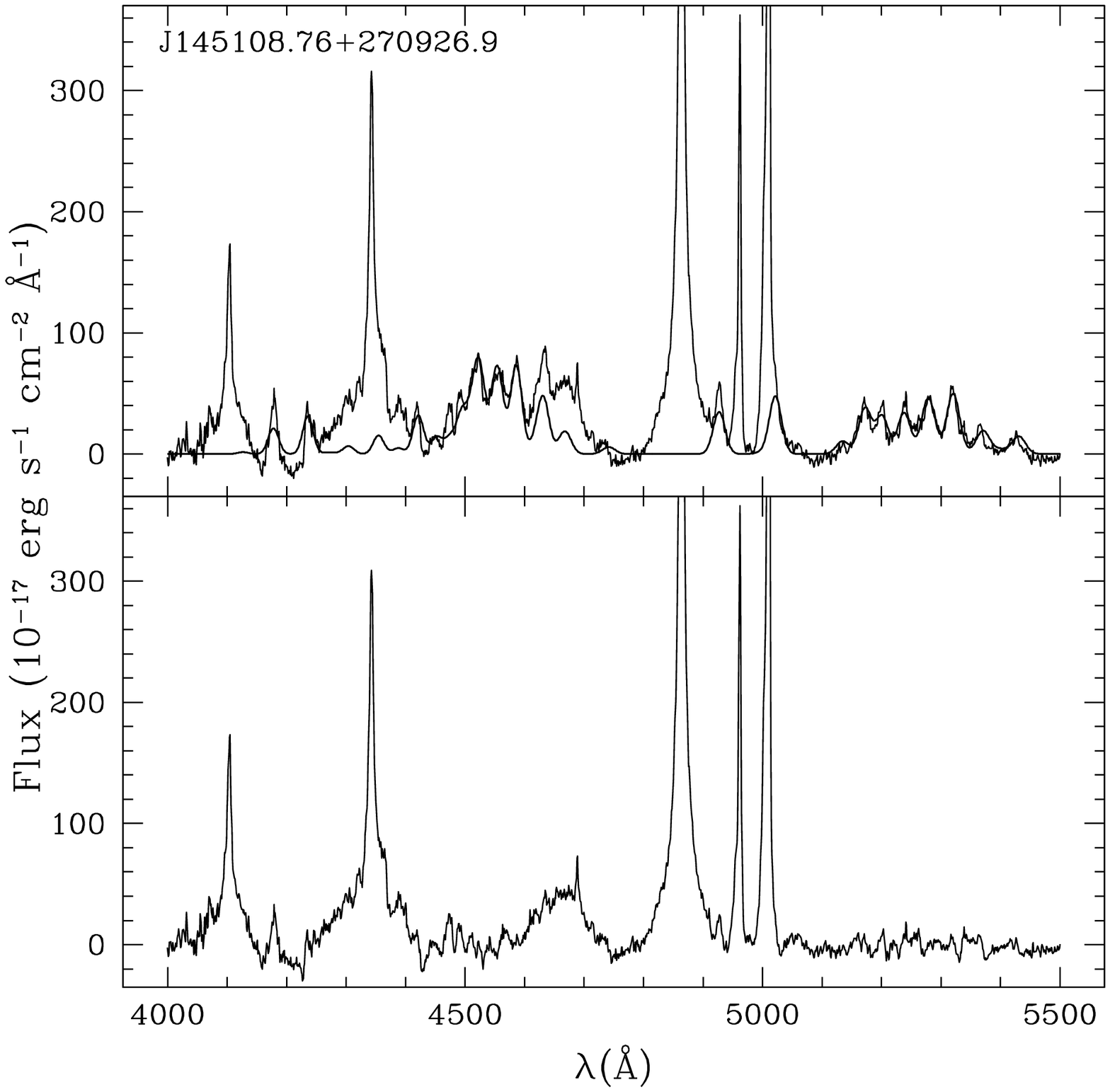} 
\includegraphics[width=\columnwidth]{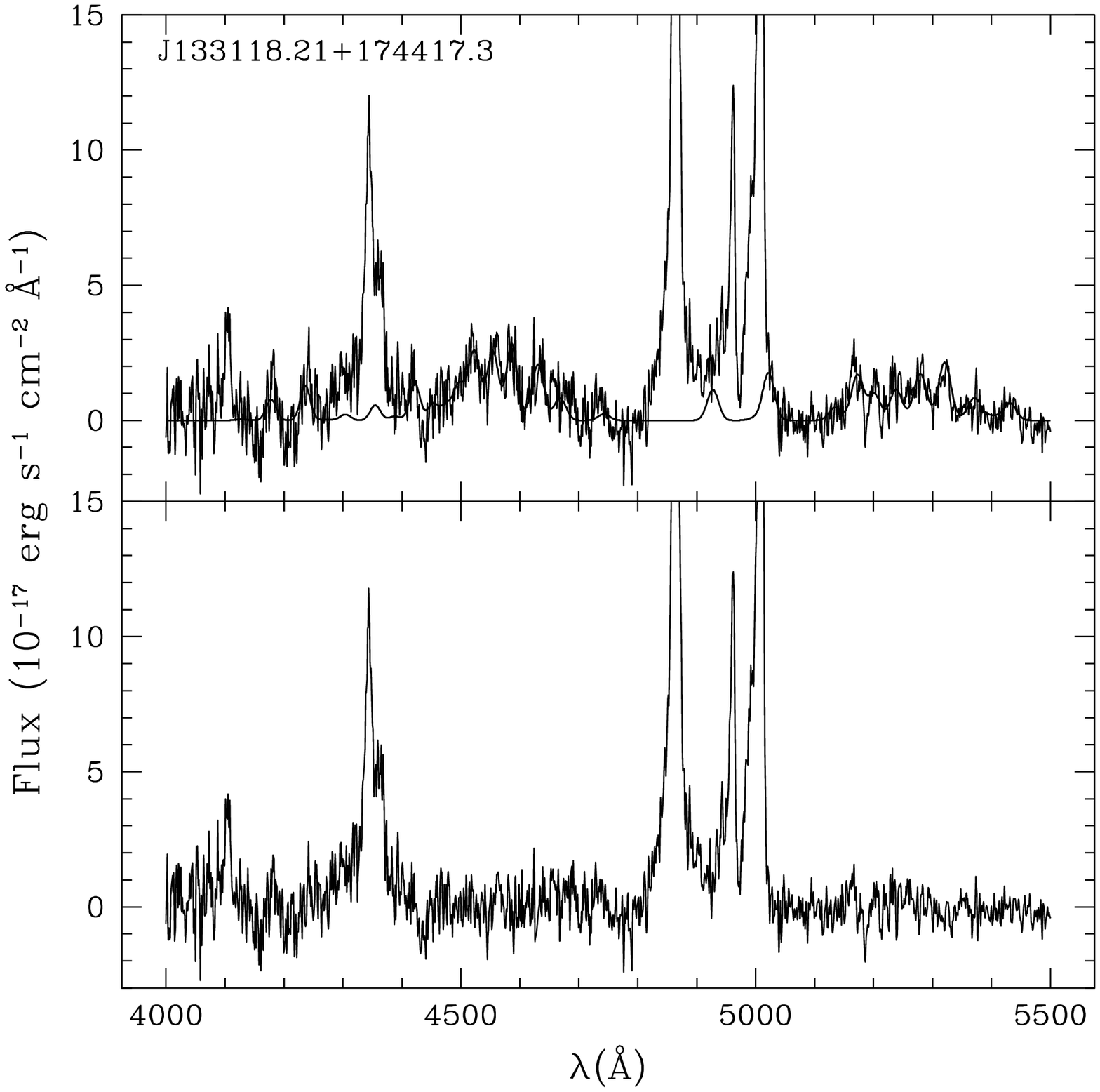} 
\caption{Two examples of the Fe\,{\sc ii} multiplets fitting. The top graph shows the case of a spectrum with S/N=50, while the bottom graph with S/N=11.
The top panel of each graph shows the original spectrum with the fitted solution overlaid, while in the bottom panel the residual spectrum is shown.}
\label{ferro_prova}
\end{figure}

For each object the outputs consist of best-fitting parameters, the spectrum of each group of Fe\,{\sc ii} multiplets and the total Fe\,{\sc ii} spectrum, which was subtracted from the original one. We evaluated the quality of the results by checking if the residuals were comparable to the noise of each spectrum. Fig.~\ref{ferro_prova} shows two examples of results for S/N = 50 (top panels) and S/N = 11 (bottom panels). The model for the high S/N spectrum is very good and also in the other case the overall trend is reproduced quite well. 
The error in flux determination due to the S/N of spectra was about 10 per cent for S/N=10 and 20 per cent for S/N=5. Since only 27 objects of our sample have a spectrum with S/N$<$10, we can reasonably assume that typical errors are lower than 10 per cent.

We point out that this software does not take into account the possible presence of Fe\,{\sc ii} narrow multiplets \citep{2004A&A...417..515V,2008ApJ...674..668W,2011ApJ...736...86D}. This simplification does not considerably affect the flux measurements, because the intensity of the narrow component is often negligible with respect to that of the broad one. Indeed, the median value of the ratio between the narrow and the broad component is 0.045 with IQR of 0.074 \citep{2011ApJ...736...86D}. By contrast, our estimate of the Fe\,{\sc ii} FWHM could be affected by the presence of narrow multiplets. We have discussed their influence on our measurements in Section \ref{errors}.

\subsubsection{O\,{\sc i} and CaT fitting}
We measured O\,{\sc i}\,$\lambda$8446, O\,{\sc i}\,$\lambda$7774 and CaT assuming that they are emitted only by the BLR, therefore we applied the same functions used for fitting the corresponding H$\beta$ broad profile obtaining a good agreement. O\,{\sc i}\,$\lambda$8446 was included in the spectral range of 67 out of 360 objects while at least the first line of CaT was included in 52 galaxies. We could reliably fit O\,{\sc i} in 41 spectra, while in the other 26 the line was at the detection limit or even not detected. O\,{\sc i}\,$\lambda$7774 was found in only 8 out of the 214 spectra with a spectral range including this line.
The Ca\,{\sc ii} lines were detected and measured in 15 objects, in 2 out of 15 we could measure Ca\,{\sc ii}\,$\lambda$8498,8542, while in 4 only Ca\,{\sc ii}\,$\lambda$8498 was found. In 4 cases the lines were detected neither in absorption nor in emission, while in 33 objects the S/N was low or the CaT was present only in absorption. 
When necessary, we fitted CaT after the subtraction of the stellar population contribution. We obtained a template using a single stellar population with solar metallicity and age of 11 Gyr \citep{2015ApJ...804..138H}. 

The narrow O\,{\sc i} emission line was not considered because its detection is not yet confirmed. It must be noticed that recently \citet{2008ApJS..174..282L} claimed to have measured for the first time the narrow component of O\,{\sc i}\,$\lambda$8446 in 8 out of 23 type 1 AGNs observed in NIR. The typical intensity ratio of narrow to broad O\,{\sc i}\,$\lambda$8446 was 0.1 and in a couple of cases even 0.2 and 0.5. In support of their result, they cited \citet{1985MNRAS.215P..57M}, who detected O\,{\sc i}\,$\lambda$8446 in two Seyfert 2 galaxies, NGC 5506 and NGC 7314. We point out that both NGC 5506 and NGC 7314 are currently believed to be peculiar NLS1s \citep{2002A&A...391L..21N,2011A&A...535A..62E}, so up to date no Seyfert 2 galaxies are known to show O\,{\sc i}\,$\lambda$8446. 
Even if we cannot exclude a priori the existence of O\,{\sc i}\,$\lambda$8446 emitted by the NLR, we believe that spectra with much higher S/N ratio are mandatory before obtaining a conclusive result on this question, especially in case of BLS1s, when O\,{\sc i}\, plus CaT profiles can be extremely difficult to fit.

\subsection{Narrow-line region}\label{NLR_fitting}

\begin{figure}
\includegraphics[width=\columnwidth]{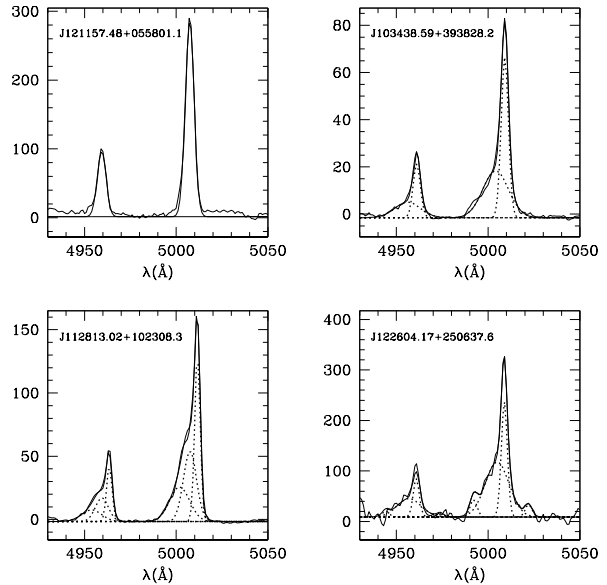} 
\caption{From the top-left to the bottom-right panel: examples of [O\,{\sc iii}] doublet fitting with one, two, three and four Gaussian functions.}
\label{sp_strani}
\end{figure}

To study  the gas conditions of the NLR we have analysed the following emission lines: [O\,{\sc ii}]$\lambda$3727, [O\,{\sc iii}]$\lambda\lambda$4959,5007, [Fe\,{\sc vii}]$\lambda$6087,  [O\,{\sc i}]$\lambda6300$ and [S\,{\sc ii}]$\lambda\lambda$6717,6731. 
Firstly, we carefully subtracted the continuum underlying each line by considering a small spectral range centred on the line. Then, for [O\,{\sc ii}], [Fe\,{\sc vii}] and [O\,{\sc i}] emission lines we used a single Gaussian function letting all the parameters free to vary.
The FWHM of the [S\,{\sc ii}] lines was forced to be the same, while the centre of [S\,{\sc ii}]$\lambda$6731 was obtained as a function of the centre of [S\,{\sc ii}]$\lambda$6717.
Finally, we measured the [O\,{\sc iii}] doublet using, as first guess, a single Gaussian function for each line with the following constraints: the same FWHM, the centre of [O\,{\sc iii}]$\lambda$4959 to be obtained as a function of the centre of [O\,{\sc iii}]$\lambda5007$, the flux ratio [O\,{\sc iii}]\,$\lambda$4959/$\lambda$5007 fixed to 1/3. Then, we checked the residuals and we applied a multi-Gaussian fitting with the same constraints, where necessary.
We obtained a good result with a single Gaussian function in only 34 out of 360 objects, while 317 objects required two Gaussian functions, one narrower (called \textit{core component}) and the other broader (called \textit{wing component}).
In seven cases, three components were used and in a single extreme case, SDSS J122604.17+250637.6, four components were applied to fit a really complicated profile. Fig.~\ref{sp_strani} shows examples of fitting of [O\,{\sc iii}] doublet with one to four Gaussian functions, from the top-left panel to the bottom-right one. Only in one case, SDSS J090236.98+353023.3, it was not possible to fit the [O\,{\sc iii}] lines, because of the low quality of the spectrum and the non-detection of these spectral features.
Summarizing, we fitted the [O\,{\sc iii}] line profiles with at least two components in 325 out of 359 spectra. 
In order to check the reliability of our measurements, we first compared the fluxes of the narrow emission lines. 
As expected, very good correlations were found between the fluxes of [O\,{\sc iii}], [O\,{\sc ii}], [O\,{\sc i}] and [Fe\,{\sc vii}] against the fluxes of the [S\,{\sc ii}] doublet, taken as a reference (the values of the linear correlation coefficient, $r$, are 0.89, 0.93, 0.92, and 0.81, respectively).
The median errors, calculated as explained in Section~\ref{errors}, are around 10--20 per cent for stronger lines, such as [S\,{\sc ii}], [O\,{\sc ii}] and [O\,{\sc iii}], up to 60--65 per cent for the weak [O\,{\sc i}].

As mentioned in Section~\ref{sample}, we need to verify whether or not the criterion of [O\,{\sc i}]$\lambda$6300 S/N $> 3$ used for the sample selection could have caused a contamination of objects ionized by shock waves. Shocks have been invoked as an additional mechanism to explain unusually high [O\,{\sc i}]$\lambda$6300/H$\beta$ ratios, even larger than 2, observed in some LINERs \citep[see e.g.][]{1983ApJ...264..105F}. They can have a role also in the NLR of AGNs affecting emission line fluxes and their profiles \citep[see e.g.][and references therein]{2001ApJS..132..211C}. However the actual contribution to the formation of the spectrum when compared to photoionization is still a matter of debate. 
In a photoionization context, [O\,{\sc i}]$\lambda$6300 is produced in the partially-ionized recombination region of an ionized gas cloud, and its strength is sensitive to the shape of the ionizing spectrum and to the cloud geometrical thickness. In particular, the larger number of high-energy photons emitted by the power-law continuum of an AGN, with respect to a blackbody stellar continuum, expands this partially-ionized region making the [O\,{\sc i}]$\lambda$6300 stronger by a factor of 10 in AGNs than in H\,{\sc ii} or star-forming regions. The predicted ratios [O\,{\sc i}]$\lambda$6300/H$\beta$ for a power-law spectrum with $\alpha=-1.5$ and assuming a gas with $N_{\rm e}=10^3$\,cm$^{-3}$ are between 0.1 and 1.0, depending on the value of the ionization parameter $U$ \citep{1998AN....319..325R}. In our sample, [O\,{\sc i}]$\lambda$6300/H$\beta$ ratios extend up to 2.25, but their distribution has median value of 0.45 and IQR of 0.4. A correction for reddening, assuming $A_V = 1$\,mag, makes this ratio 25 per cent lower. 
Since the H$\beta$ narrow flux was obtained from a fit of a broad-line dominated profile, we additionally compared [O\,{\sc i}]$\lambda$6300/[O\,{\sc iii}]$\lambda$5007 and [O\,{\sc ii}]$\lambda$3727/[O\,{\sc iii}]$\lambda$5007 ratios of our targets with the O$_{123}$ diagnostic diagram published by \citet{2012MNRAS.427.1266V}. Our values fall in the region occupied by the NLR of Seyfert galaxies, well distant from that of LINERs. Therefore, we can state that even if it is not possible to exclude a priori the presence of shocks, our sample is not biased by the criterion we have adopted, since the ionization of the NLR of our targets is dominated by the AGN power-law continuum. Furthermore, the above mentioned correlation between [O\,{\sc i}] and [S\,{\sc ii}] fluxes confirms that photoionization is the prevalent mechanism at work.

The electron density was estimated by using the observed [S\,{\sc ii}]$\lambda6717$/[S\,{\sc ii}]$\lambda6731$ ratio and the 5-level atom approximation implemented in the {\sc iraf} task {\sc temden}. This intensity ratio works effectively in the range 0.5--1.4, that is between $N_{\rm e}=10$ cm$^{-3}$ and $N_{\rm e}=10^4$ cm$^{-3}$, and it is weakly dependent on the electron temperature. Unfortunately, we could not directly determine the value of the temperature from the usual ratio [O\,{\sc iii}]$\lambda$5007+[O\,{\sc iii}]$\lambda$4959/[O\,{\sc iii}]$\lambda$4363, because [O\,{\sc iii}]$\lambda$4363 was not clearly detectable. We decided to adopt a temperature of $T = 10^4$~K. We obtained values of density for 293 out of 360 objects, and for 238 out of the 296 NLS1s and for 55 BLS1s. In the other cases the intensity ratios were outside the range or the lines were too weak to be measured. The median electron density is 360 cm$^{-3}$ with an IQR of 349 cm$^{-3}$ for the whole sample. The error in density, measured as explained in Section~\ref{errors}, is about 25 per cent. In order to check the reliability of our result, we first calculated the density values for $T=1.5\times10^4$~K and $T=2\times10^4$~K. The medians of the distributions are 420 cm$^{-3}$ and 470 cm$^{-3}$, respectively, well within the uncertainty. Therefore, we decided to use the values obtained with $T = 10^4$ K. Then, we calculated the S/N of the [S\,{\sc ii}] doublet and we obtained that it is larger than 5 in 232 out of 293 spectra the S/N([S\,{\sc ii}]). Thus, the lines were strong enough to be detected and measured. Finally, we evaluated the possible dependence of the density values on the S/N of the continuum at 5100 \AA\ and on redshift, and we did not find any correlation or trend. In conclusion, no systematic errors or biases are expected. 
 
We measured the ratio between the [O\,{\sc ii}]$\lambda$3727 and [O\,{\sc iii}]$\lambda$5007 fluxes and we calculated the ionization parameter $U$ using the formula given by \citet{1990A&A...236...53P}:
\begin{equation}
\log U = -2.74 - \log[{\rm O\,II}]/[{\rm O\,III}] .
\end{equation}
The distribution of $\log U$ values ranges between $-3$ and $-1.6$ with a median value of $-2.1$.
Since we did not apply any internal extinction correction, we tried to quantify the effect of reddening on the [O\,{\sc ii}]/[O\,{\sc iii}] ratio. In particular, the changes between the intrinsic ratio ($R_{\rm int}$) and the observed one ($R_{\rm obs}$) were calculated as a function of different values of extinction. According to \citet{2012MNRAS.427.1266V} the distribution of the extinction for intermediate-type Seyfert is peaked at about $A_V=1$. Therefore, values of extinction lower than 1 are expected. By applying the \citet{1989ApJ...345..245C} extinction law, the ratio $R_{\rm int}/R_{\rm obs}$ varies between 1.21 for $A(V)=0.5$, to 1.47 for $A_V=1.0$, which turns into a decrease of $\log U$ between 0.08 and 0.17 dex.

The FWHM of narrow emission lines were calculated after the correction for instrumental resolution (R = 1800). Their widths are consistent and significantly correlated, both considering low-ionization lines and the core component of [O\,{\sc iii}] ($r$ is between 0.34 and 0.59, with p-value $\ll 10^{-8}$). Median and IQR values for the distributions of each emission line are comparable: 305 and 263 km\,s$^{-1}$, 305 and 157 km\,s$^{-1}$, 273 and 133 km\,s$^{-1}$, for [O\,{\sc ii}], [S\,{\sc ii}] and [O\,{\sc iii}], respectively. [O\,{\sc ii}] has a distribution shifted toward higher values with a median FWHM of 442 km\,s$^{-1}$ and an IQR of 105 km\,s$^{-1}$, probably because the line profile is the overlap of two unresolved components. 
We measured the errors of low and high ionization lines using [S\,{\sc ii}] and [O\,{\sc iii}], as explained in Section~\ref{errors}, finding errors of 5 up to 10 per cent. 

The [O\,{\sc iii}] profile was analysed by means of the asymmetry parameter A20 \citep{1985MNRAS.213....1W}. We carried out the calculations of values and relative errors ($\sigma_{A20}$) as in \citet{2011MNRAS.418.2630C} and \citet{2012MNRAS.427.1266V}. The asymmetry is defined as:
\begin{equation}
A20 = \frac{a-b}{a+b} ,
\end{equation}
where $a=\lambda_{50}-\lambda_{10}$ and $b=\lambda_{90}-\lambda_{50}$, and $\lambda_{10}$, $\lambda_{50}$, $\lambda_{90}$ are the wavelengths corresponding to the 10, 50 and 90 per cent of the spectral line area. Negative values indicate a red asymmetry, while positive values a blue asymmetry. 

We measured the shift of the [O\,{\sc iii}] core component with respect to the velocity of the low-ionization line [S\,{\sc ii}]$\lambda6717$, $\Delta v = v_1 - v_{\rm [S\,II]}$, and also the shift of the other emission lines having increasing ionization degree: [O\,{\sc i}]$\lambda6300$, [O\,{\sc ii}]$\lambda3727$ and [Fe\,{\sc vii}]$\lambda6087$. In this case, negative values indicate blueshifts, while positive values indicate redshifts.

\section{Results on broad line region} \label{BLR}

\subsection{H$\beta$ emission}

We fitted 291 NLS1s and 44 BLS1s with a Lorentzian function, while the double-Gaussian was used only in 5 NLS1s and in 20 BLS1s. This result suggests that NLS1s typically have symmetric profiles, which can be successfully fitted by a Lorentzian function, and it confirms that a Lorentzian function is suited for type 1 AGNs having FWHM(H$\beta_{\rm b}$) $< 4000$ km s$^{-1}$ \citep{2000ApJ...536L...5S}. 

\begin{figure}
\includegraphics[width=\columnwidth]{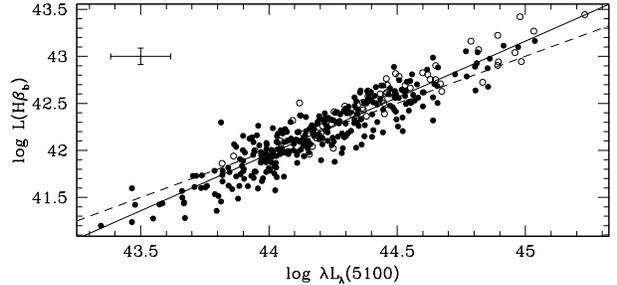} 
\caption{The relation between $\lambda$L$_{\lambda}(5100)$ and $L({\rm H\beta_b})$. Solid circles are NLS1s, while open circles are BLS1s. The solid line is the best fitting, the dashed line is the fitting with slope fixed at 1.}
\label{Lc_LHb}
\end{figure}

We plotted the $\lambda$L$_{\lambda}(5100)$ against the $L({\rm H\beta_b})$ (Fig.~\ref{Lc_LHb}) and we obtained the following strong correlation ($r=0.91$):

\begin{equation}\label{eq:Lc_Lhb}
\log L({\rm H\beta_b}) = {\rm A_1} \times \log \lambda L_{\lambda}(5100) + {\rm B_1} ,
\end{equation}
where A$_1=1.203\pm 0.026$ and B$_1=- 10.980\pm 1.159$.
This result is expected, in case of photoionization and for optical spectra not dominated by beamed continuum associated with the radio-jet. It is in good agreement with \citet{2006ApJS..166..128Z} and also with \citet{2002MNRAS.337..275C}, who found that the Balmer line equivalent widths (EWs) are positively correlated with continuum luminosity. 
We also found a correlation between the luminosity of H$\beta_{\rm b}$ and its EW, with $r_{\rm s}=0.69$ and p-value $=1.4\times10^{-51}$, stronger than in \citet{2001A&A...372..730V}, who obtained $r=0.43$.  
\begin{figure}
\includegraphics[width=\columnwidth]{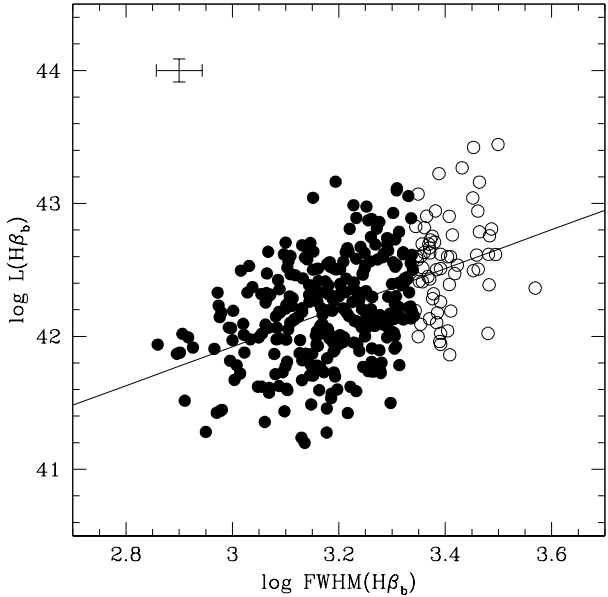} 
\caption{The  H$\beta_{\rm b}$ luminosity plotted against the H$\beta_{\rm b}$FWHM. Symbols are the same as in Fig.~\ref{Lc_LHb}. The solid line is the best fitting.}
\label{fig10a_veron}
\end{figure}

Finally, we plotted the FWHM(H$\beta_{\rm b}$) against the $L({\rm H\beta_b})$ and we obtained a significant but weak correlation ($r_{\rm s}=0.46$, p-value $= 8.1\times10^{-20}$), which could be possibly related to our limited range of luminosities (Fig.~\ref{fig10a_veron}). Indeed, \citet{2001A&A...372..730V} found a stronger correlation ($r=0.76$), probably because of their larger range of luminosities.
This correlation is somehow expected. In fact, the H$\beta_{\rm b}$ luminosity is proportional to the continuum luminosity, which in turn depends on the $M_{\rm BH}$ and accretion rate ($\dot{m}$), and the kinematics of the BLR is dominated by the $M_{\rm BH}$.
The observed scatter could be  caused by the additional contributions of BLR radius and $\dot{m}$.

\subsection{Black hole masses and Eddington ratios}

\begin{figure}
\includegraphics[width=\columnwidth]{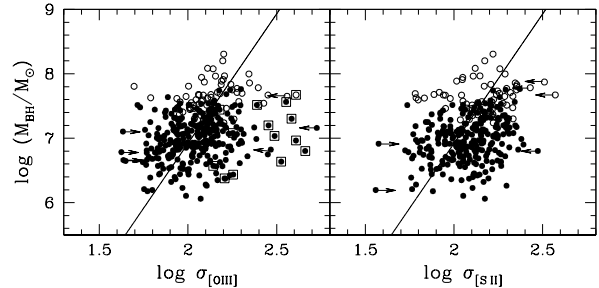} 
\caption{$M_{\rm BH}$-$\sigma$ relation obtained using the velocity dispersion of the core component of [O\,{\sc iii}]$\lambda5007$ (left panel), and of [S\,{\sc ii}]$\lambda6717$ (right panel). The solid line is the \protect\citet{2002ApJ...574..740T} relation. Solid circles are NLS1s, while open circles are BLS1s. Open squares indicate the blue outliers. Arrows represent lower or upper limit values (see text for details). }
\label{BHsigma}
\end{figure}

The BH masses of our sample of NLS1s range from $\log (M_{\rm BH}/{\rm M_{\sun}}) =$ 6.1 to $\log (M_{\rm BH}/ {\rm M_{\sun}})=$ 7.8, with a median value of 7.00 and an IQR of 0.43, while for our sample of BLS1s they range from 7.2 to 8.3, with  median and IQR values of 7.65 and 0.33. We plotted the $M_{\rm BH}$--$\sigma$ relation for our data using the gas velocity dispersion obtained by means of both the [S\,{\sc ii}] and the core component of [O\,{\sc iii}], in place of the stellar velocity dispersion (Fig.~\ref{BHsigma}).
Our data are in agreement with those published by  \citet{2006ApJS..166..128Z} and \citet{2007ApJ...667L..33K}, who found similar ranges, and are distributed around the \citet{2002ApJ...574..740T} relation. 
Their spread was estimated by analysing the distributions of their distances from this correlation. The median values are about 0.11 dex both for [O\,{\sc iii}] and [S\,{\sc ii}], and the IQRs are 0.13 dex for [O\,{\sc iii}] and 0.14 dex for [S\,{\sc ii}]. In Fig~\ref{BHsigma}, arrows indicate lower and upper limits, while squares indicate the \textit{blue outliers} (see Section \ref{bshift}). The four upper limits of [S\,{\sc ii}] lines are due to actually broad profiles, whose correct fitting would imply a double-Gaussian approach for each line of the doublet. However, the S/N of these spectra does not allow to properly fit the [S\,{\sc ii}] lines with two Gaussian functions, so we decided to use a single Gaussian even if this choice results in a broader FWHM. The [O\,{\sc iii}] lines show only two upper limits that are too noisy to be correctly fitted with two Gaussian functions each, resulting in FWHMs larger than 850 km\,s$^{-1}$. The other upper limits are due to red or blue outliers. In these cases, according to \citet{2007ApJ...667L..33K}, radial motions are at work, causing a broadening of the line whose FWHM cannot be related only to the galaxy gravitational potential.   
For both lines, lower limits are typically spectra showing complex emission line profiles due to a low S/N.  

In agreement with previous results \citep[see e.g.][and references therein]{2007ApJ...670...60X}, we found high accretion rates in our sample of NLS1s: 154 out of 296 are beyond the Eddington limit. The distribution of $\frac{L_{\rm bol}}{L_{\rm Edd}}$  ratios for NLS1s ranges from 0.23 to 4.75, with a median value of 1.03 and an IQR of 0.84. On the contrary, the ratios for BLS1s are confined to lower values: they vary from 0.17 to 1.04, with a median of 0.45 and an IQR of 0.21. Recently, \citet{2012AJ....143...83X} found values of the Eddington ratio for NLS1s between 0.25 and 2, while for BLS1s between 0.04 and 0.5. For comparison, the Eddingtion ratios of about 105000 quasar published by \citet{2011ApJS..194...45S} span a wider range from about 0.001 to about 10. 

On the basis of the different broad line shapes found in NLS1s with respect to BLS1s, we decided to investigate if the H$\beta$ profile is related to the Eddington ratio. We considered the distributions of $L_{\rm bol}/L_{\rm Edd}$ values of the samples of 335 galaxies fitted by a Lorentzian and the 25 galaxies fitted by a double-Gaussian. The median ratios of  $\log L_{\rm bol}/L_{\rm Edd}$ are $-0.03$ and $-0.3$, respectively. The K-S test confirms that the two distributions are different (p-value $= 2\times10^{-7}$). Since 291 out of 335 are NLS1s, this result reinforces the idea that NLS1s are characterized by both Lorentzian shape profiles and high Eddington ratios.

\subsection{Fe\,{\sc ii} emission} \label{FeII_emission}

\begin{figure}
\includegraphics[width=\columnwidth]{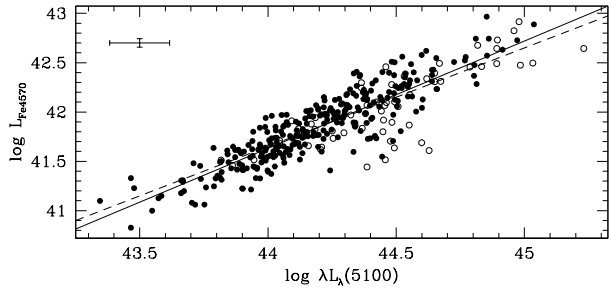} 
\caption{The relation between the luminosity of Fe4570 and the $\lambda L_{\lambda}(5100)$. Symbols and lines are the same as in Fig.~\ref{Lc_LHb}. The solid line is the best fitting, the dashed line is the fitting with slope fixed at 1.}
\label{Lc_LFe4570}
\end{figure}

We integrated the Fe\,{\sc ii} spectrum in two intervals: the first, called Fe4570, between 4434 and 4684 \AA, the second, called Fe5200, between 5190 and 5320 \AA. The luminosity of Fe4570 turns out to be strongly correlated with the continuum luminosity ($r = 0.88$) (Fig.~\ref{Lc_LFe4570}):

\begin{equation}
\log L_{\rm Fe4570} = {\rm A_3} \times \log \lambda L_{\lambda}(5100) + {\rm B_3} ,
\end{equation}
where A$_3=1.090\pm 0.031$ and B$_3=- 6.310\pm 1.358$, 

\begin{figure}
\includegraphics[width=\columnwidth]{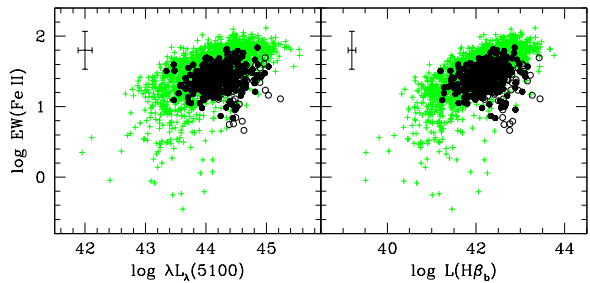} 
\caption{The equivalenth width of Fe\,{\sc ii} plotted against the luminosity of both continuum at 5100 \AA\ (left panel) and the broad component of H$\beta$ (right panel). Solid circles are NLS1 galaxies, while open circles are BLS1 galaxies. Green plus symbols are data from \citet{2006ApJS..166..128Z}. }
\label{baldwin_zhou}
\end{figure}

We plotted EW(Fe4570) against $\lambda L_{\lambda}(5100)$ and $L({\rm H\beta_b})$, finding correlation coefficients of $r_{\rm s} = 0.18$ with p-value $\sim 6\times10^{-4}$. We cannot confirm the results from \citet{2006ApJS..166..128Z}, who found positive correlations between EW(Fe4570) and L(5100) and L(H$\beta_{\rm b}$), probably because our selection criteria limited the luminosity values to a smaller range. However, as visible in Fig.~\ref{baldwin_zhou}, our data are in agreement with theirs, falling within their range and showing a similar spread.
\citet{2006ApJS..166..128Z} called their result \textit{inverse Baldwin effect} and explained it in terms of a hardening of the ionizing spectrum with increasing luminosity. The existence of these correlations does not support the shock hypothesis, because photoionization would be able to sufficiently heat the weakly-ionized region where Fe\,{\sc ii} is expected to be present. 

We explored the dependence of EW(Fe4570) on the continuum slope.
The continuum slope was measured by means of the {\sc iraf} task {\sc nfit1d}, using a power-law $F(\lambda)=k_*(\lambda/\lambda_{\rm ref})^{\alpha}$,  where $\lambda_{\rm ref}=4800$ \AA, $k_*$ is the flux at $\lambda=\lambda_{\rm ref}$ and $\alpha$ is the spectral index. 
The errors, estimated as explained in Section~\ref{errors}, are between 10 and 50 per cent and larger values are found for nearly flat slopes.
This is expected because, in these cases, small changes can result in a significant difference in the measured values.
The distribution of the spectral index $\alpha$ is quite broad. It ranges between $-1.81$ and 0.09 with a median value of $-0.81$ and an IQR of 0.57 for the whole sample. The median value for the 296 NLS1s is $-0.78$ with an IQR of 0.52, and for the 64 BLS1s is $-1$ with an IQR of 0.7. By comparison, \citet{2000ApJS..126...63R} found typical values of $-0.59$, $-0.90$ and $-0.79$, for NLS1s, Seyfert 1s and the whole sample, respectively. 

We did not find any correlation ($r_{\rm s}=-0.09$, p-value $=0.074$) between $\alpha$  and EW(Fe4570) that is, the strength of Fe\,{\sc ii}  may not depend on the shape of the nuclear spectrum, if we can extrapolate from our results on the optical continuum and this contradicts the idea that photoionization can justify by itself the Fe\,{\sc ii} emission. However, to correctly estimate the slope, we should consider the contribution of both the internal extinction and the host galaxy, which are very difficult to evaluate in type 1 AGNs. The tests we made (see Section~\ref{errors}) indicate that spectra with flatter slopes show higher values of EW(Ca\,K), that is they are more affected by the host galaxy and require larger correction factors, which make them steeper. A correction for internal extinction  acts in the same direction but independently on the slope. In conclusion, both these corrections contribute to make the distribution of $\alpha$ values narrower, confirming the lack of a correlation with EW(Fe4570).

We calculated the R4570 ratio between Fe4570 and the total H$\beta$ flux \citep{2001A&A...372..730V}. The distribution of R4570 for the whole sample ranges between 0.05 and 1.21, and has a median of 0.45 with an IQR of 0.30. The same values for the 296 NLS1s are 0.49 and 0.29, and for the 64 BLS1s are 0.32 and 0.25. Similar ranges were reported by \citet{2007ApJ...670...60X}, who used the same R4570 definition and found a larger average value of 0.75 for NLS1s, but 0.30 for BLS1s. \citet{2001A&A...372..730V} found values mostly lower than 1.4 with a median of 0.63.
To evaluate the influence of Fe4570 and H$\beta$ on R4570 values, we divided our sample in two sub-samples, the first one including the objects having R4570 $\geq$ 0.5 (sample A, 153 objects) and the second one including those having R4570 $<$ 0.5 (sample B, 207 objects). The median value of  EW(Fe4570) for the sample A is $31.5$ \AA\ and for the sample B is $21.3$ \AA. We applied the K--S test to the two distributions and we found that they are significantly different (p-value $\sim 10^{-15}$). Similarly, the median value of EW(H$\beta$) for the sample A is $47.5$ and for the sample B is $69.3$. Again, the distributions are significantly different, according to the values of K--S test (p-value $\sim 10^{-16}$). We can conclude that high values of R4570 are a combination of both high EW(Fe4570) and low EW(H$\beta$), while the contrary happens for objects with low values of R4570.  Actually, we obtained significant correlations between R4570 and both EW(Fe\,{\sc ii}) with $r_{\rm s} = 0.64$ (p-value $= 1.5 \times 10^{-43}$), and EW(H$\beta$) with $r_{\rm s} = -0.60$ (p-value $=4.1 \times 10^{-37}$). Therefore, NLS1 galaxies with stronger Fe\,{\sc ii} emission seem to have also weaker H$\beta$ emission. This is partially in disagreement with \citet{1985ApJ...291..112G}, who suggested that large R4570 ratios are due to weak Balmer emission, rather than to strong Fe\,{\sc ii} emission.

\begin{figure}
\includegraphics[width=\columnwidth]{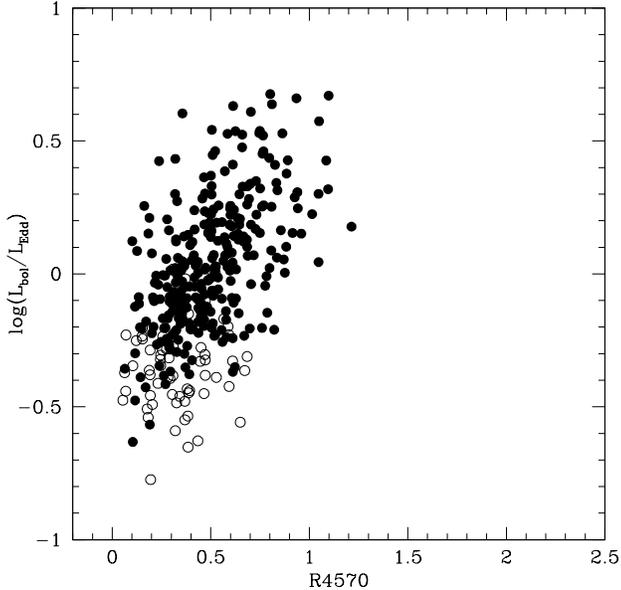} 
\caption{The values of R4570 are plotted against the logarithm of the Eddington ratio. Symbols are the same as in Fig.~\ref{Lc_LHb}.}
\label{R4570_Ledd}
\end{figure}

Fig.~\ref{R4570_Ledd} shows the plot of R4570 versus the Eddington ratio. We found that these two quantities appear well correlated, as confirmed by the values of the statistical test, $r_{\rm s} = 0.56$ with p-value $=1.8 \times 10^{-30}$. 
On the basis of a sample of almost 10000 type 1 SDSS AGNs having  $ z < 0.75$, \citet{2007ApJ...654..754N} obtained already a significant correlation between R4570 and $L/L_{\rm Edd}$ and they suggested that since the metallicity of the BLR seems to depend on the accretion rate, R4570 could be affected by the iron abundance. 

\begin{figure}
\includegraphics[width=\columnwidth]{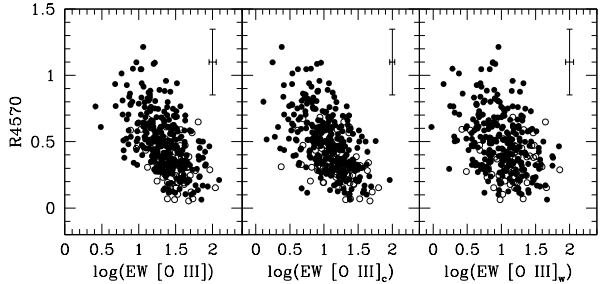} 
\caption{The values of R4570 are plotted against the logarithm of the equivalent widths of  [O\,{\sc iii}]$\lambda$5007. $c$ and $w$ indicate the core and the wing component, respectively. Symbols are the same as in Fig.~\ref{Lc_LHb}.}
\label{R4570_EWO3}
\end{figure}

We obtained a clear anti-correlation between R4570 and EW([O\,{\sc iii}]$\lambda 5007$), the well-known Eigenvector 1 (EV1) \citep[e.g.,][]{1992ApJS...80..109B}, and between R4570 and the equivalent widths of the core and wing components of [O\,{\sc iii}] (Fig.~\ref{R4570_EWO3}). This result indicates that EV1 applies well also to our sample.
This is confirmed by the statistical test: the correlations exist for both the total and the [O\,{\sc iii}] core component ($r_{\rm s}=-0.52$ and p-value $=1.6 \times 10^{-26}$, $r_{\rm s}=-0.53$ and p-value $=2.9 \times 10^{-27}$, respectively), while for the wing the correlation is weaker ($r_{\rm s}=-0.34$ and p-value $=3.3 \times 10^{-10}$).
We found also a weak but significant anti-correlation between R4570 and FWHM(H$\beta$), both considering the whole sample ($r_{\rm s} =-0.46$, p-value $=6.1 \times 10^{-20}$) and the NLS1s ($r_{\rm s} = -0.40$, p-value $=5.7 \times 10^{-13}$).

\begin{figure}
\includegraphics[width=\columnwidth]{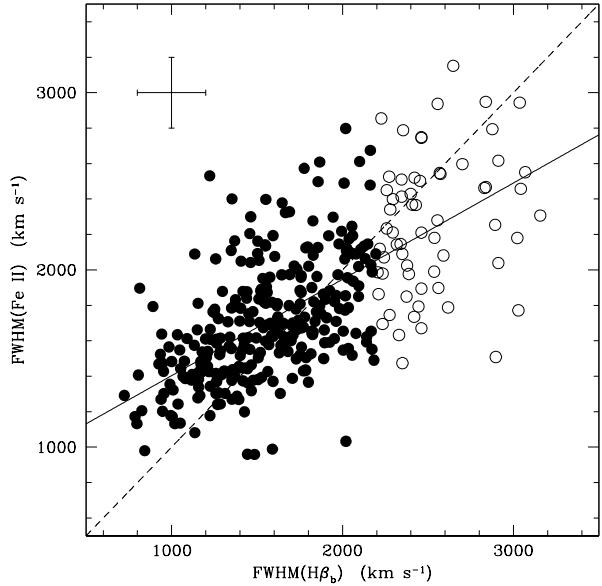} 
\caption{The FWHM of H$\beta_{\rm b}$ plotted against the FWHM of Fe\,{\sc ii}. The solid line is the best fitting, the dashed line is the 1:1 relation. Symbols are the same as in Fig.~\ref{Lc_LHb}.}
\label{FW_veron}
\end{figure}

Finally we plotted the FWHM(H$\beta_{\rm b}$) versus the FWHM(Fe\,{\sc ii}) and we found a significant correlation ($r=0.66$ and p-value $=2 \times 10^{-46}$,  see Fig.~\ref{FW_veron}). 
The data are distributed around the 1:1 relation, suggesting that the kinematics of iron and hydrogen are in agreement. 
When the width of the BLR lines is approximately lower than 1500 km s$^{-1}$, the FWHM of Fe\,{\sc ii} appears to be larger than that of H$\beta_{\rm b}$. In these cases the Fe\,{\sc ii} multiplets are showing narrower lines with a Lorentzian shape, therefore, a Gaussian fitting leads to an overestimate of the Fe\,{\sc ii} FWHM. 
For larger values, the fitting errors are larger and, as explained in Section \ref{errors}, the software cannot correctly fit the spectra and we expected to obtain smaller values of the FWHM.

\subsection{O\,{\sc i}\,$\lambda$8446 and Ca\,{\sc ii} triplet}
 
\begin{table}
\caption{O\,{\sc i}\,$\lambda8446$ sample. (1)\,Name of the object; (2)\,ratio between the O\,{\sc i} and H$\beta_{\rm b}$ fluxes; (3)\,ratio between the Fe4570 and H$\beta_{\rm b}$ fluxes; (4)\,FWHM of O\,{\sc i} in km\,s$^{-1}$; (5)\,FWHM of the broad component of H$\beta$ in km\,s$^{-1}$.}
\label{tabOI}
\begin{tabular}{ccccc}
\hline
  \multicolumn{1}{c}{Name} &
  \multicolumn{1}{c}{O\,{\sc i}/H$\beta_{\rm b}$} &
  \multicolumn{1}{c}{Fe4570/H$\beta_{\rm b}$} &
  \multicolumn{1}{c}{O\,{\sc i}} &
  \multicolumn{1}{c}{H$\beta_{\rm b}$} \\
\hline
   J014644.82$-$004043.1 & 0.10 & 0.45 & 701 & 1353\\
   J030417.78+002827.3 & 0.16 & 0.75 & 662 & 1253\\
   J075245.60+261735.7 & 0.11 & 0.66 & 574 & 1281\\
   J081324.00+542236.7 & 0.27 & 0.64 & 1057 & 1647\\
   J083949.65+484701.4 & 0.09 & 0.51 & 854 & 1533\\
   J090821.73+522103.0 & 0.15 & 0.94 & 557 & 1150\\
   J092343.00+225432.6 & 0.10 & 0.21 & 966 & 2276\\
   J092603.26+124403.7 & 0.09 & 0.33 & 1520 & 2234\\
   J093643.12+505249.5 & 0.13 & 1.09 & 363 & 1504\\
   J093701.04+010543.7 & 0.11 & 0.36 & 990 & 1350\\
   J094842.67+502931.4 & 0.36 & 0.56 & 1010 & 1850\\
   J095707.16+243316.1 & 0.15 & 0.82 & 363 & 1166\\
   J101645.11+421025.5 & 0.15 & 0.49 & 1084 & 1511\\
   J101718.25+291434.0 & 0.21 & 0.51 & 1033 & 1717\\
   J101906.79+231837.7 & 0.10 & 0.38 & 786 & 1441\\
   J103438.59+393828.2 & 0.37 & 0.69 & 802 & 814\\
   J105151.03+213726.4 & 0.07 & 0.49 & 761 & 1619\\
   J110424.06+314727.7 & 0.18 & 0.63 & 800 & 1439\\
   J112813.02+102308.3 & 0.19 & 0.67 & 739 & 1226\\
   J113246.66+062624.4 & 0.18 & 0.62 & 527 & 990\\
   J113629.37+213545.8 & 0.16 & 0.45 & 1352 & 2561\\
   J113913.91+335551.1 & 0.29 & 0.54 & 1493 & 1268\\
   J120719.81+241155.8 & 0.26 & 0.72 & 923 & 1019\\
   J122355.78+282701.8 & 0.12 & 0.50 & 456 & 934\\
   J122604.17+250637.6 & 0.22 & 0.72 & 466 & 724\\
   J122903.50+294646.0 & 0.20 & 0.50 & 501 & 925\\
   J130803.04+035114.6 & 0.12 & 0.57 & 599 & 1081\\
   J132255.43+080941.6 & 0.21 & 0.81 & 1127 & 1614\\
   J140234.44+215951.5 & 0.17 & 0.56 & 755 & 1515\\
   J140259.03+172056.0 & 0.18 & 1.25 & 567 & 1204\\
   J144431.62+153643.2 & 0.11 & 0.26 & 1295 & 1907\\
   J144932.70+002236.3 & 0.18 & 0.64 & 587 & 1242\\
   J145108.76+270926.9 & 0.17 & 0.69 & 472 & 1167\\
   J151108.27+045116.7 & 0.20 & 0.62 & 1046 & 1639\\
   J153141.25+201929.9 & 0.13 & 0.42 & 831 & 1210\\
   J154257.49+030653.3 & 0.36 & 0.83 & 471 & 891\\
   J155909.63+350147.5 & 0.18 & 0.92 & 441 & 797\\
   J160334.48+203546.8 & 0.29 & 0.99 & 845 & 1116\\
   J162012.75+400906.1 & 0.10 & 0.39 & 636 & 1348\\
   J171304.46+352333.5 & 0.19 & 1.05 & 571 & 1393\\
   J221918.53+120753.1 & 0.16 & 1.13 & 314 & 1053\\
\hline\end{tabular}
\end{table}

The ratios between the O\,{\sc i}\,$\lambda$8446 and H$\beta_{\rm b}$ fluxes are distributed in the range 0.07--0.37, with a median of 0.17 and an IQR of 0.08. 
Table~\ref{tabOI} reports O\,{\sc i}/H$\beta_{\rm b}$ and Fe4570/H$\beta_{\rm b}$ ratios, and the FWHM of both O\,{\sc i} and H$\beta_{\rm b}$.
In Table~\ref{tabOItot}, the properties of O\,{\sc i}\,$\lambda7774$ are listed. The O\,{\sc i}\,7774/8446 ratios vary between 0.04 and 0.42 with a median value of 0.1. 
\begin{table}
\caption{ O\,{\sc i}\,$\lambda7774$ sample. (1)\,Name of the object; (2)\,ratio between the O\,{\sc i}\,$\lambda7774$ and of O\,{\sc i}\,$\lambda8446$ fluxes; (3)\,ratio between the O\,{\sc i}\,$\lambda7774$ and H$\beta_{\rm b}$ fluxes.}
\label{tabOItot}
\begin{tabular}{ccc}
\hline
  \multicolumn{1}{c}{Name} &
  \multicolumn{1}{c}{O\,{\sc i}\,7774/O\,{\sc i}\,8446} &
  \multicolumn{1}{c}{O\,{\sc i}\,7774/H$\beta_{\rm b}$} \\
\hline
   J092343.00+225432.6 & 0.42 & 0.04\\
   J093701.04+010543.7 & 0.10 & 0.01\\
   J101645.11+421025.5 & 0.26 & 0.04\\
   J105007.75+113228.6 & -- & 0.01\\
   J113246.66+062624.4 & 0.23 & 0.04\\
   J121342.81+422743.0 & -- & 0.01\\
   J132255.43+080941.6 & 0.26 & 0.06\\
   J133205.28+312736.4 & -- & 0.05\\
   J145108.76+270926.9 & 0.06 & 0.01\\
   J153141.25+201929.9 & 0.04 & 0.01\\
   J155909.63+350147.5 & 0.10 & 0.02\\
   J171304.46+352333.5 & 0.07 & 0.01\\
\hline\end{tabular}
\end{table}

\begin{figure}
\centering
\includegraphics[width=\columnwidth]{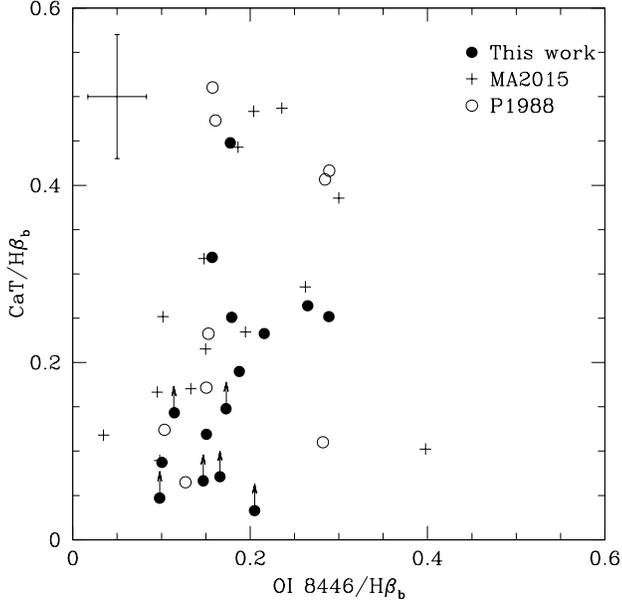} 
\caption{Plot of O\,{\sc i}\,$\lambda8446$/H$\beta_{\rm b}$ versus CaT/H$\beta_{\rm b}$ ratios. Solid circles are the data from this work, empty circles are the data taken from \citet{1988ApJ...330..751P} and plus symbols are the data taken from \citet{2015ApJS..217....3M}. The arrow indicates a lower limit.}
\label{oicat}
\end{figure}

\begin{table}
\caption{Ca\,{\sc ii} triplet sample. (1)\,Name of the object; (2)\,ratio between the Ca\,{\sc ii} triplet and H$\beta_{\rm b}$ fluxes; (3)\,ratio between the O\,{\sc i}\,$\lambda8446$ and H$\beta_{\rm b}$ fluxes; (4)\,FWHM of the Ca\,{\sc ii} triplet in km\,s$^{-1}$; (5)\,FWHM of the broad component of H$\beta$ in km\,s$^{-1}$.}
\label{tabCaII}
\begin{tabular}{ccccc}
\hline
  \multicolumn{1}{c}{Name} &
  \multicolumn{1}{c}{CaT/H$\beta_{\rm b}$} &
  \multicolumn{1}{c}{O\,{\sc i}/H$\beta_{\rm b}$} &
  \multicolumn{1}{c}{CaT} &
  \multicolumn{1}{c}{H$\beta_{\rm b}$} \\
\hline
J014644.82$-$004043.1	& 0.05$^1$ & 0.10 &	1324 &	1353 \\
J030417.78+002827.3	 & 0.32\phantom{$^1$}	& 0.16	& 1218 &	1253 \\
J075245.60+261735.7	 & 0.14$^1$ &	0.11 &	1253 &	1281 \\
J092343.00+225432.6 &	0.09\phantom{$^1$}	& 0.10 &	2101 &	2276 \\
J095707.16+243316.1	 & 0.07$^1$ & 	0.15 &	617 &	1166 \\
J101645.11+421025.5	 & 0.12\phantom{$^1$}	& 0.15 &	1520 &	1511 \\
J112813.02+102308.3 & 0.19\phantom{$^1$} & 0.19 & 1003 & 1226\\
J113913.91+335551.1 & 0.25\phantom{$^1$} & 0.29 & 1080 & 1268\\
J120719.81+241155.8 & 0.26\phantom{$^1$} & 0.26 & 816 & 1019\\
J122604.17+250637.6 &	0.23\phantom{$^1$} &	0.22 &	784 &	724 \\ 
J140234.44+215951.5 &	0.15$^2$ &	0.17 &	1395 &	1515 \\
J140259.03+172056.0	 & 0.45\phantom{$^1$} &	0.18 &	1209 &	1204 \\
J145108.76+270926.9 & 0.07$^2$ & 0.17 & 683 & 1167\\
J151108.27+045116.7 &	0.03$^1$ &	0.21 &	1395 &	1639 \\
J155909.63+350147.5 & 0.25\phantom{$^1$} & 0.18 & 813 & 797\\
\hline
\end{tabular}

\medskip
$^1$The flux was measured only for Ca\,{\sc ii}\,$\lambda8498$; $^2$the flux was measured only for Ca\,{\sc ii}\,$\lambda8498,8542$.
\end{table}
The CaT/H$\beta_{\rm b}$ ratio was obtained for 15 objects. The average ratio, excluding objects where we could measure only one or two lines of the triplet, is 0.24. 
CaT/H$\beta_{\rm b}$ with the corresponding O\,{\sc i}\,$\lambda8446$/H$\beta_{\rm b}$ ratios, and FWHMs of CaT and H$\beta_{\rm b}$ are listed in Table~\ref{tabCaII}.
We plotted O\,{\sc i}/H$\beta_{\rm b}$ versus CaT/H$\beta_{\rm b}$, adding data of 14 quasars from \citet{2015ApJS..217....3M} and of 9 out of 40 Seyfert 1 galaxies and QSOs for which \citet{1988ApJ...330..751P} claimed to have a clear detection of CaT  (Fig.~\ref{oicat}). The arrows indicate lower limits of CaT/H$\beta_{\rm b}$ in objects for which only one or two lines of the CaT could be measured (see Table~\ref{tabCaII}).
It seems that no correlation exists between these two quantities, but we need new good quality data to draw any conclusion.
However, this plot suggests that a strong CaT is always coupled to a strong O\,{\sc i}.
Obviously, these values are reddening dependent. By assuming $A_V=0.5$ and the extinction law by \citet{1989ApJ...345..245C}, as in Section \ref{FeII_emission}, these ratios become lower by an average factor of 25 per cent.

\begin{figure}
\includegraphics[width=\columnwidth]{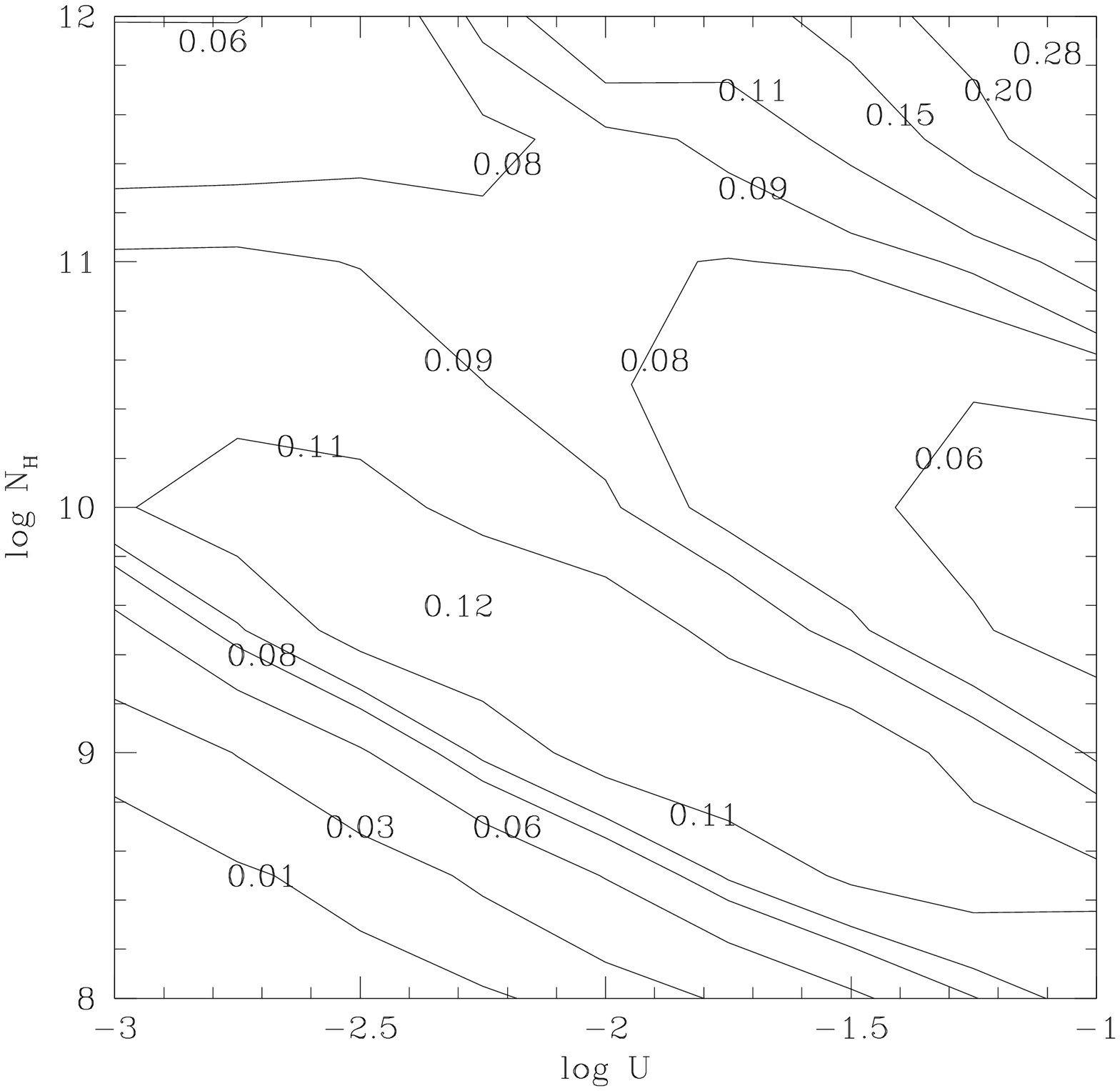} 
\includegraphics[width=\columnwidth]{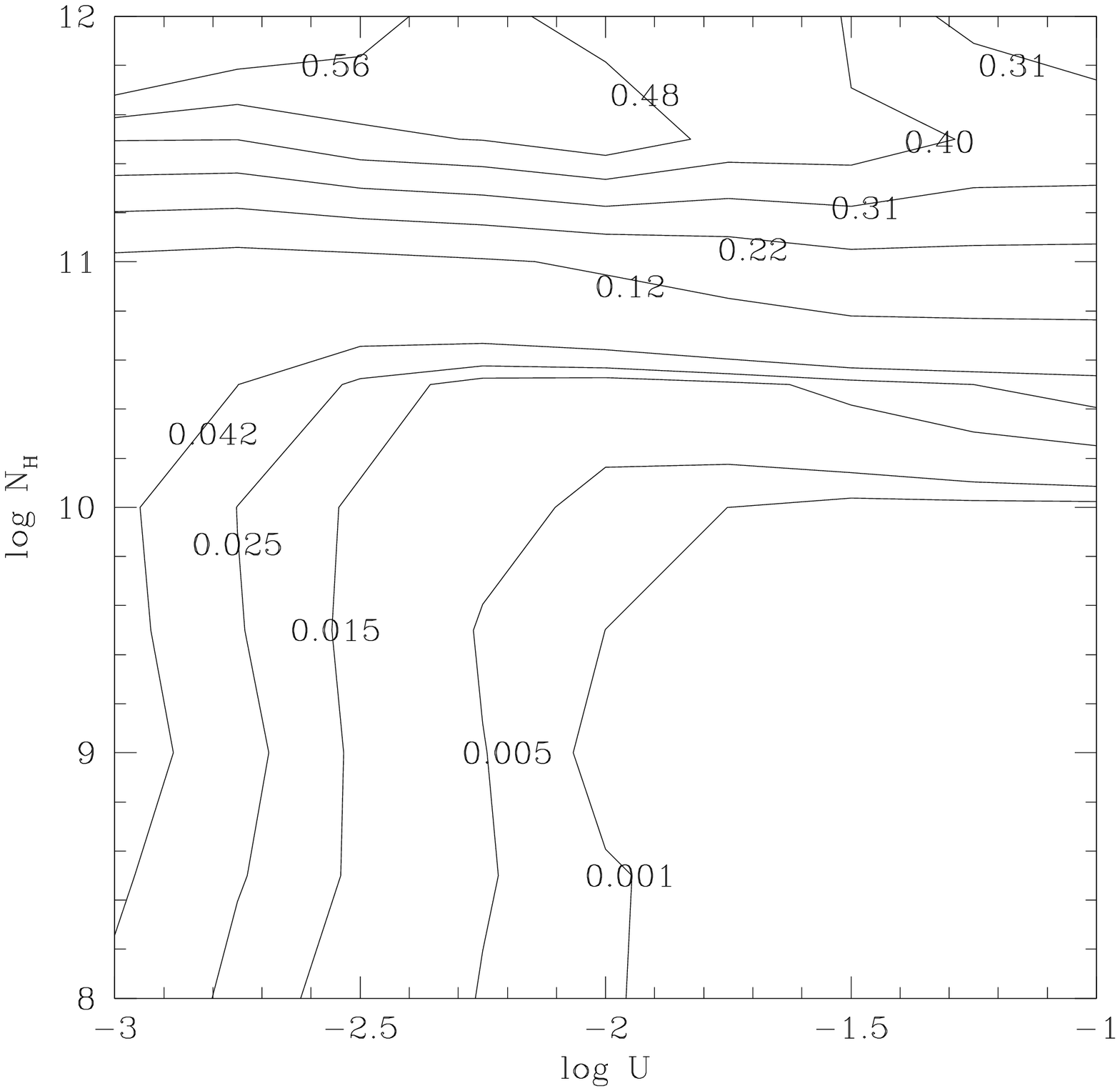} 
\caption{The solid lines indicate the values of the O\,{\sc i}/H$\beta_{\rm b}$ (upper panel) and CaT/H$\beta_{\rm b}$ (lower panel) ratios as a function of density and ionization parameter, obtained with {\sc cloudy} as explained in the text.}
\label{modelli}
\end{figure}

We decided to re-explore the gas physical conditions necessary to emit these lines through photoionization models obtained with the {\sc cloudy} code, version 13.03 \citep{2013RMxAA..49..137F}. As input parameters, we assumed a typical AGN continuum, solar abundances \citep{1991sia..book.1227G}, and we let free to vary the ionization parameter,  $-3 \leq \log U \leq -1$ with step of 0.25, and the hydrogen density, $8 \leq \log N_{\rm H} \leq 12$ with step of 0.5, in order to produce a grid of models. 
Fig.~\ref{modelli} shows how the O\,{\sc i}/H$\beta$ (upper panel) and CaT/H$\beta$ (lower panel) ratios vary as a function of $\log U$ and $\log N_{\rm H}$. Our results are in agreement with \citet{2007ApJ...663..781M} and \citet{2015ApJS..217....3M}. 
It appears evident that a high density, $\log N_{\rm H} \simeq 11$, is mandatory to have a CaT/H$\beta > 0.1$, almost independently from the value of the ionization parameter. Only the large ratios (CaT/H$\beta > 0.5$) require $\log U < -2.5$. 
More puzzling is the situation of O\,{\sc i}/H$\beta$. Many different combinations of $\log U$ in the considered interval and $\log N_{\rm H}$ between 8.5 and 10.5 can reproduce the ratios around 0.1. Viceversa, densities higher than 10$^{11}$ cm$^{-3}$ and $\log U > -1.5$ are a necessary condition to justify the ratios larger than 0.15--0.20. 

\begin{figure}
\centering
\includegraphics[width=\columnwidth]{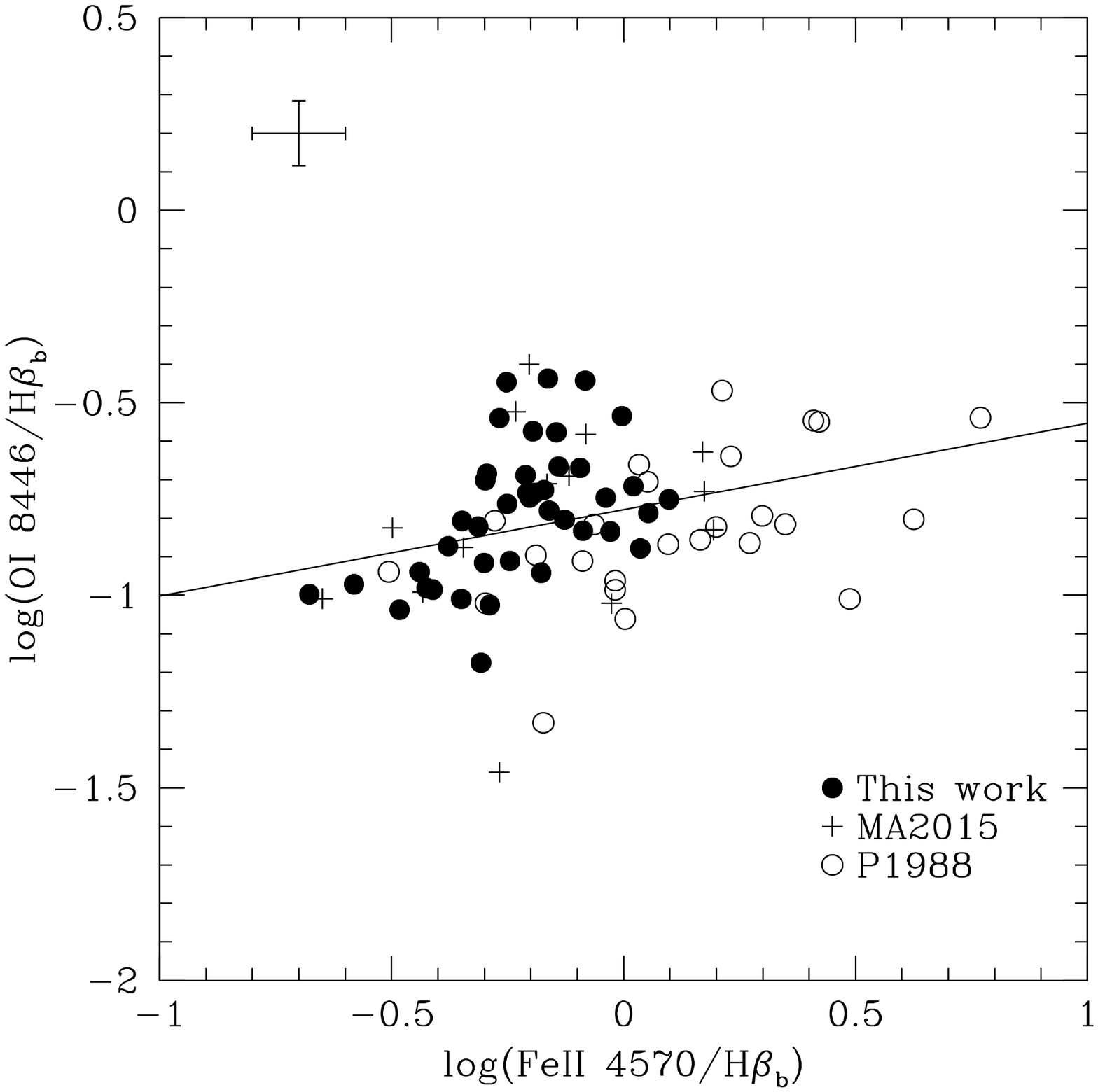} 
\includegraphics[width=\columnwidth]{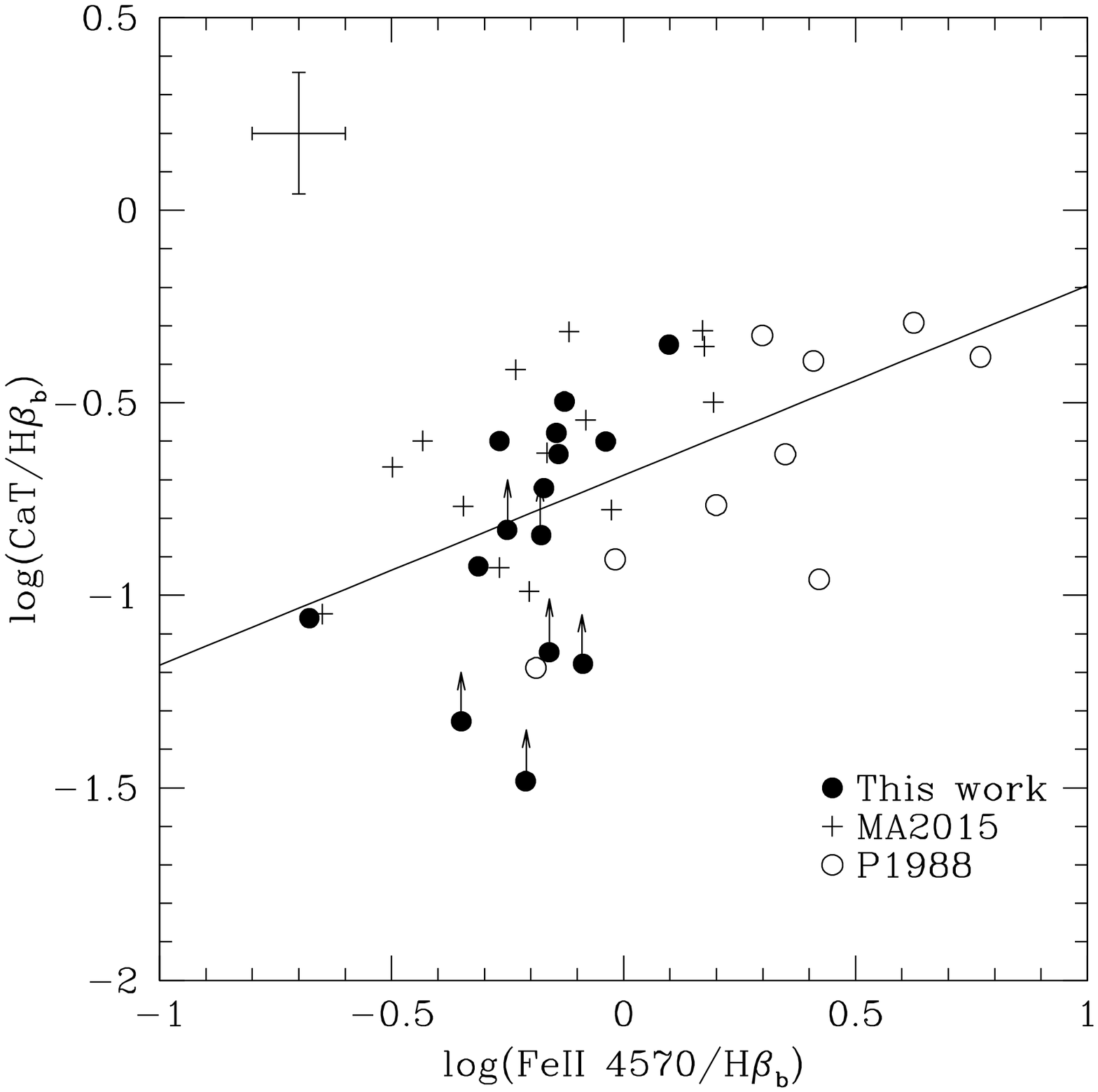} 
\caption{The Fe\,{\sc ii}\,$\lambda$4570/H$\beta_{\rm b}$ ratio plotted against the O\,{\sc i}/H$\beta_{\rm b}$ (upper panel) and CaT/H$\beta_{\rm b}$ (lower panel) ratios. The solid line is the best fitting. Symbols are the same as in Fig.~\ref{oicat}.}
\label{FeO1CaT}
\end{figure}

We plotted the logarithm of Fe\,{\sc ii}\,$\lambda$4570/H$\beta_{\rm b}$ versus the logarithm of O\,{\sc i}/H$\beta_{\rm b}$, adding the data of 14 quasars from \citet{2015ApJS..217....3M} and of the 25 Seyfert 1 galaxies and QSOs from \citet{1988ApJ...330..751P}, who however used Fe\,{\sc ii}\,$\lambda$5200/H$\beta_{\rm b}$ in place of Fe\,{\sc ii}\,$\lambda$4570/H$\beta_{\rm b}$ (Fig.~\ref{FeO1CaT}, upper panel). 
To include the \citet{1988ApJ...330..751P} data, we obtained the relation between Fe5200/H$\beta_{\rm b}$ and Fe4570/H$\beta_{\rm b}$, using our values: Fe\,{\sc ii}\,$\lambda$4570/H$\beta_{\rm b} = 2.4 \times$ Fe\,{\sc ii}\,$\lambda$5200/H$\beta_{\rm b}$. This relation was applied to the \citet{1988ApJ...330..751P} data, obtaining Fe4570/H$\beta_{\rm b}$ values to compare with ours. 
The result indicates that our data are well in agreement with \citet{2015ApJS..217....3M} and also with \citet{1988ApJ...330..751P}, who selected targets with a strong Fe\,{\sc ii} emission. A positive correlation does exist, with $r_{\rm s} = 0.5$ and p-value $= 2\times10^{-6}$.
A similar result was obtained by plotting Fe\,{\sc ii}\,$\lambda$4570/H$\beta_{\rm b}$ against CaT/H$\beta_{\rm b}$ (Fig.~\ref{FeO1CaT}, lower panel). Also in this case our data are in agreement with \citet{2015ApJS..217....3M} and with 9 objects out of 40, for which \citet{1988ApJ...330..751P} claimed to have a clear detection of CaT. A positive correlation again exists, with $r_{\rm s} = 0.75$ and p-value $= 1.2\times10^{-9}$.
In conclusion, we can confirm that strong Fe\,{\sc ii} in AGNs is very likely to be associated with strong O\,{\sc i} and CaT.

\begin{figure}
\centering
\includegraphics[width=\columnwidth]{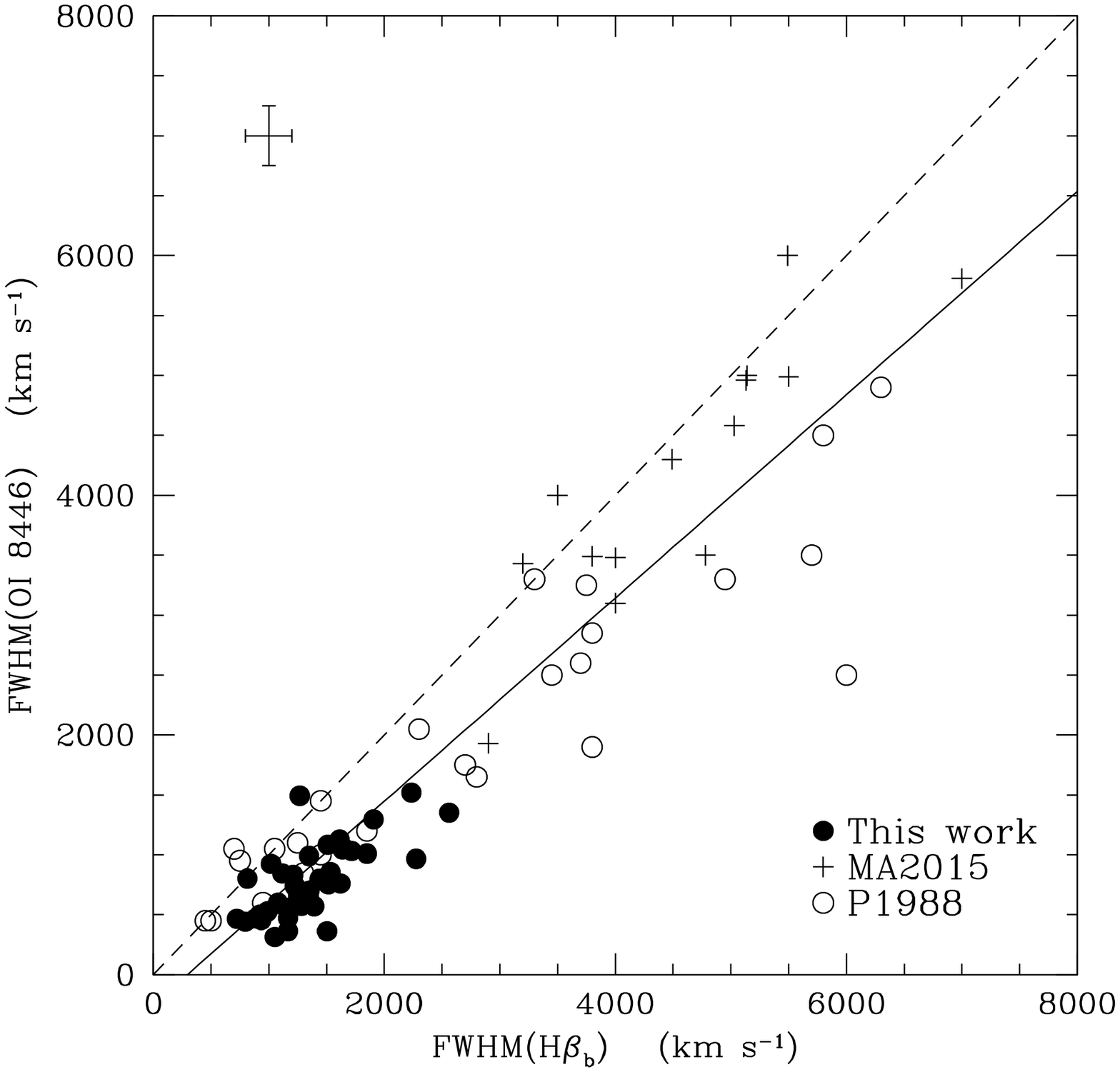} 
\includegraphics[width=\columnwidth]{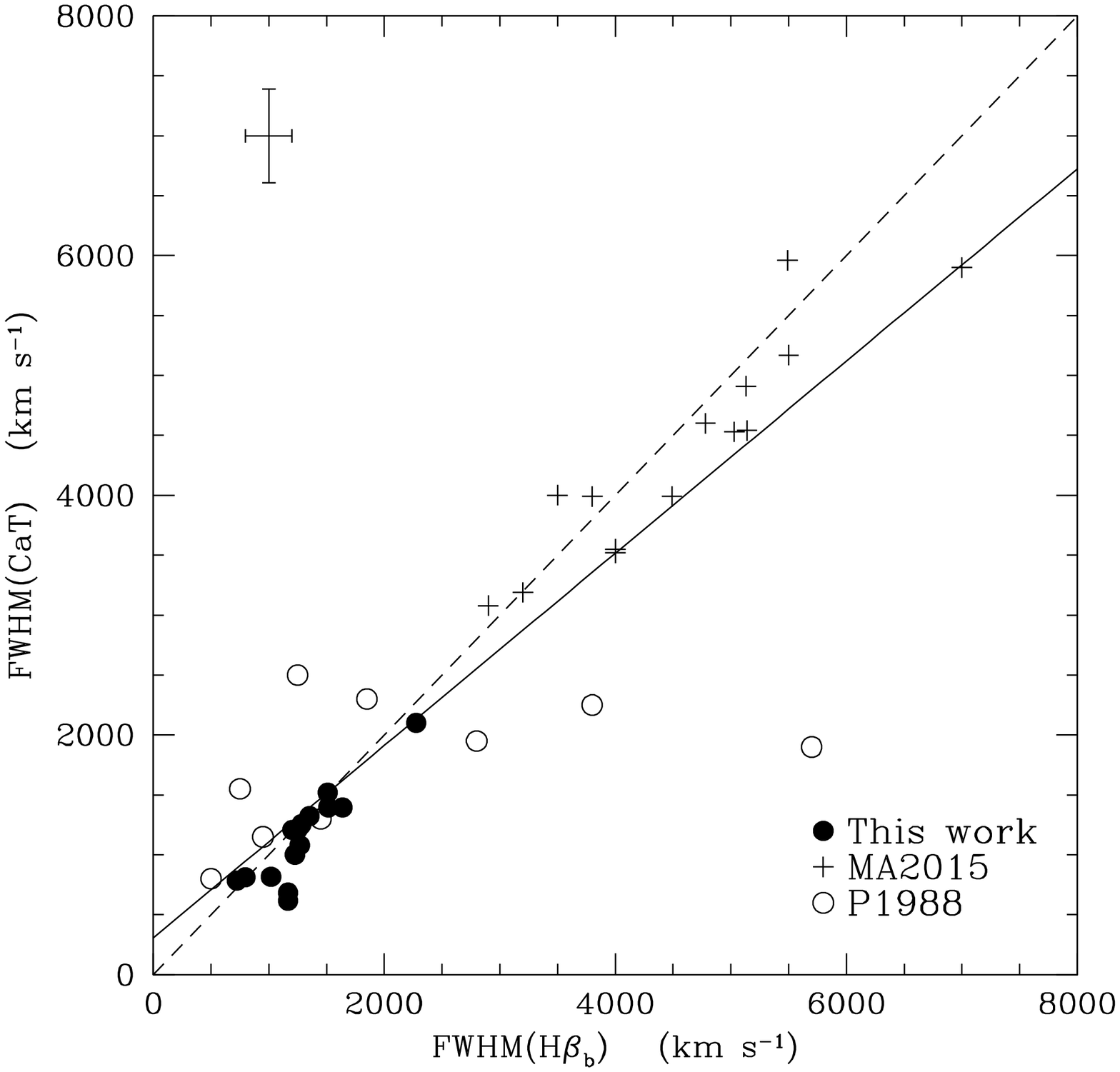} 
\caption{Comparison between the FWHM of H$\beta_{\rm b}$ and the FWHM of O\,{\sc i}\,$\lambda$8446 (upper panel) and CaT (lower panel). The solid line is the best fitting, the dashed line is the 1:1 relation. Symbols are the same as in Fig.~\ref{oicat}.}
\label{FWHMO1CaT}
\end{figure}

We compared the O\,{\sc i} FWHM with the H$\beta_{\rm b}$ one (Fig.~\ref{FWHMO1CaT}, upper panel). To have a larger range of values we added the measurements by \citet{2015ApJS..217....3M} and \citet{1988ApJ...330..751P}. The plot shows that most of the data are below the 1:1 line, forming a sort of parallel sequence with a systematically lower O\,{\sc i} FWHM, as also indicated by the best-fitting line. This suggests that O\,{\sc i} is emitted by gas with a different kinematics than the H$\beta_{\rm b}$ emitting one, likely located more distant from the SMBH, if we assume the virial case.
We also compared the CaT FWHM with the H$\beta_{\rm b}$ one (Fig.~\ref{FWHMO1CaT}, lower panel), adding again the data from \citet{2015ApJS..217....3M}, and the 9 out of 13 objects from \citet{1988ApJ...330..751P} with measured FWHM(CaT).
The result is interesting, because all the values are well distributed around the 1:1 line, therefore the velocity dispersions are in agreement. 
On the basis of these data we can conclude that while Fe\,{\sc ii} and CaT are emitted in a region whose kinematical broadening is consistent with the one of the H$\beta_{\rm b}$ component emitting region, the same does not seem to happen for O\,{\sc i}. However, a more careful analysis indicates that the values of Martinez-Aldama et al. and those of Persson having FWHM(H$\beta_{\rm b}$) $< 2000$ km s$^{-1}$ are distributed around the diagonal line, while our values, apart from very few cases, are below the line. The discrepancy could be due to data having different quality, therefore it is likely that a conclusive result should be reached through a much larger sample of spectra with a high S/N ratio in O\,{\sc i}.

\section{Results on Narrow line region} \label{NLR}

\subsection{Electron density and ionization degree}

\begin{figure}
\centering
\includegraphics[width=\columnwidth]{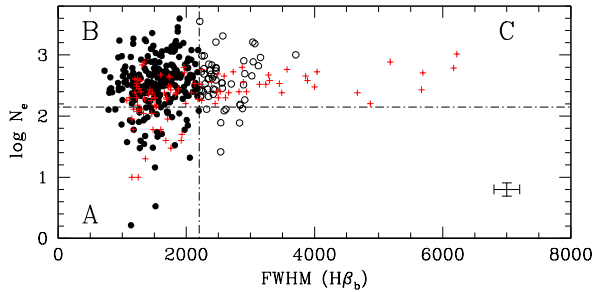} 
\caption{NLR density plotted against the H$\beta_{\rm b}$ FWHM. The dot-dashed vertical line is the separation between BLS1s and NLS1s. The dot-dashed horizontal line is the density limit $N_{\rm e}= 140$ cm$^{-3}$ found by \citet{2007ApJ...670...60X}. Solid circles are NLS1s, open circles are BLS1s and red plus symbols are data from  \citet{2007ApJ...670...60X}. A, B and C zones are explained in the text.}
\label{ne_gauss}
\end{figure} 

Fig.~\ref{ne_gauss} shows a plot of density against H$\beta_{\rm b}$ FWHM with \citet{2007ApJ...670...60X} data overlaid for comparison.
These authors found that their sample of 38 BLS1s avoids low densities, showing $N_{\rm e}>140$ cm$^{-3}$, while their sample of 55 NLS1s have a larger scatter in density and 17 of them (31 per cent) have $N_{\rm e}<140$ cm$^{-3}$.
In our case, 40 out of 238 NLS1s (17 per cent) show $N_{\rm e}<140$ cm$^{-3}$ (zone A), while 198 (83 per cent) show $N_{\rm e}>140$ cm$^{-3}$ (zone B). Among BLS1s, 48 out of 55 (87 per cent) have $N_{\rm e}>140$ cm$^{-3}$ (zone C), and the remaining 7 (13 per cent) fall in the region called \textit{zone of avoidance} by  \citet{2007ApJ...670...60X}. 
Our data indicate that the probability of finding galaxies with $N_{\rm e}<140$ cm$^{-3}$ is similar for NLS1s and BLS1s, in disagreement with \citet{2007ApJ...670...60X}.
Moreover, the K--S test applied to the $N_{\rm e}$ distributions for our NLS1s and BLS1s confirms that no difference exists in the NLR density between the two types of AGNs at any level of statistical significance (p-value $= 0.56$). Of course, these numbers are sensitive to the limit used to separate NLS1s from BLS1s, to the adopted lower limit in density and possibly to the small sample size. We matched our sample and the \citet{2007ApJ...670...60X} one, finding 13 NLS1s in common. By considering the whole sample of 93 BLS1s and 280 NLS1s slightly different percentages are found: 55 out of 280 NLS1s ($\sim20$ per cent) have $N_{\rm e}<140$ cm$^{-3}$, while only our 7 BLS1s ($\sim8$ per cent) show these low values of density. Hence, NLS1s with low density seem to be more frequent than BLS1s, but when the K--S test  was applied to these samples, similar density distributions was found again (p-value $=0.16$).  We believe that additional data of BLS1s are mandatory before drawing a final conclusion.

\begin{figure}
\includegraphics[width=\columnwidth]{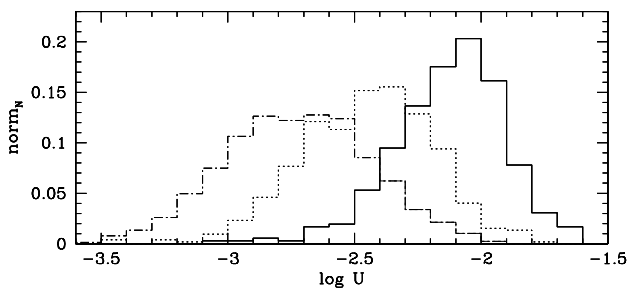} 
\caption{Histogram refers to the distributions of $\log U$ for our whole sample (solid line). For comparison, the $\log U$ distributions for Seyfert 2 (dot-dashed line) and intermediate-type Seyfert galaxies (dotted line) by \protect\citet{2012MNRAS.427.1266V} are shown. }
\label{logU}
\end{figure}

We compared our values of $\log U$ with \citet{2012MNRAS.427.1266V} finding, as expected, that the ionization parameter of NLS1s is higher than that of intermediate-type Seyfert and Seyfert 2 galaxies (Fig.~\ref{logU}, solid line, dotted line and dot-dashed line, respectively).
As discussed in Section~\ref{NLR_fitting} even with A$(V)=1.0$, the distribution of $\log U$ for NLS1s is still peaked at higher values. In addition, we estimated the ionization degree with the ratio between the flux of [O\,{\sc iii}]$\lambda$5007 and of the H$\beta$ narrow component, which is not sensitive to reddening. The distribution ranges from 2.3 to 21.5, with a median ratio for the whole sample of 9.8 and an IQR of 6.2. As a comparison, the median and the IQR values calculated for Seyfert 2 and intermediate-type Seyfert galaxies are 4.5 and 4.4, and 7.4 and 5.4, respectively \citep{2012MNRAS.427.1266V}. The K--S test applied to the three samples of AGNs confirms that the difference between these distributions are real (p-value $< 10^{-18}$).

\subsection{[O\,{\sc iii}]$\lambda$5007 emission line profile}

The A20 distribution ranges between $-0.05$ and $0.57$, with a median value of 0.25 and an IQR of 0.17. 96 per cent of the sample shows $\sigma_{A20}  <0.05$ and 70 per cent has $\sigma_{A20} <0.02$. According to these values, the sample was divided into objects with a red asymmetry ($A20\leq-0.02$), objects with a symmetric profile ($-0.02<A20<0.02$) and objects with a blue asymmetry ($A20\geq0.02$). 338 galaxies out of 359 (94 per cent) show a blue asymmetry, 18 have a symmetric profile and only 3 show a red asymmetry.

\begin{figure}
\includegraphics[width=\columnwidth]{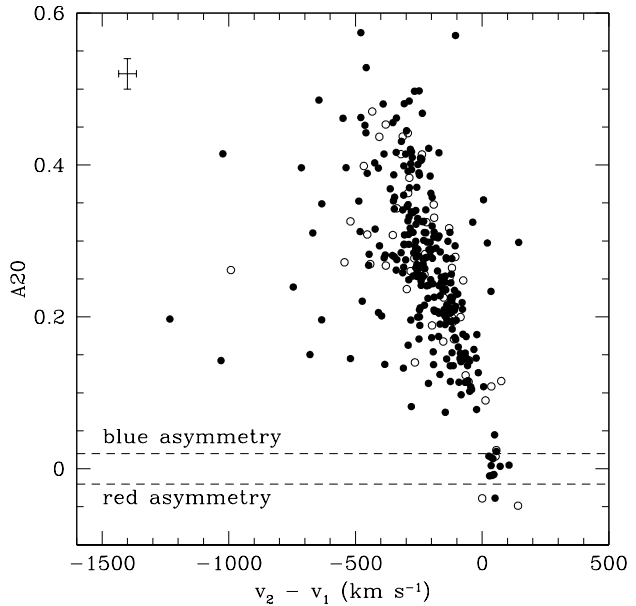} 
\caption{The difference between wing and core velocity ($v_2 - v_1$) plotted against the asymmetry parameter (A20). The dashed lines indicate the chosen limits of {$\pm 0.02$} to define red and blue asymmetry. Symbols are the same as in Fig.~\ref{ne_gauss}.}
\label{asim_FWHM}
\end{figure}

A20 can depend on several factors, such as the widths of the wing and the core and the difference in velocity between the wing and core components ($v_2-v_1$).  In our data, we did not find any correlation between $A20$ and the FWHM of the core component, while we found a weak but significant correlation with the width of the wing. In particular, the wings show larger FWHM as the asymmetry increases ($r_{\rm s}=0.41$, p-value $=3\times10^{-14}$). We found also a significant correlation ($r_{\rm s}=-0.66$, p-value $=2\times10^{-41}$) between A20 and $v_2-v_1$ (Fig.~\ref{asim_FWHM}).
Finally, cores and wings contribute almost equally to the total emission of [O\,{\sc iii}]: 51 and 49 per cent, on average, respectively, in agreement with \citet{2011ApJ...737...71Z}, \citet{2013MNRAS.433..622M} and \citet{2014RAA....14..913P}. Therefore, we decided to explore the possible effect of the flux ratio between the components on $A20$, finding only a trend ($r_{\rm s}=-0.29$, p-value$=3 \times 10^{-8}$).
Being the $A20$ versus $v_2 - v_1$ the strongest correlation among those analysed, we can conclude that the shape of [O\,{\sc iii}] profiles are mostly influenced by the difference in velocity between the wing and the core component. However, we cannot exclude that the asymmetric profiles can be due to a combination of the previously discussed factors.

Out of the 325 objects fitted with at least 2 Gaussian functions, 299 have blue wings, while only 26 have red wings.
A NLS1, SDSS J134206.56+050523.8, shows both a blue and a red wing. 
271 NLS1s (92 per cent) have a wing and 252 (85 per cent) show a blue wing. This confirms that asymmetric profiles are a common phenomenon at least for the highly-ionized emission lines of the NLR. 
The inconsistency between the number of objects with a red asymmetry and with a red wing is due to the difference in the methods. Objects showing weak red wings has small velocity shifts: most of them have wing to core flux ratios lower than 1 and a velocity shift lower than 150\,km\,s$^{-1}$. $A20$ is not sensitive to weak wings, indeed it does not correlate with the velocity shift if we consider only red wings ($r_s=-0.17$, p-value=0.40), while it correlates in case of blue wings ($r_s=-0.60$, p-value=$7.5\times10^{-31}$).

The median values of the differences between the velocity of the components with respect to the core are $v_2 - v_1 = -210$ km~s$^{-1}$ and $v_3 - v_1 = -583$ km~s$^{-1}$. 
The core component has a median FWHM of 273 km s$^{-1}$ while the blue and red wings have median FWHM of 838 km s$^{-1}$ and 796 km s$^{-1}$, in agreement with the expected values. The distribution of the FWHM of the core component is narrower than those of the wings: the IQRs are $\sim$130 km\,s$^{-1}$, $\sim$400 km\,s$^{-1}$ and $\sim$670 km\,s$^{-1}$.

\begin{figure}
\includegraphics[width=\columnwidth]{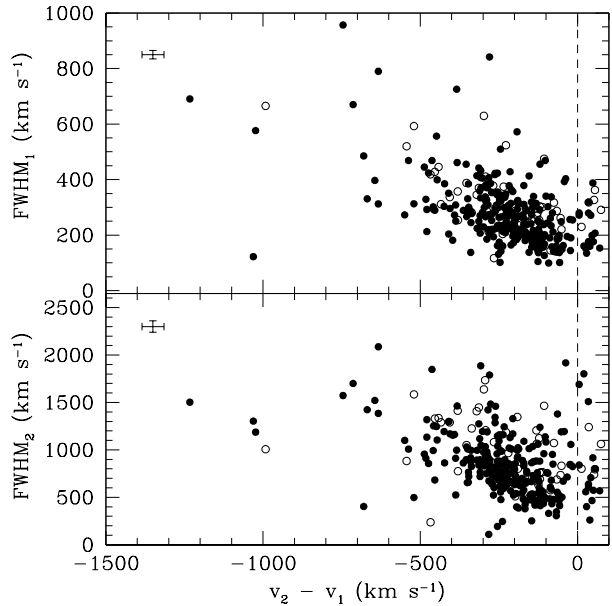} 
\caption{$v_2 - v_1$ plotted against the FWHM of the core (top panel) and the wing (bottom panel) components. Symbols are the same as in Fig.~\ref{ne_gauss}.}
\label{v1_v2_FWHM}
\end{figure}

We found a weak but significant correlation between $v_2-v_1$ and the FWHM of the wing (Fig.~\ref{v1_v2_FWHM}, lower panel), in agreement with \citet{2011ApJ...739...28X}.
This correlation ($r_{\rm s}=-0.39$, p-value $=3\times10^{-13}$) suggests that the faster an outflow is moving through a medium the more turbulent the gas should become. 
At the same time, the outflow should affect the kinematics of the medium. Indeed, a similar correlation ($r_{\rm s}=-0.46$, p-value $=1.4\times10^{-18}$) exists with the FWHM of the core component (Fig.~\ref{v1_v2_FWHM}, upper panel). 

\begin{figure}
\includegraphics[width=\columnwidth]{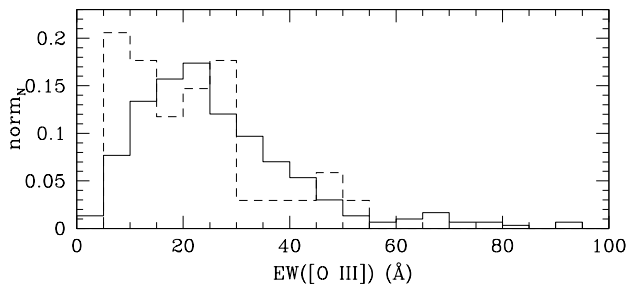} 
\caption{The histogram refers to the distribution of EW([O\,{\sc iii}]) values for two sub-samples: objects with blue wings (solid line) and with symmetric profiles (dashed line).}
\label{asym_EW_O3}
\end{figure}

\begin{figure}
\includegraphics[width=\columnwidth]{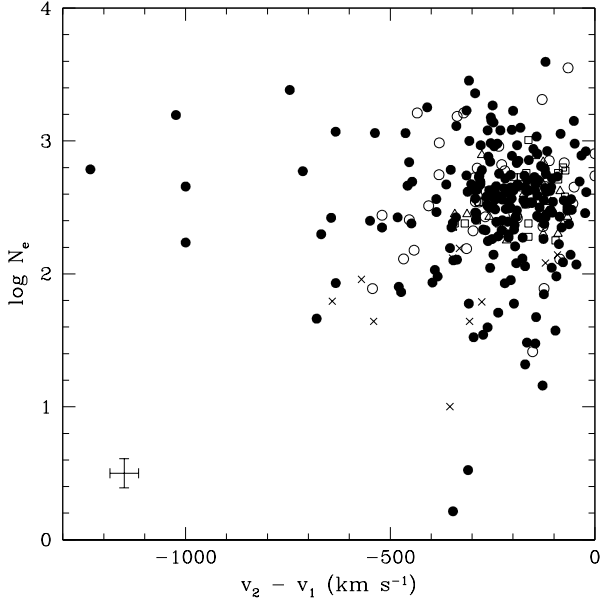} 
\caption{The difference in velocity between wing and core component plotted against the density. 
Solid circles are our NLS1s and open circles are our BLS1s, open triangles are the \citet{2007ApJ...670...60X} NLS1s, open squares are BLS1s from \citet{2007ApJ...670...60X}, finally crosses are low-density objects found by \citet{2007ApJ...670...60X}.}
\label{wing_ne}
\end{figure}

In our sample, the trend between EW([O\,{\sc iii}]) and $v_2-v_1$ ($r_{\rm s}=0.26$, p-value $=7\times10^{-6}$) suggests that faster outflows are found in objects with lower EWs, in agreement with \citet{2011ApJ...737...71Z} and \citet{2012ApJ...756...51L}. However, by plotting the distributions of EW([O\,{\sc iii}]) for objects with and without wings (Fig.~\ref{asym_EW_O3}) and then by applying the K--S test, we cannot reject the hypothesis that the distributions are similar (p-value $=0.18$). In addition and in disagreement with \citet{2007ApJ...670...60X}, who had a smaller sample, no correlation appears to exist between $N_{\rm e}$ and highly blueshifted wings (Fig.~\ref{wing_ne}).

\begin{figure}
\includegraphics[width=\columnwidth]{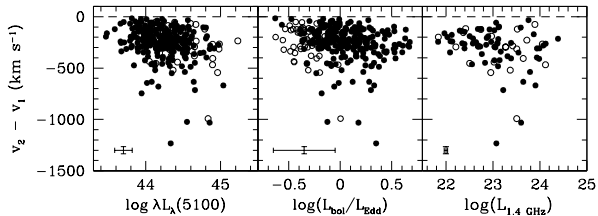} 
\caption{$v_2 - v_1$ plotted against the continuum luminosity at 5100 \AA\ (left panel), the Eddington ratio (middle panel) and the radio luminosity at 1.4 GHz in units of W Hz$^{-1}$ (right panel). Symbols are the same as in Fig.\ref{ne_gauss}.}
\label{v2_v1_lumO3}
\end{figure}

We did not find any correlation between $\lambda {\rm L}_{\lambda} (5100)$ and $v_2 - v_1$ (see Fig.~\ref{v2_v1_lumO3}, left panel), but faster outflows are present only in powerful AGNs or equally less luminous AGNs can only have outflows with smaller velocities. 
A slight increase in the strength of the blue wing with increasing luminosity was also found by \citet{2013MNRAS.433..622M}. 

NLS1s are characterized by high Eddington ratios, therefore a possible reason for the blue wings is the strong radiation pressure which can accelerate the NLR gas in these AGNs \citep{2005ApJ...627..721G}. 
Our data show a trend ($r_{\rm s}=-0.24$, p-value $=10^{-5}$) suggesting that faster outflows are only observed in higher Eddington ratio objects (Fig.~\ref{v2_v1_lumO3}, middle panel).
A similar result was found by considering the shift of the wing components with respect to [S\,{\sc ii}]$\lambda 6717$ ($r_{\rm s}=-0.23$, with p-value $= 3\times 10^{-5}$), in agreement with \citet{2011ApJ...737...71Z}.


Since it does not seem that the power of the AGN or its accretion rate are directly responsible for the outflow velocity, and that viceversa the surrounding interstellar medium can play a significant role, another hypothesis to explain the blue wings is the interaction between the radio jet and the NLR, which can disturb the diffuse gas around AGNs. Indeed, broad [O\,{\sc iii}]$\lambda$5007 profiles (FWHM $>$ 1000 km s$^{-1}$) were found in objects with radio luminosity at 1.4 GHz ($L_{\rm 1.4\,GHz}$) between $10^{23}$ and $10^{25}$ W Hz$^{-1}$ \citep{2013MNRAS.433..622M} and in compact steep-spectrum radio sources, which are believed to be AGNs in a younger evolutionary stage \citep{2008MNRAS.387..639H, 2011ApJ...741...50W}. Unfortunately, this effect is much more evident in radio-loud AGNs, while NLS1s are typically radio-quiet objects. However, the radio-jet might be present also in radio-quiet NLS1s \citep{2011nlsg.confE..24F}.
We cross-correlated our whole sample with the Faint Images of the Radio Sky at Twenty-centimeters (FIRST) survey \citep{1994ASPC...61..165B} and with the NRAO VLA Sky Survey \citep[NVSS,][]{1998AJ....115.1693C}(NVSS) finding 94 galaxies which are present in at least one of the two samples (70 NLS1s and 24 BLS1s). In case of double measurements, the highest value of flux was used. Our $L_{1.4\,\rm GHz}$ values range between $\sim 10^{22}$ and $10^{24}$~W~Hz$^{-1}$. The radio-loudness $R=F_{\rm 5 GHz}/F_{B_{\rm band}}$ \citep{1989AJ.....98.1195K} was calculated assuming  a spectral index $\alpha=-0.5$ to extrapolate the 5 GHz flux and integrating each spectrum multiplied by a $B$-filter transmission curve to obtain the $B-$band flux. We found 12 radio-loud sources, 11 of them are NLS1s, which represent the 3.7 per cent of our total sample and the 15.7 per cent of those radio detected. 
We did not find any correlation between $L_{\rm 1.4\,GHz}$ and the asymmetry of the [O\,{\sc iii}] or the velocity shift of the wing relative to the core component (Fig.~\ref{v2_v1_lumO3}, right panel), but our data seem to suggest that high velocity outflows are not typically associated to low values of $L_{\rm 1.4\,GHz}$. Viceversa, a weak trend ($r_s=0.23$ and p-value$=0.009$) with the FWHM of the wing component was found.

\subsection{[O\,{\sc iii}]$\lambda5007$ blueshift} \label{bshift}

\begin{figure}
\includegraphics[width=\columnwidth]{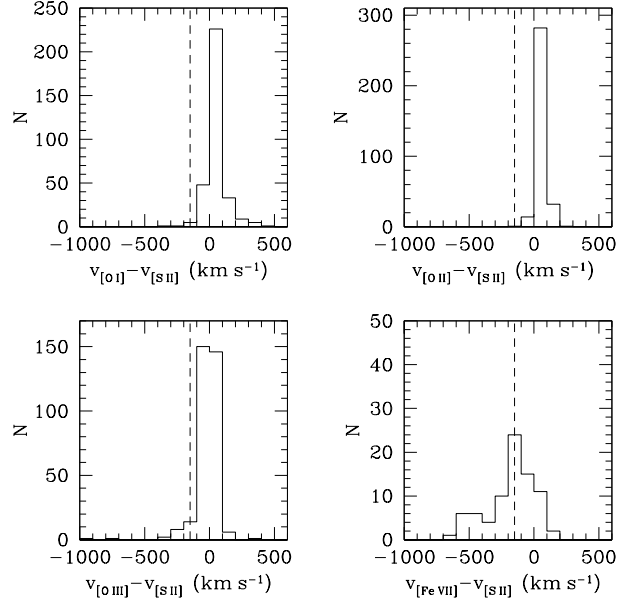} 
\caption{Comparison between the distribution of the shift in velocity for NLR emission lines with different ionization potentials. From top-left to bottom-right: $v_{\rm[O\,I]}-v_{\rm[S\,II]}$, $v_{\rm[O\,II]}-v_{\rm[S\,II]}$, $v_{\rm[O\,III]}-v_{\rm[S\,II]}$, $v_{\rm[Fe\,VII]}-v_{\rm[S\,II]}$. The dashed line indicates the limit of $-150$ km\,s$^{-1}$ \citep{2008ApJ...680..926K} adopted to define blue outliers (see text).
}
\label{shift_ionization_pot}
\end{figure}

In Fig.~\ref{shift_ionization_pot}, the histograms of the velocity shift distributions are shown: low-ionization lines are distributed around zero, showing a positive median velocity shift of 36 km s$^{-1}$ for [O\,{\sc i}] and 61 km s$^{-1}$ for [O\,{\sc ii}], while both the histograms of [O\,{\sc iii}] and [Fe\,{\sc vii}]  show a clear wing extended towards negative values.
The existence of a relation between velocity shift and ionization potential, found by \citet{1996A&A...315..335M} and then confirmed by other authors \citep{2000ApJ...532..256K, 2000ApJ...530..733L, 2008ApJ...680..926K, 2009ApJ...702L..42S, 2009PASJ...61..259O, 2012ApJ...756...51L}, supports the hypothesis of outflowing gas from an inner and compact region, where the ionization is dominated by the central source.

\begin{figure}
\includegraphics[width=\columnwidth]{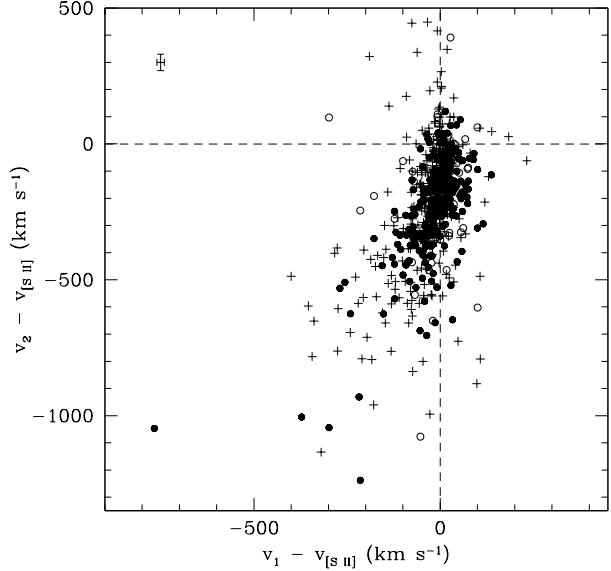} 
\caption{Plot of the velocities of the core components and of the wings calculated with respect to the velocity of the host galaxy. Solid circles are NLS1 galaxies, while open circles are BLS1 galaxies. Cross symbols are data from \citet{2011ApJ...737...71Z}.}
\label{v1_vS2_vs_v2}
\end{figure}

We did not find higher values of asymmetry in objects with higher [O\,{\sc iii}] blueshifts ($\Delta v = v_1 - v_{\rm [S II]}$), then it does not seem that the profile and the shift of [O\,{\sc iii}] emission lines are connected ($r_{\rm s}=-0.12$, p-value $=0.025$), in agreement with \citet{2005AJ....130..381B}, but in contrast with \citet{2005ApJ...618..601A} and \citet{2011ApJ...741...50W}. 
In Fig.~\ref{v1_vS2_vs_v2}, the shift of the core component against the shift of the wing was plotted. The higher is the blueshift, the higher is also the shift of the wing ($r_{\rm s}=0.52$ and p-value $=10^{-21}$), suggesting that the outflow is involving both the wing and the core components. As a comparison, we added also the sample of \citet{2011ApJ...737...71Z} (the cross symbols), which confirms and reinforces our result.

\begin{figure}
\includegraphics[width=\columnwidth]{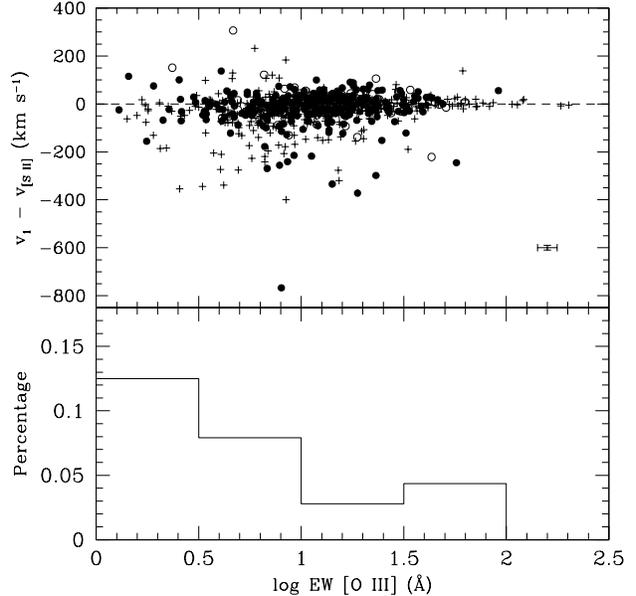} 
\caption{The EW of [O\,{\sc iii}] plotted against the shift of the core component with respect to [S\,{\sc ii}]. Symbols are the same as in Fig.~\ref{v1_vS2_vs_v2} (upper panel). The histogram refers to the fraction of objects with velocity shift higher than $-150$ km s$^{-1}$ for different bins of EW, as explained in the text (lower panel).}
\label{EW_Dv}
\end{figure}

We found a weak but significant correlation between the blueshift of [O\,{\sc iii}] and the FWHM of the core component ($r_{\rm s}=-0.36$, p-value $=3\times10^{-11}$), but only a  trend with the FWHM of the wing component ($r_{\rm s}=-0.23$, p-value $=3\times10^{-5}$). On the contrary, no correlation was found between the velocity shift and the EWs of the core, wing or total [O\,{\sc iii}]. This is in disagreement with \citet{2011ApJ...737...71Z}, who found that the EW decreases as the blueshift increases. 
Even when considering the two samples together the correlation coefficient still remains very low ($r_{\rm s}=0.18$, p-value $=10^{-6}$). 
The upper panel of Fig.~\ref{EW_Dv}  shows the plot of EW of the core against the velocity shift for our sample and for the \citet{2011ApJ...737...71Z} data. 
The entire sample was divided into five bins of EW: $\log$ EW[O\,{\sc iii}] lower than 0.5, between 0.5 and 1.0, between 1.0 and 1.5, between 1.5 and 2.0 and higher than 2.0. For each bin we calculated the fraction of objects showing blueshifts higher than $-150$ km s$^{-1}$. In the lower panel of Fig.~\ref{EW_Dv} the histogram of these fractions was plotted for each bin. This plot shows that the fraction of objects decreases slightly as the EW increases, in agreement with \citet{2011ApJ...737...71Z}. 

No significant correlation was found between $\Delta v$ and $L_{\rm bol}/L_{\rm Edd}$, $\lambda L_{\lambda}(5100)$ and $L_{\rm 1.4\,GHz}$. We stress that the $\Delta v$ -- $L_{\rm bol}/L_{\rm Edd}$ relation shows the highest value of the correlation coefficient ($r_{\rm s}=-0.18$, with p-value $=10^{-3}$). This is consistent with \citet{2011ApJ...737...71Z}, who suggested that the Eddington ratio is the most important parameter affecting the shift of [O\,{\sc iii}] ($r_s=-0.35$, with p-value$=3\times10^{-12}$).

Among the 332 objects for which [S\,{\sc ii}] was available, only 6 are blue outliers and 2 are red outliers, according to \citet{2002ApJ...576L...9Z} criterion, while 14 are blue outliers and 3 are red outliers when considering the less-restrictive limit of $-150$ km\,s$^{-1}$ \citep{2008ApJ...680..926K}. We adopted this last value as threshold since it is close to the instrumental resolution of our data. 
In both cases, the number of these outliers is very small. 
In order to enlarge the sample and to better understand the properties of these objects, we added the previously known blue outliers from literature, obtaining a total sample of 69 objects (see Table~\ref{tabBO}).
\begin{table*}
\begin{minipage}{125mm}
\caption{Blue outliers. (1)\,Name of the object; (2)\,right ascension at J2000; (3)\,declination at J2000; (4)\,blueshift of [O\,{\sc iii}] in units of km s$^{-1}$, calculated according to references in col(11); (5)\,FWHM of [O\,{\sc iii}] in units of km s$^{-1}$; (6)\,FWHM of H$\beta_{\rm b}$ in units of km s$^{-1}$; (7)\,logarithm of the continuum luminosity at $\lambda=5100$ \AA; (8)\,logarithm of the black hole mass; (9)\,logarithm of the Eddington ratio; (10)\,logarithm of the luminosity at 1.4 GHz in units of W Hz$^{-1}$; (11)\,references for the data: (1) \protect\citet{2002ApJ...576L...9Z}; (2) \protect\citet{2003MNRAS.345.1133M}; (3) \protect\citet{2005ApJ...618..601A}; (4) \protect\citet{2005AJ....130..381B}; (5) \protect\citet{2005MNRAS.364..187B}; (6) \protect\citet{2008ApJ...680..926K}; (7) \protect\citet{2011ApJ...737...71Z}; (8) \protect\citet{2015arXiv151202642Z}; (9) this work.}
\label{tabBO}
\begin{tiny}
\centering
\begin{tabular}{|l|l|l|c|c|c|c|c|c|c|c|c|}
\hline
  \multicolumn{1}{|c|}{Name} &
  \multicolumn{1}{c|}{$\alpha$} &
  \multicolumn{1}{c|}{$\delta$} &
  \multicolumn{1}{c|}{\phantom{$-$}$\Delta$v} &
  \multicolumn{1}{c|}{[O\,{\sc iii}]} &
  \multicolumn{1}{c|}{H$\beta_{\rm b}$} &
  \multicolumn{1}{c|}{$\lambda L_{5100}$} &
  \multicolumn{1}{c|}{$ M_{\rm BH}$} &
  \multicolumn{1}{c|}{\phantom{$-$}$\frac{L_{\rm bol}}{L_{\rm Edd}}$} &
  \multicolumn{1}{c|}{$ L_{\rm 1.4\,GHz}$}&
  \multicolumn{1}{c|}{Ref.} \\
  \multicolumn{1}{|c|}{(1)} &
  \multicolumn{1}{c|}{(2)} &
  \multicolumn{1}{c|}{(3)} &
  \multicolumn{1}{c|}{(4)} &
  \multicolumn{1}{c|}{(5)} &
  \multicolumn{1}{c|}{(6)} &
  \multicolumn{1}{c|}{(7)} &
  \multicolumn{1}{c|}{(8)} &
  \multicolumn{1}{c|}{(9)} &
  \multicolumn{1}{c|}{(10)}&
  \multicolumn{1}{c|}{(11)} \\
\hline
  J001340.73-111100.6     & 00:13:40.74 & $-$11:11:00.7     & $-$320 & --    & 6052$^{a}$ & 44.24$^{a}$ & 8.28$^{b}$ & $-$1.19$^{b}$ & -- & 7\\
  IZw1                    & 00:53:34.94 & +12:41:36.2       & $-$640 & 1440$^{f}$  & 1090 & 44.83 & 7.34 & \phantom{$-$}0.49   & 22.65 & 1\\
  J010226.31-003904.6     & 01:02:26.32 & $-$00:39:04.6     & $-$931 & 647   & 1680 & 45.04 & 7.86 & \phantom{$-$}0.03 & --    & 5\\
                          &  &                              & $-$880 & 1320  & 1500 & 45.28 & 7.93 & \phantom{$-$}0.35 & --    & 3\\
  J015910.04+010514.6     & 01:59:10.06 & +01:05:14.5       & $-$227 & 691$^{c}$   & 3276$^{a}$ & 44.29$^{a}$ & 7.77$^{b}$ & $-$0.63$^{b}$ & -- & 7\\
  J011929.06-000839.7     & 01:19:29.07 & $-$00:08:39.7     & $-$220 & 380   & 1220 & --    & 6.20 & $-$0.05           & --    & 6\\
                          &  &                              & $-$178 & 379   & 1009 & 43.86 & 6.38 & \phantom{$-$}0.44 &--     & 9\\
                          &  &                              & $-$219 & 488   & 715  & --    & 6.12 & \phantom{$-$}1.19 &--     & 4\\
  J013521.68-004402.2     & 01:35:21.68 & $-$00:44:02.1     & $-$576 & 711   & 1181 & 43.45 & 6.44 & $-$0.14           & 22.68 & 5\\
                          &  &                              & $-$240 & 620   & 1710 & --    & 6.50 & $-$0.30           &--     & 6\\
  RXJ0136.9-3510          & 01:36:54.40 & $-$35:09:52.0     & $-$380 & 900   & 1050 & 44.73 & 7.24 & \phantom{$-$}0.50 &--     & 2\\
  J015249.76+002314.7     & 01:52:49.77 & +00:23:14.6       & $-$501 & 1277  & 1852 & 44.71 & 7.72 & $-$0.16           &--     & 5\\
  J024037.89+001118.9     & 02:40:37.88 & +00:11:19.0       & $-$458 & 381   & 1789 & 44.19 & 7.32 & $-$0.28           &--     & 5\\
  J032606.75+011429.9     & 03:26:06.77 & +01:14:29.7       & $-$180 & 530   & 1230 & --    & 6.30 & \phantom{$-$}0.0  &--     & 6\\
  J035219.08+010934.7     & 03:52:19.08 & +01:09:34.7       & $-$168 & 502$^{c}$   & 2189$^{a}$ & 44.36$^{a}$ & 7.52$^{b}$ & $-$0.31$^{b}$ & -- & 7\\
  RXJ0439.7-4540          & 04:39:44.84 & $-$45:40:42.0     & $-$580 & 1020  & 1020 & 45.04 & 7.43 & \phantom{$-$}0.62 &--     & 2\\
  IRAS04416+1215          & 04:44:28.78 & +12:21:11.7       & $-$300 & 650$^{e}$  & 1470 & 45.11 & 7.79 & \phantom{$-$}0.32 &--     & 3\\
  IRAS04576+0912          & 05:00:20.77 & +09:16:55.6       & $-$300 & 380$^{e}$  & 1210 & 43.68 & 6.62 & \phantom{$-$}0.06 &--     & 3\\
  PKS0736+01              & 07:39:18.04 & +01:37:04.6       & $-$430 & 720$^{f}$   & 3260 & 44.98 & 8.39 & $-$0.41           &--     & 1\\
  J074644.79+294059.0     & 07:46:44.80 & +29:40:59.0       & $-$183 & 575$^{c}$   & 2245$^{a}$ & 44.39$^{a}$ & 7.38$^{b}$ & $-$0.13$^{b}$ & -- & 7\\
  J075620.07+304535.4     & 07:56:20.08 & +30:45:35.3       & $-$339 & 822$^{c}$   & 3426$^{a}$ & 44.67$^{a}$ & 7.98$^{b}$ & $-$0.46$^{b}$ & -- & 7\\
  J080131.59+354436.4     & 08:01:31.58 & +35:44:36.4       & $-$275 & --    & 1979$^{a}$ & 44.01$^{a}$ & 7.00$^{b}$ & $-$0.13$^{b}$ & -- & 7\\
  J080829.17+440754.1     & 08:08:29.17 & +44:07:54.1       & $-$174 & 563$^{c}$   & 2649$^{a}$ & 44.40$^{a}$ & 7.64$^{b}$ & $-$0.38$^{b}$ & -- & 7\\
  PG0804+761              & 08:10:58.60 & +76:02:42.5       & $-$305 & 780$^{f}$   & 3300 & 44.97 & 8.40 & $-$0.42           &--     & 1\\
  J081803.70+434442.1     & 08:18:03.71 & +43:44:42.2       & $-$283 & 884$^{d}$   & 1537$^{a}$ & 44.07$^{a}$ & 6.65$^{b}$ & \phantom{$-$}0.28$^{b}$ &--     & 7\\
  J082942.94+283521.5     & 08:29:42.95 & +28:35:21.5       & $-$274 & 557$^{c}$   & 5002$^{a}$ & 44.20$^{a}$ & 8.13$^{b}$ & $-$1.08$^{b}$ & -- & 7\\
  J083049.34+172826.3     & 08:30:49.35 & +17:28:26.4       & $-$256 & 355   & 2004 & 44.3  & 7.44 & $-$0.09           &--     & 9\\
  J083448.48+015921.1     & 08:34:48.48 & +01:59:21.1       & $-$937 & 2811  & 4952 & -- & 8.69$^{g}$ & -- & -- & 8 \\
  J091313.73+365817.2     & 09:13:13.73 & +36:58:17.2       & $-$150 & 350   & 1680 & --    & 6.50 & $-$0.30           &22.39  & 6\\
  J092247.03+512038.0     & 09:22:47.03 & +51:20:38.0       & $-$430 & 720   & 1250 & --    & 6.70 & \phantom{$-$}0.18 &--     & 6\\
                          &             &                   & $-$344 & --    & 1763$^{a}$ & 44.0$^{a}$  & 6.84$^{b}$ & \phantom{$-$}0.01$^{b}$ &--     & 7\\ 
  J092309.87+453046.      & 09:23:09.87 & +45:30:46.4       & $-$354 & 639$^{c}$   & 1903$^{a}$ & 44.33$^{a}$ & 7.31$^{b}$ & $-$0.12$^{b}$ & -- & 7\\
  J093031.10+183454.4     & 09:30:31.10 & +18:34:54.5       & $-$372 & 790   & 1120 & 43.91 & 6.64 & \phantom{$-$}0.28 &--     & 9\\
  J093939.69+375705.8     & 09:39:39.69 & +37:57:05.8       & $-$205 & 425$^{c}$   & 2549$^{a}$ & 44.34$^{a}$ & 7.51$^{b}$ & $-$0.31$^{b}$ & -- & 7\\
  J100402.61+285535.3     & 10:04:02.62 & +28:55:35.4       & $-$680 & 960$^{f}$   & 1760 & 45.6  & 8.29 & \phantom{$-$}0.31 &--     & 1\\
  J100420.13+051300.4     & 10:04:20.14 & +05:13:00.5       & $-$179 & 822$^{c}$   & 1890$^{a}$ & 44.34$^{a}$ & 7.51$^{b}$ & $-$0.31$^{b}$ & -- & 7\\
  J101314.86-005233.5     & 10:13:14.87 & $-$00:52:33.7     & $-$314 & 991   & 1578 & 44.29 & 7.28 & $-$0.14           &--     & 5\\
  J110428.48+115613.8     & 11:04:28.48 & +11:56:13.9       & $-$196 & 646$^{d}$   & 1457$^{a}$ & 44.04$^{a}$ & 6.70$^{b}$ & \phantom{$-$}0.20$^{b}$ &--     & 7\\
  J111006.95+612521.4     & 11:10:06.95 & +61:25:21.4       & $-$155 & 575$^{c}$   & 2203$^{a}$ & 44.15$^{a}$ & 7.43$^{b}$ & $-$0.42$^{b}$ & -- & 7\\
  J113246.66+062624.4     & 11:32:46.66 & +06:26:24.4       & $-$152 & 424   & 990  & 43.73 & 6.44 & \phantom{$-$}0.25 &--     & 9\\
  J115533.50+010730.6     & 11:55:33.50 & +01:07:30.5       & $-$283 & 919   & 1628 & 43.76 & 6.94 & $-$0.33           &--     & 5\\
                          &  &                              & $-$330 & 780   & 1510 & --    & 6.70 & $-$0.05           &--     & 6\\
                          &  &                              & $-$186 & --    & 2037$^{a}$ & 43.89$^{a}$ & 7.01$^{b}$ & $-$0.27$^{b}$ & -- & 7\\
  J120226.76-012915.3     & 12:02:26.76 & $-$01:29:15.3     & $-$170 & 340   & 1460 & --    & 6.80 & \phantom{$-$}0.04 & 23.88 & 6\\
  J120945.20+321701.1     & 12:09:45.21 & +32:17:01.1       & $-$269 & 422   & 1286 & 44.35 & 7.08 & \phantom{$-$}0.25 & 23.09 & 9\\
  J123220.11+495721.7     & 12:32:20.12 & +49:57:21.8       & $-$242 & 822$^{c}$   & 1836$^{a}$ & 44.46$^{a}$ & 7.31$^{b}$ & \phantom{$-$}0.002$^{b}$ & -- & 7\\
  J123241.73+091209.3     & 12:32:41.73 & +09:12:09.3       & $-$1520& 4971 & 4990 & -- & 8.62$^{g}$ & -- & -- & 8 \\ 
  J124441.41+585626.8     & 12:44:41.42 & +58:56:26.9       & $-$400 & 524$^{c}$   & 2205$^{a}$ & 43.87$^{a}$ & 7.06$^{b}$ & $-$0.33$^{b}$ & -- & 7\\
  J124635.25+022208.8     & 12:46:35.25 & +02:22:08.8       & $-$159 & 392   & 751  & --    & 6.16 & \phantom{$-$}1.15 & 22.09 & 4\\
                          &  &                              & $-$150 & 300   & 1200 & --    & 6.20 & $-$0.05           &--     & 6\\
  J124904.15-022235.9     & 12:49:04.16 & $-$02:22:35.9     & $-$298 & 957   & 1866 & 43.83 & 6.96 & $-$0.18           &--     & 9\\
  IRAS13224-3809          & 13:25:19.38 & $-$38:24:52.7     & $-$370 & 810   & 650  & 45.14 & 7.10 & \phantom{$-$}1.04 &--     & 3\\
  J134206.56+050523.8     & 13:42:06.57 & +05:05:23.9       & $-$767 & 576   & 1786 & 44.65 & 7.56 & \phantom{$-$}0.14 & 24.07 & 9\\
  J140516.21+255534.1     & 14:05:16.21 & +25:55:34.1       & $-$300 & 900$^{f}$   & 1940 & 45.19 & 8.09 & \phantom{$-$}0.10 & 22.74 & 1\\
  J140827.82+240924.6     & 14:08:27.83 & +24:09:24.7       & $-$218 & 670   & 1567 & 44.52 & 7.20 & \phantom{$-$}0.30 & 23.22 & 9\\
  J141700.82+445606.4     & 14:17:00.83 & +44:56:06.4       & $-$600 & 660$^{f}$   & 2560 & 44.96 & 8.17 & $-$0.21           & 22.51 & 1\\
  J141944.76+391317.4     & 14:19:44.77 & +39:13:17.5       & $-$206 & 501$^{c}$   & 2285$^{a}$ & 44.18$^{a}$ & 7.39$^{b}$ & $-$0.35$^{b}$           & --    & 7\\
  J143230.99-005228.9     & 14:32:30.99 & $-$00:52:28.9     & $-$288 & 1222  & 1559 & 43.99 & 7.06 & $-$0.22           & --    & 5\\
  J144240.79+262332.5     & 14:42:40.80 & +26:23:32.6       & $-$156 & 327   & 947  & 44.28 & 6.57 & \phantom{$-$}0.67 & 23.04 & 9\\
  J144359.75+012956.2     & 14:43:59.75 & +01:29:56.3       & $-$214 & 577   & 1953 & 44.33 & 7.52 & $-$0.12           &--     & 9\\
  J150521.92+014149.7     & 15:05:21.92 & +01:41:49.8       & $-$166 & 538   & 649  & --    & 6.28 & \phantom{$-$}1.38 & 23.02 & 4\\
  J154259.49+030653.3     & 15:42:57.49 & +03:06:53.3       & $-$162 & 515   & 712  & --    & 5.86 & \phantom{$-$}1.09 &--     & 4\\
  J154530.24+484609.1     & 15:45:30.24 & +48:46:09.1       & $-$950 & --    & 1560 & --    & --   & --                & 24.08 & 1\\
                          &  &                              & $-$1150& 2200  & 1630 & 45.59 & 8.05 &  0.41             &--     & 3\\
  J160518.50+375653.5     & 16:05:18.50 & +37:56:53.4       & $-$152 & 614$^{c}$   & 1993$^{a}$ & 44.2$^{a}$  & 7.28$^{b}$ & $-$0.23$^{b}$ & -- & 7\\
  J160732.85+484620.0     & 16:07:32.86 & +48:46:20.1       & $-$221 & 958   & 2647 & 44.29 & 7.67 & $-$0.39           &22.94  & 9\\
  J161259.83+421940.3     & 16:12:59.84 & +42:19:40.3       & $-$245 & 1079  & 1154 & 44.31 & 6.80 & \phantom{$-$}0.54 &23.77  & 9\\
  J163631.58+461704.3     & 16:36:31.58 & +46:17:04.3       & $-$220 & 580$^{c}$   & 2373$^{a}$ & 44.18$^{a}$ & 7.41$^{b}$ & $-$0.38$^{b}$ & -- & 7\\
  J170116.08+354945.2     & 17:01:16.08 & +35:49:45.2       & $-$210 & --    & 1783$^{a}$ & 43.77$^{a}$ & 6.74$^{b}$ & $-$0.12$^{b}$ & -- & 7\\ 
  J171207.44+584754.5     & 17:12:07.46 & +58:47:54.5       & $-$190 & 593   & 2193$^{a}$ & 44.18$^{a}$ & 7.49$^{b}$ & $-$0.46$^{b}$ & -- & 7\\  
  J171304.46+352333.5     & 17:13:04.47 & +35:23:33.6       & $-$241 & 725   & 1393 & 44.23 & 7.03 & \phantom{$-$}0.15 &23.31  & 9\\
  J171829.01+573422.4     & 17:18:28.99 & +57:34:22.4       & $-$150 & 470   & 1760 &  --   & 6.60 & $-$0.30           &--     & 6\\
  J204527.70-003236.2     & 20:45:27.71 & $-$00:32:36.2     & $-$334 & 904   & 1632 & 44.30 & 7.30 & \phantom{$-$}0.06 &--     & 9\\
                          &             &                   & $-$276 & 540$^{c}$   & 2053$^{a}$ & 44.17$^{a}$ & 7.26$^{b}$ & $-$0.24$^{b}$ & -- & 7\\
  J212347.83-082842.9     & 21:23:47.84 & $-$08:28:43.0     & $-$171 & 526$^{c}$   & 3231$^{a}$ & 44.13$^{a}$ & 7.72$^{b}$ & $-$0.73$^{b}$ & -- & 7\\
  RXJ2217.9-5941          & 22:17:56.64 & $-$59:41:31.0     & $-$330 & 1140  & 1370 & 44.92 & 7.59 & \phantom{$-$}0.32 &--     & 2\\
  J232326.17-010033.1     & 23:23:26.17 & $-$01:00:33.1     & $-$769 & 2625  & 3683 & -- & 8.22$^{g}$ & -- & -- & 8\\
  RXJ2340.6-5329          & 23:40:23.19 & $-$53:28:56.1     & $-$490 & 780   & 1230 & 44.99 & 7.55 & \phantom{$-$}0.44 &--     & 2\\
  MS2340.9-1511           & 23:43:28.17 & $-$14:55:40.6     & $-$420 & 780   & 970  & 44.50 & 7.01 & \phantom{$-$}0.50 &--     & 2\\
\hline\end{tabular}
\end{tiny}

\medskip
$^{a}$Data taken from \citet{2011ApJ...736...86D}; $^{b}$data obtained using our formula and the FWHM of H$\beta_{\rm b}$ and the $\lambda L_{\lambda}(5100)$ taken from \citet{2011ApJ...736...86D}; $^{c}$data obtained from \citet{2008MNRAS.383..581D}; $^{d}$data obtained from \citet{2012ApJ...755..167D}; $^{e}$FWHM of the core component obtained from \citet{2001A&A...372..730V}; $^{f}$FWHM obtained from \citet{2003MNRAS.345.1133M}; $^{g}M_{\rm BH}$ obtained using our formula and the FWHM(H$\beta$) and $L$(H$\beta$) taken from \citet{2015arXiv151202642Z}.
\end{minipage}
\end{table*}
In case of objects with multiple measurements, we used our values or the values obtained with the method more similar to ours. 

\begin{figure}
\includegraphics[width=\columnwidth]{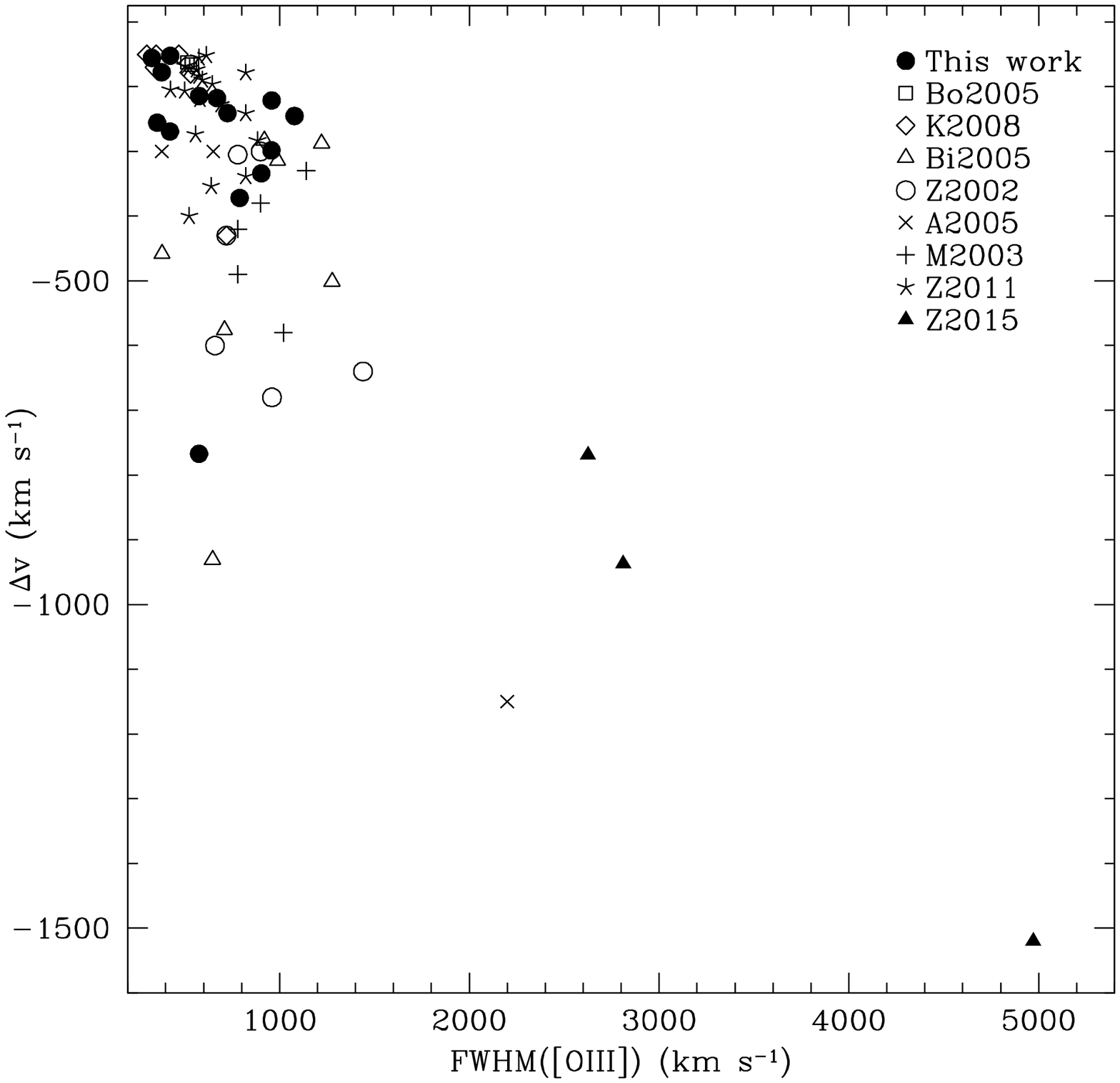} 
\caption{The [O\,{\sc iii}] FWHM plotted against the [O\,{\sc iii}] velocity shift. Solid circles are our data, open squares are the data by \protect\citet{2005AJ....130..381B}, open diamonds are \protect\citet{2008ApJ...680..926K} data, open triangles are the data from \protect\citet{2005MNRAS.364..187B}, open circles are the objects from \protect\citet{2002ApJ...576L...9Z}, crosses are the data by \protect\citet{2005ApJ...618..601A}, plus symbols are the data from \protect\citet{2003MNRAS.345.1133M}, stars are the data from \citet{2011ApJ...737...71Z} and solid triangles are the data from \citet{2015arXiv151202642Z}.}
\label{fw_shift_O3_blue}
\end{figure}

Higher blueshifts of [O\,{\sc iii}] are associated to broader [O\,{\sc iii}] emission lines \citep{2005AJ....130..381B, 2005ApJ...618..601A, 2008ApJ...680..926K}. This trend was observed also for neon lines \citep{2009ApJ...702L..42S} and for the C{\sc iv}$\lambda$1549 line \citep{2013ApJ...769...30G}. 
In Fig.~\ref{fw_shift_O3_blue} the [O\,{\sc iii}] FWHM was plotted against the blueshift  for the whole sample.
The result is a significant correlation ($r_{\rm s}=-0.64$, p-value $=5\times10^{-9}$) with a large spread, confirming that the velocity dispersion of the NLR gas is larger in objects showing a higher blueshift and reinforcing the idea that the blueshift could be the result of outflowing gas coming from the inner NLR \citep{2015arXiv151107138M}. 
We stress that we used the FWHM of the core component, like  \citet{2001A&A...372..730V} and \citet{2008ApJ...680..926K}, while the other authors used the width of the whole profile. Obviously, in this latter case larger values of FWHM are found. A less significant correlation ($r_{\rm s}=-0.57$, p-value $=6\times10^{-3}$) was obtained considering only the FWHM of the core component, possibly due to the smaller number of objects. 

\begin{figure}
\includegraphics[width=\columnwidth]{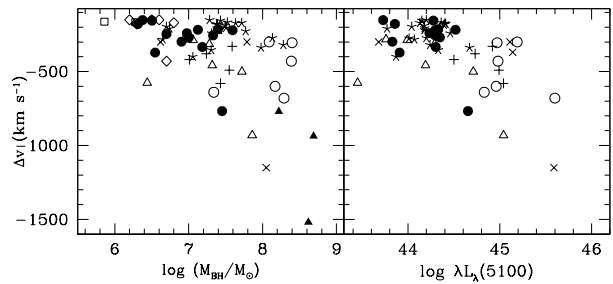} 
\caption{Left panel: the black hole mass against the velocity shift for the whole sample of blue outliers. Right panel: $\lambda L_{\lambda}(5100)$ against the velocity shift for the whole sample of blue outliers. Symbols are as in Fig.~\ref{fw_shift_O3_blue}.}
\label{Mbh_Deltav}
\end{figure}

According to \citet{2005ApJ...618..601A}, the necessary condition to produce a blue outlier is a combination of high Eddington ratio ($\log (L_{\rm bol}/L_{\rm Edd}) > 0.1$), large black hole mass (M$_{\rm BH} > 10^7$ M$_{\sun}$) and also large continuum luminosity ($\log \lambda L_{\lambda}(5100\,{\rm \AA})>44.6$ erg s$^{-1}$).
Weak but significant correlations were found both with the black hole mass and with the continuum luminosity ($r_{\rm s}=-0.48$, p-value $=3\times10^{-5}$ and $r_{\rm s}=-0.46$, p-value $=3\times10^{-4}$, respectively), suggesting that blue outliers tend to show higher blueshifts as the black hole mass or the continuum luminosity increases (Fig.~\ref{Mbh_Deltav}). Furthermore, the last two criteria of \citet{2005ApJ...618..601A} seem to be well fulfilled by the higher-velocity blue outliers.

Finally, no correlation appears to exist between $\Delta v$ and $L_{\rm 1.4 GHz}$ ($r_{\rm s}=-0.17$ and p-value $= 0.002$), and this can be due to the lack of powerful radio sources in our sample. Indeed, \citet{2015arXiv150605800B} found more blue outliers in radio-loud NLS1s. Therefore a large sample of radio-loud NLS1s showing high [O\,{\sc iii}] blueshifts should be analysed to check if the radio emission can significantly affect the NLR kinematics when the radio power is high, as suggested by the work of \citet{2015arXiv150605800B}.

\section{Discussion}
\subsection{H$\beta_{\rm b}$ profile}
The question of the shape of the permitted emission lines in NLS1s is still open. NLS1s seem to have symmetric emission lines with Lorentzian rather than Gaussian shapes. Indeed, many authors applied a Lorentzian function to fit the profile of H$\beta$ \citep{1996ApJS..106..341M, 1999ApJS..125..317L, 1999A&A...341..662G, 2001A&A...372..730V, 2006ApJS..166..128Z, 2012AJ....143...83X}. \citet{1999ApJS..125..317L} affirmed that a Lorentzian profile in NLS1s provides a substantially better description of the H$\beta$ shape than a single Gaussian. Lorentzian profiles give, as a result, narrow emission line ratios in good agreement with those observed in the NLR of Seyfert 2 galaxies as found by  \citet{2001A&A...372..730V} who suggested also that the broad Balmer lines could be emitted by a large disk. \citet{2012AJ....143...83X} affirmed that a single Lorentzian profile is essentially equivalent to a double-Gaussian profile when fitting NLS1s, while it is not appropriate to fit BLS1s.

On the contrary, \citet{2005ApJ...623..700D} found that Lorentzian profiles yield less satisfying results than employing two Gaussian components having different FWHMs. \citet{2007ApJ...654..754N} used two broad Gaussians to fit the H$\beta$ profile, while \citet{2008MNRAS.385...53M} and \citet{2011ApJ...737...71Z} applied a multi-Gaussian fitting. The double Gaussian can be justifed by assuming that the BLR is made of two kinematically different regions placed at different distances from the central source: an intermediate-line region and a very broad-line region \citep[see e.g.][and references therein]{2010ApJS..189...15K}. A multi-Gaussian approach could suggest that the BLR is made of a continuum of different regions.

Our analysis of the H$\beta$ broad emission line in a sample of 296 NLS1s and 64 BLS1s suggests that the Lorentzian profiles are predominant in NLS1s. More in general, a Lorentzian profile seems to be more suited for objects having ${\rm FWHM}<4000$ km~s$^{-1}$ \citep{2002ApJ...566L..71S}. Our BLS1s appear to be transition objects between NLS1s and the BLS1s family sharing common properties with NLS1s, in agreement with \citet{2000ApJ...536L...5S} and \citet{2000NewAR..44..503L} who declared that NLS1s show a clear continuity with BLS1s and that they do not really represent a disjoint class of AGNs. Furthermore, our upper limit to define a NLS1 (FWHM(H$\beta_{\rm b}$) $<$ 2200 km s$^{-1}$) depends on the spectral resolution of our data rather than on a different distribution of FWHM for the two samples. Therefore, it seems that the main difference between the BLS1 and the NLS1 families is not the widths of the broad line profiles, but their shapes. Indeed, in our sample there is a continuity in the distribution of profiles going from the Lorentzian symmetric shapes of almost all NLS1s (with the exception of 5 cases) to the double-Gaussian and often asymmetric profiles of about a third of our BLS1s. This effect can be explained if we assume that the BLR motion is due to a turbulent component kinematically represented by a Lorentzian profile combined with a rotational component represented by a Gaussian profile \citep{2013A&A...549A.100K}. NLS1s would be slow-rotating AGNs, in which narrower Lorentzian profiles are the effect of the dominant role of the turbulent motions \citep{2012MNRAS.426.3086G}. This, in turn, can be due to the presence of less massive BHs in NLS1s than in BLS1s, even if an inclination effect of a flat distribution of BLR clouds cannot be completely ruled out \citep{2008MNRAS.386L..15D} and it could explain the spread in the FWHM distribution, as suggested by \citet{2014Natur.513..210S}. However, other authors \citep[see e.g.][]{2005MNRAS.356..789B, 2005ApJ...627..721G, 2007ApJ...667L..33K, 2011ApJ...739...28X, 2015ApJ...801...38W} have shown that lower BH masses in NLS1s are associated to lower stellar velocity dispersion. This is in agreement with the results of \citet{2015A&A...575A..13F} and \citet{2015A&A...578A..28B}, who measured the BH masses with the second-order momentum of the line, which is independent of the inclination. Therefore, we do not believe that, in general, NLS1s show narrower permitted lines because of an inclination effect of a disk-like BLR. 

Lower BH masses imply higher normalized accretion rates in NLS1s with respect to BLS1s, as confirmed also by our data. Indeed, the objects fitted by a Lorentzian function show also higher values of Eddington ratio than the objects fitted by a double-Gaussian profile. Since they are mostly NLS1s, this result reinforces the idea that NLS1s are characterized by both Lorentzian shape profiles and high Eddington ratios. 
\citet{2007ApJ...654..754N} suggested that the Eddington ratio could be used to define NLS1s in place of the classic criterion of FWHM(H$\beta)<2000$\,km\,s$^{-1}$. In particular, a ratio $\frac{L_{\rm bol}}{L_{\rm Edd}}\geq 0.25$ should imply NLS1s. Before comparing physical quantites from different works, it is necessary to take into account that the values of the Eddington ratios depend significantly on the methods used to measure the BH mass and to calculate the bolometric luminosity. Since \citet{2007ApJ...654..754N} followed a different approach, we re-calculated both our BH masses and Eddington ratios by means of their formulas, finding, on average, BH masses 0.2 dex higher and Eddington ratios $0.37$ dex lower. This last value moves the proposed criterion to $\frac{L_{\rm bol}}{L_{\rm Edd}}\geq 0.6$. The median FWHM(H$\beta$) is about 2000 km s$^{-1}$ around $\frac{L_{\rm bol}}{L_{\rm Edd}} = 0.6$ and 261 out of 296 NLS1s are above this limit, while 54 out of 64 BLS1s are below. 
This confirms that, also in our case, the Eddington ratio can be effectively used to classify NLS1s.

\subsection{Fe\,{\sc ii} properties}
 
\begin{figure}
\includegraphics[width=\columnwidth]{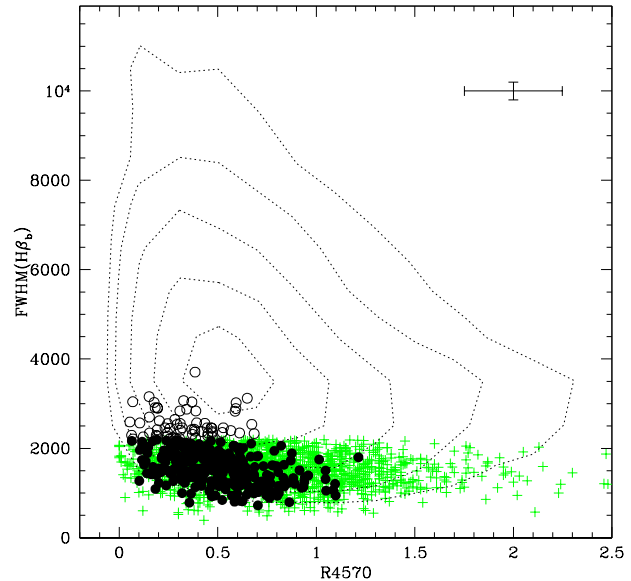} 
\caption{The values of R4570 are plotted against the FWHM of the broad component of H$\beta$. 
Solid circles are NLS1s, while open circles are BLS1s. Green plus symbols are the data from \citet{2006ApJS..166..128Z}, while the contours are adapted from fig.~1 of \citet{2014Natur.513..210S}. }
\label{R4570_FW_Hb}
\end{figure}

\begin{figure}
\includegraphics[width=\columnwidth]{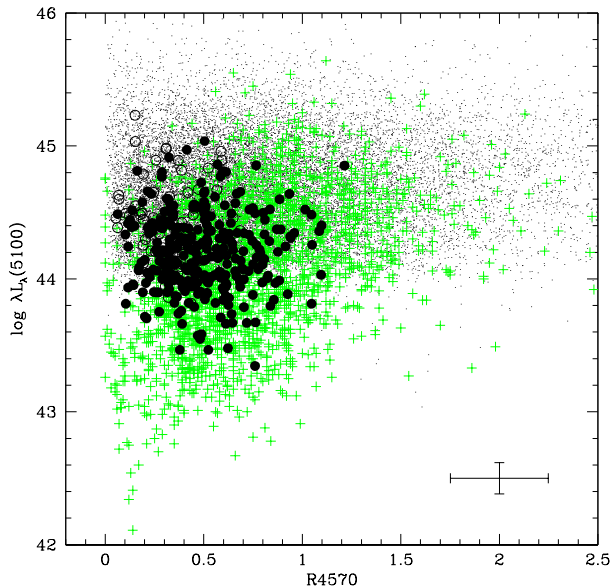} 
\caption{The values of R4570 are plotted against the $\lambda L_{\lambda}$(5100). Symbols are the same as in Fig.~\ref{R4570_FW_Hb}, except for black dots, that are data from \citet{2011ApJS..194...45S}.} 
\label{R4570_L5100}
\end{figure}

NLS1s are known to be strong Fe\,{\sc ii} emitters. Indeed,  the highest values of R4570 are observed in objects having FWHM(H$\beta)< 4000$ km~s$^{-1}$ \citep{2014Natur.513..210S}. However, it is also true that NLS1s  show a wide range of R4570 values, therefore this cannot be a distinctive property of these AGNs.
\citet{2006ApJS..166..128Z}, \citet{2007ApJ...654..754N} and \citet{2014Natur.513..210S} calculated the ratio between Fe4570 and H$\beta$ broad and obtained ranges extended up to more than 2, as it can be seen in Fig.~\ref{R4570_FW_Hb}, where we plotted R4570 versus FWHM(H$\beta_{\rm b}$) finding that our data are confined to a smaller range of R4570 values compared to those from \citet{2006ApJS..166..128Z} (green plus symbols) and \citet{2014Natur.513..210S} (black contours). 
The data from \citet{2007ApJ...654..754N} were not available. It is difficult to give a convincing answer to this difference.
First of all, we must remind that we used the total flux of H$\beta$, however the negligible contribution of the narrow component does not affect the following analysis.
Our sample was matched with the \citet{2006ApJS..166..128Z} one, finding 120 objects in common.
If we consider only this sample, the resulting distribution is consistent with ours: the median value of R4570 measured by Zhou et al. is 0.54, perfectly in agreement with the median value of our sample of NLS1s. 
Both \citet{2006ApJS..166..128Z} and \citet{2014Natur.513..210S} chose an upper limit in redshift higher than ours, $z < 0.8$ and $z < 0.9$ respectively, therefore we checked whether the luminosity could cause this difference. \citet{2004ApJ...614..558N} suggested that Fe\,{\sc ii}/H$\beta$ depends on the luminosity and, in particular, that the most extreme values of this ratio require very high luminosity. Our values show a range similar to those of \citet{2006ApJS..166..128Z} and \citet{2014Natur.513..210S}, $\lambda L_{\lambda}$(5100) $\sim 10^{42.5-45}$ and $\sim 10^{44-45.5}$ respectively. By plotting R4570 versus $\lambda L_{\lambda}$(5100) we observed that our measurements are included within those of Zhou et al., and, mostly, that their higher values of R4570 belong to the upper end of their luminosity range, but they are not the highest ones (Fig.~\ref{R4570_L5100}). Therefore, a bias in luminosity can be excluded.
We estimated both EW(H$\beta$) and EW(Fe4570) from the measurements of \citet{2006ApJS..166..128Z} and \citet{2014Natur.513..210S}. 
The K--S test shows that the two distributions are significantly different from ours (with p-values $<< 10^{-7}$). 
Indeed, there is a higher fraction of objects having low EW(H$\beta$) and high EW(Fe4570) in the sample of Zhou et al. and Shen \& Ho, which could justify the excess of high R4570 ratios. Rejecting the hypothesis that our limited range of R4570 values is caused by our lower limit of S/N ratio, which should affect the sample in the opposite direction, the last hypothesis is the less constrained selection criteria applied by \citet{2006ApJS..166..128Z} and \citet{2014Natur.513..210S}, which allowed them to extract and then analyse samples 10--100 times larger than ours.

\citet{2004ApJ...614..558N}  found that R4570 correlates with $L/L_{\rm Edd}$ ($r_{\rm s}=0.48$), and suggested that the increase of R4570 with the Eddington ratio is likely caused by the decrease of EW(H$\beta$) with $L/L_{\rm Edd}$. 
Our data confirm this correlation: our NLS1s show larger ranges of both R4570 and Eddington ratio, than our BLS1s which are confined to lower values. We agree with their hypothesis, furthermore our data suggest that R4570 correlates simultaneously with the equivalent  widths of Fe\,{\sc ii} ($r_s = 0.64$) and H$\beta$ ($r_s=-0.60$) at a high level of significance. This is in agreement with similar previous findings by \citet{1992ApJS...80..109B} and \citet{1996A&A...309...81W}. 
The Eddington ratio is also considered the main driver to explain the R4570--EW([O\,{\sc iii}]) anti-correlation and, more generally, all the correlations forming the so-called EV1 \citep[see e.g.,][]{1994ApJ...435..611L,2000NewAR..44..503L,2000ApJ...536L...5S,2001ApJ...558..553M,2002ApJ...565...78B,2004MNRAS.350L..31B,2004AJ....127.1799G,2004ApJ...614..558N,2009ApJ...703L...1D}. On the other hand, other physical properties have been proposed as underlying drivers, like the black hole mass and inclination angle \citep[see e.g.,][]{1994ApJ...435..611L,1996A&A...309...81W}. The orientation hypothesis was already rejected by some authors \citep[see e.g.,][and references therein]{2000ApJ...542..631K,2004AJ....127.1799G}. The idea that the Eddington ratio is the most important factor was later confirmed and re-inforced by \citet{2014Natur.513..210S}, who analysed about 20 000 broad line AGNs extracted from the SDSS. EV1 was successfully tested with our data: R4570 and EW([O\,{\sc iii}]) are well and significantly anti-correlated. The analysis was extended and the [O\,{\sc iii}] core and wing components were separated. Clear anti-correlations were found, but in case of the wing component the scatter is larger. Similarly,  \citet{2014Natur.513..210S} decomposed the [O\,{\sc iii}]$\lambda$5007 profile by fitting each line with core plus wing components, but they found that while the core component anti-correlates strongly with R4570, the same does not occur for the wing component. 
They argued that the excitation of the core component is dominated by photoionization, as expected, while the outflowing gas of the wing component is likely excitated by additional mechanisms, as shocks. 
Finally, it is worth to discuss the existence of the anti-correlation between R4570 and FWHM(H$\beta_{\rm b}$) ($r = -0.55$), that belongs to the EV1 correlations firstly reported by \citet{1992ApJS...80..109B} and later confirmed in different works \citep{1996A&A...309...81W,2000ApJ...536L...5S,2001A&A...372..730V,2012AJ....143...83X}. On the contrary, \citet{2006ApJS..166..128Z} obtained only a trend between these two quantities, with $r_{\rm s} =-0.23$. 
Our results indicate that if an anti-correlation exists between R4570 and FWHM(H$\beta$), it is relatively weak even if significant, both for NLS1s and for the whole sample of 360 objects. However, the recent plot by \citet{2014Natur.513..210S} (see their fig.~1) approximately reproduced by means of contours in our Fig.~\ref{R4570_FW_Hb}, changed our view on this question. Indeed, it confirms what \citet{2000ApJ...536L...5S} noticed about the lack of AGNs with very broad lines and high R4570 ratios, but also shows that the anti-correlation was an apparent effect caused probably by the limited number of available data or the limited ranges of the two physical quantities involved. In the R4570--FWHM(H$\beta$) plane, the distribution of the broad line AGNs forms an horizontal sequence driven by the R4570-EW([O\,{\sc iii}]) anti-correlation, which in turn depends on the Eddington ratio, as already suggested also by \citet{2004ApJ...614..558N} and \citet{2007ApJ...654..754N}.

\subsection{O\,{\sc i} and CaT properties}

The origin of the O\,{\sc i}\,$\lambda$8446 was debated for long time and it is not yet completely understood \citep[see e.g.][]{1980ApJ...238...10G, 2008ApJS..174..282L}. This line seems to be produced by Ly-$\beta$ fluorescence, that excites electrons which return to the ground state through a series of transitions, producing photons at 11287, 8446 and 1302 \AA. Other possible mechanisms are collisional excitation and recombination, which can be evaluated on the basis of the presence and intensity of O\,{\sc i}\,$\lambda$7774: the ratio O\,{\sc i}\,7774/8446 should be $\sim$0.3 in case of collision \citep{1980ApJ...238...10G} and $\sim$1.1 in case of recombination \citep{2008ApJS..174..282L}. The observations show that the Ly$\beta$ fluorescence cannot explain the observed intensity of O\,{\sc i}\,$\lambda$8446 \citep{2002ApJ...572...94R,2008ApJS..174..282L,2007ApJ...663..781M} and that an additional process needs to be invoked. 
The continuum fluorescence, proposed by \citet{1976ApJ...207..713O}, was excluded by \citet{1980ApJ...238...10G}, \citet{2002ApJ...572...94R} and \citet{2007ApJ...663..781M} because the expected O\,{\sc i} transitions at 7254, 7990 and 13165 \AA were too weak or absent. \citet{2008ApJS..174..282L} reported to have measured the O\,{\sc i}\,$\lambda$13165 in 5 objects, even if in their figures this line is, in fact, mildly visible only in two cases. 
\citet{1980ApJ...238...10G} did not see O\,{\sc i}\,$\lambda$7774 in their sample of 16 Seyfert 1 galaxies. \citet{2002ApJ...572...94R} found this line only in one of their seven AGNs, while \citet{2008ApJS..174..282L} measured it in only 7 out of 23 sources and ascribed it to recombination because the O\,{\sc i}\,7774/8446 ratio was always larger than 1, but in one case. They concluded that Ly$\beta$ fluorescence contributes to O\,{\sc i}\,$\lambda$8446 for 50 per cent and more in 20 out of 23 objects and that the additional contribution is given mostly by recombination. 

Our data confirm that O\,{\sc i} is mostly caused by Ly$\beta$ fluorescence with an additional but apparently rather uncommon contribution given by collisional excitation. Indeed, in our sample O\,{\sc i}\,$\lambda$7254,7990, potentially visible in 307 and 161 out of 360 spectra respectively, are absent or at least too faint to be detected (O\,{\sc i}\,$\lambda$13165 is obviously not detectable in our spectral range). This excludes the continuum fluorescence as additional mechanism. O\,{\sc i}\,$\lambda$7774 was detected and measured in only 12 out of 214 spectra. The O\,{\sc i}\,7774/8446 ratios have a median value of 0.1 and range between 0.04 and 0.42, suggesting the contribution of collisional excitation. In conclusion O\,{\sc i} emission depends more on the radiation field, than on the density of the gas.

CaT is a transition onto a metastable triplet produced by collisional excitation and it is known to be emitted by a high-density, optically-thick neutral gas. \citet{1988ApJ...330..751P}, \citet{2007ApJ...663..781M}, \citet{2013AstRv...8d...4M} and \citet{2015ApJS..217....3M} claimed that $\log N_{\rm H}$ between 11 and 12 and $\log U$ between $-2.5$ and $-1.5$ are required to justify CaT emission. These physical values seem to be appropriate also for O\,{\sc i}\,$\lambda$8446 \citep{2007ApJ...663..781M} but, according to \citet{2013AstRv...8d...4M} and \citet{2015ApJS..217....3M}, it is likely that O\,{\sc i} needs a higher photon flux than CaT and that it can be emitted in less extreme conditions, where CaT does not appear. 

We found that the intensity of Fe\,{\sc ii} is well correlated with the intensity of Ca\,{\sc ii} and slightly less with O\,{\sc i}, which could be emitted in a region with different physical properties. In fact, our models indicate that Ca\,{\sc ii} emission requires high density gas to be detected, independently on its ionization degree, and these conditions support the collisional excitation mechanism invoked to explain the frequently observed strong Fe\,{\sc ii} emission in type 1 AGNs and especially in NLS1s. High density and high ionization are mandatory for strong O\,{\sc i} emissions, while lower values of these physical parameters can explain weaker emissions. 
It is interesting to note that O\,{\sc i}\,$\lambda$8446 was detected and measured in 41 objects out of 67 having a spectral range including this line, while CaT only in 15 out of 52. Therefore 26 out of 41 objects with O\,{\sc i}\,$\lambda$8446 have no detection of CaT. In 11 cases, O\,{\sc i}/H$\beta_{\rm b}$ is low ($<0.15$, non reddening corrected) and these small values could require different combination of density and ionization parameter, such as lower density which would weaken CaT up to make it non detectable or hardly measurable. The other 15 objects have high values of O\,{\sc i}/H$\beta_{\rm b}$ ($\geq 0.15$) which imply, according to our models, a high density ($\log N_{\rm e}>11.5$). In these conditions, high values of CaT/H$\beta_{\rm b}$ are expected, as well. CaT can be observed in absorption due to the stellar component and this could explain the lack of detection, however NLS1s are characterized by low stellar velocity dispersion and narrow absorption lines which cannot suppress the quite strong and broad CaT emission lines.  
In conclusion, while on the one hand it is possible to justify why we can observe O\,{\sc i} more frequently than CaT, on the other hand the question about the apparent lack of CaT when O\,{\sc i} is strong is still open and it requires new high S/N spectra to be solved.

The question of the location of the gas emitting Fe\,{\sc ii}, O\,{\sc i} and CaT is also still open and debated for long time.
Our analysis shows that Fe\,{\sc ii} and H$\beta$ share a similar kinematics, even if with a large scatter. This result is in agreement with \citet{2003ApJS..145..199M}, who found a correlation between FWHM(H$\beta_{\rm b}$) and FWHM(Fe4570) in objects with FWHM(H$\beta_{\rm b}$) $< 4000$ km\,s$^{-1}$, and with \citet{2010ApJS..189...15K}, who used two Gaussians for fitting H$\beta_{\rm b}$ and showed that the Gaussian corresponding to an intermediate broad line region shares the same kinematics of Fe\,{\sc ii}. 
Recently \citep{2015ApJ...804..138H} confirmed this result through a reverberation-mapping analysis of a small number of AGNs and contradicted their previous findings \citep{2008ApJ...687...78H} that the kinematics of Fe\,{\sc ii} and H$\beta_{\rm b}$ are well correlated, but the FWHM(Fe\,{\sc ii}) is systematically lower than the FWHM(H$\beta_{\rm b}$).
On the contrary \citet{2002ApJ...572...94R} found that Fe\,{\sc ii} and O\,{\sc i} have similar profiles, while Pa$\beta$ is much broader, suggesting that they both come from outer regions of the BLR. Similarly \citet{2016arXiv160205159M} claimed that Fe\,{\sc ii}, O\,{\sc i} and also CaT are emitted by an outer part of the BLR with respect to hydrogen \citep{2015ApJS..221...35K}.
\citet{1980ApJ...238...10G} affirmed that O\,{\sc i} and H$\alpha$ profiles are very similar and excluded that O\,{\sc i} is emitted in the region of neutral hydrogen. But in fact, the FWHMs of these lines were in agreement only in 3 out of 8 cases for which the FWHM of O\,{\sc i} could be measured.
We obtained a good agreement between the FWHMs of CaT and H$\beta$, adding data from \citet{1988ApJ...330..751P} and \citet{2015ApJS..217....3M}, while O\,{\sc i} FWHMs appear to be systematically lower than those of H$\beta_{\rm b}$. In other words, our results suggest that  Fe\,{\sc ii}, CaT and hydrogen are emitted by BLR gas located at the same distance, while O\,{\sc i} seems to originate from a region located at a larger distance. This is in agreement with \citet{1988ApJ...330..751P} who found that the FWHMs of O\,{\sc i} and H$\beta$ correlate, but the FWHMs of O\,{\sc i} grow more slowly, and with \citet{2013AstRv...8d...4M} who said that O\,{\sc i} probably comes from regions not exactly coincident with those emitting CaT.

\subsection{[O\,{\sc iii}] asymmetry and blue wings}
The majority of our sample (94 per cent) shows [O\,{\sc iii}] with asymmetric profiles and no difference appears to exist between NLS1s and our BLS1s, as also found by \citet{2013MNRAS.433..622M}. \citet{2012MNRAS.427.1266V} found blue wings in 73 per cent of their sample of intermediate-type Seyfert galaxies and in 68 per cent of their sample of Seyfert 2. A slightly smaller percentage (43 per cent) was found by \citet{2016ApJ...817..108W} who claimed that outflows are common in very bright type 2 AGNs. These decreasing percentages, going from type 1 to type 2 AGNs, are in agreement with the Unified Model and suggest that the gas kinematics of the NLR is more turbulent in its inner part. From our analysis we can infer that this asymmetry is very likely caused by the presence of outflowing gas from  the inner regions of the active nucleus, which interacts with the surrounding medium both by transferring kinetic energy and making it more turbulent and by reducing the EW of [O\,{\sc iii}] as its velocity increases. Indeed, in their analysis of the properties of [O\,{\sc iii}] emission line profiles, \citet{2011ApJ...737...71Z} found that the shift of the wing component correlates with the EW of the [O\,{\sc iii}] whole emission ($r_{\rm s}=0.38$, p-value $=10^{-14}$). \citet{2012ApJ...756...51L} confirmed that AGNs with blue wings in [O\,{\sc iii}] are characterized by lower values of EW and claimed that the radial motion of the clouds can decrease the covering factor of the NLR, resulting in a lower EW. 

The high Eddington ratio in NLS1s is in principle expected to affect also the NLR, as the resulting radiation pressure accelerates the NLR gas \citep{2005ApJ...627..721G}. However, it seems that the presence and the strength of a blue wing are not correlated with the Eddington ratio \citep{2011ApJ...739...28X,2012ApJ...756...51L}, even if high values of $L_{\rm bol}/L_{\rm Edd}$ ($>0.1$) could be associated with more prominent blue wings \citep{2013MNRAS.433..622M}. We found that the outflow velocity does not seem to depend on the AGN power and only weakly on the accretion rate and the radio luminosity. Nonetheless, we did not observe fast outflows in objects showing low luminosities and low accretion rates. 
Indeed, according to \citet{2011ApJ...737...71Z}, while the AGN determines the starting values of the outflow velocity, its final speed is more related to the density of the ISM: high density ISM implies more [O\,{\sc iii}] emission and decelerates the outflow more efficiently. 
Unexpectedly, no correlation was found between the velocity of the outflow and the electron density of the medium, in disagreement with \citet{2007ApJ...670...60X}, but in fact we measured the average density of the low-ionization gas, which is not representative of the whole medium and does not take into account possible small size and overdense regions \citep{2012MNRAS.427.1266V}. In conclusion, in our case, it is not clear whether a physical connection related to the ISM properties and in particular to its density really exists between the [O\,{\sc iii}] emission and the presence of a wing. On the other hand, high radio luminosities or accretion rates seem to be a necessary, but not sufficient condition, because the velocity of the outflowing gas and the strength of its emission must depend on the environment through which it moves. 

\subsection{[O\,{\sc iii}] blueshift}
The radial motion of the gas seems to be responsible also for the observed blueshift of the core component of [O\,{\sc iii}]. Our analysis shows that the faster is the outflow the higher is the blueshift. In this picture the so-called blue outliers are only the objects at the extreme tail of the blueshift distribution, as already suggested by \citet{2011ApJ...737...71Z}. The continuous distribution in Fig.~\ref{v1_vS2_vs_v2} justifies the small number of observed blue outliers, since high blueshifts can be detected only in objects with very fast outflows ($v_2 - v_{\rm [S\,II]} < -400$ km~s$^{-1}$). Of course, this is a first result which must be definitely proved with spectra having higher S/N and higher spectral resolution than SDSS spectra.

Fast outflows make the gas more turbulent, both considering the core and the wing components. This is more evident in blue outliers, whose FWHMs are found to increase with increasing blueshift. 
This trend was observed not only for [O\,{\sc iii}] lines \citep[e.g.][]{2008ApJ...680..926K}, but also for other NLR lines \citep[e.g.][]{2009ApJ...702L..42S} and for the C\,{\sc iv}\,$\lambda1549$ line \citep[e.g.][]{2013ApJ...769...30G}.
High-ionization NLR lines show more frequently large shifts, as found for [O\,{\sc iii}] and [Fe\,{\sc vii}] in this work, for [Ne\,{\sc iii}] and [Ne\,{\sc v}] by \citet{2009ApJ...702L..42S} and for lines up to [Fe\,{\sc x}] by \citet{2008ApJ...680..926K}. Thus, the frequency and the strength of the shift increase with increasing ionization potential, confirming the scenario of an outflow moving through a stratified ISM ionized by the nuclear source.   
Moreover, the high-ionization BLR C\,{\sc iv}\,$\lambda1549$ line is found to be blueshifted by about 600 km\,s$^{-1}$ \citep{1982ApJ...263...79G} and to correlate with [O\,{\sc iii}] blueshift \citep{2002ApJ...576L...9Z}. These correlations and other similarities between [O\,{\sc iii}] and C\,{\sc iv} shift suggest that there could be a kinematic connection between BLR and NLR \citep{2002ApJ...576L...9Z,2011ApJ...737...71Z}. \citet{2013ApJ...769...30G} affirmed that an explanation for the observed blueshift could be the Rayleigh scattering associated to gas inflowing towards the SMBH. The kinematic connection between BLR and NLR gas could imply that also NLR blueshifts are due to inflow and scattering. 

If the radial motion of the clouds is expected to decrease the ionized gas covering factor and to reduce its emissivity, we should observe for the core component of [O\,{\sc iii}] a similar relation found between the velocity shift of the wing and the EW of the line. But this was not the case, since we  did not find any correlation showing that larger blueshifts are preferentially detected in objects having weaker [O\,{\sc iii}] lines, in disagreement with \citet{2011ApJ...737...71Z}. We only observed a slight decrease of the fraction of objects with blueshifts larger than $-150$ km\,s$^{-1}$ as the EW increases. 

The radial motion could be caused by the accretion disk wind and therefore we should expect to find a correlation with the Eddington ratio, as found by \citet{2005ApJ...618..601A}. Indeed, \citet{2011ApJ...737...71Z} claimed that the blueshift is mainly affected by the Eddington ratio. With our data this hypothesis cannot be confirmed, because we did not find a correlation between $\Delta v$ and $L_{\rm bol}/L_{\rm Edd}$. The blueshift seems to weakly depend on the Eddington ratio, but high blueshifts are not detected at low values of this parameter. Interestingly, even the blue outliers appear not to depend on the Eddington ratio, but rather on the BH mass and the luminosity of the active nucleus. However, the Eddington ratio is calculated by means of these two quantities, therefore, if the correlations of both continuum luminosity and BH mass with the velocity shift of [O\,{\sc iii}] are similar, the dependency on the Eddington ratio could disappear. 

\begin{figure}
\includegraphics[width=\columnwidth]{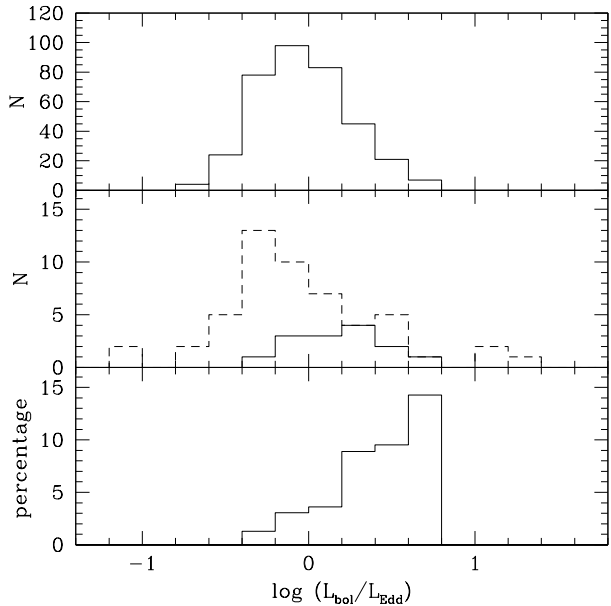} 
\caption{Comparison between the distribution of the Eddington ratios for our whole sample (top panel), the distribution of the same ratio for our blue outliers (middle panel, solid line) and for the whole sample of blue outliers (middle panel, dashed line). The bottom panel shows the frequency of blue outliers.}
\label{hist_eddington}
\end{figure}

Our blue outliers show Eddington ratios distributed within the range of values of the whole sample and they are not those AGNs with the highest values of this quantity (see top and middle panels of Fig.~\ref{hist_eddington}). However, the frequency of blue outliers increases with increasing Eddington ratio (see bottom panel of Fig.~\ref{hist_eddington}). Indeed, if it is true that blue outliers are often characterized by high $L_{\rm bol}/L_{\rm Edd}$ ratios \citep{2003MNRAS.345.1133M,2008ApJ...680..926K}, not all high Eddington ratio AGNs with narrow H$\beta_{\rm b}$ are blue outliers \citep{2005AJ....130..381B, 2005ApJ...618..601A}.   

Finally, the blueshift of [O\,{\sc iii}] core component does not seem to be affected by the radio emission, as already found for the wing component. We believe that a powerful radio emission is mandatory to significantly affect the NLR kinematics. Since only 11 NLS1s in our sample are radio-loud, the lack of a correlation with radio properties is expected. Indeed, \citet{2009ApJ...702L..42S} found a correlation between the [Ne\,{\sc iii}] linewidth and the radio luminosity for high luminosity objects and claimed that the interaction of the radio jet with the ISM could explain the observed kinematics. Furthermore, \citet{2015arXiv150605800B} found that blue outliers are more frequent in radio-loud sources likely because of an interaction between the NLR and the relativistic jets.

\section{Summary}

In this work we analysed the spectra of 360 NLS1--candidates, selected from a sample of 2372 objects extracted from the SDSS archive. 296 out of 360 turned out to be real NLS1s. We classified the remaining 64 as BLS1s and we used them to show that NLS1s and BLS1s are AGNs with contiguous properties. 
In the following the main results of this paper are summarized:
\begin{itemize}
\item[--] The broad H$\beta$ line was successfully fitted with a Lorentzian function in 291 out of 296 NLS1s and 44 out of 64 BLS1s. This kind of profile is typical in NLS1s and more in general in type 1 AGNs having FWHM(H$\beta$) $<$ 4000 km s$^{-1}$. This seems to be a more substantial difference between NLS1s and classic BLS1s rather than the different width of their broad lines.

\item[--] The Eddington ratio can be used in place of FWHM(H$\beta$) as a criterion to define NLS1s.
Therefore an AGN having broad lines with a Lorentzian profile and a high Eddington ratio is very likely a NLS1.

\item[--] R4570 in NLS1s covers a wide range of ratios, which depends on the Eddington ratio, in agreement with the Eigenvector 1. NLS1s are commonly known to have strong Fe\,{\sc ii} and indeed the highest values of R4570 are observed in AGNs with FWHM(H$\beta$) $<$ 4000 km s$^{-1}$, but in fact they can equally have weak Fe\,{\sc ii} emission. Those NLS1s showing high R4570 ratios are simultaneously characterized by strong Fe\,{\sc ii} and weak H$\beta$.

\item[--] The Fe\,{\sc ii}, calcium and hydrogen emission regions seem to share the same location in the BLR, showing a similar kinematics. The high density ($N_{\rm H} \sim 10^{11}$ cm$^{-3}$) required by Ca\,{\sc ii} triplet emission is in agreement with the collisional excitation mechanism invoked to explain the  strong Fe\,{\sc ii} emission frequently observed in NLS1s.

\item[--] Fe\,{\sc ii} is often associated to O\,{\sc i}, which seems to be emitted by a region located farther away from the SMBH and likely having different physical properties. O\,{\sc i} does not always require high density as Ca\,{\sc ii} triplet does, and it is mainly caused by Ly$\beta$ fluorescence with an additional, but rare, contribution of collisional excitation.

\item[--] 94 per cent of our entire sample shows [O\,{\sc iii}] lines with an asymmetric profile. No difference appears to exist between NLS1s and BLS1s with respect to this property. The asymmetry is produced by gas outflowing from the inner regions of the AGN. This gas is turbulent and affects the surrounding medium making it turbulent,  as well. In addition, the outflow is likely responsible for the observed blueshift of the core component of [O\,{\sc iii}] line.
The velocity of the outflow does not seem to depend on the AGN power or on the radio emission and only very weakly on the accretion rate. Nonetheless, fast outflows are not observed in low-luminosity and low-accretion rate objects.

\item[--] Blue outliers are rare likely because very fast outflowing gas is mandatory to make the [O\,{\sc iii}] core component strongly blueshifted. The amount of the shift in blue outliers seems to be correlated both with the mass of the SMBH and with the continuum luminosity, but not with the Eddington ratio. However, blue outliers are high-Eddington-ratios AGNs, therefore having high Eddington ratios is likely a necessary but not a sufficient condition. 

\end{itemize}

\section*{Acknowledgments}

We thank the referee for useful comments and suggestions.

Funding for the SDSS and SDSS-II has been provided by the Alfred P. Sloan Foundation, the Participating Institutions, the National Science Foundation, the U.S. Department of Energy, the National Aeronautics and Space Administration, the Japanese Monbukagakusho, the Max Planck Society, and the Higher Education Funding Council for England. The SDSS Web Site is http://www.sdss.org/.

The SDSS is managed by the Astrophysical Research Consortium for the Participating Institutions. The Participating Institutions are the American Museum of Natural History, Astrophysical Institute Potsdam, University of Basel, University of Cambridge, Case Western Reserve University, University of Chicago, Drexel University, Fermilab, the Institute for Advanced Study, the Japan Participation Group, Johns Hopkins University, the Joint Institute for Nuclear Astrophysics, the Kavli Institute for Particle Astrophysics and Cosmology, the Korean Scientist Group, the Chinese Academy of Sciences (LAMOST), Los Alamos National Laboratory, the Max-Planck-Institute for Astronomy (MPIA), the Max-Planck-Institute for Astrophysics (MPA), New Mexico State University, Ohio State University, University of Pittsburgh, University of Portsmouth, Princeton University, the United States Naval Observatory, and the University of Washington.

This research has made use of the NASA/IPAC Extragalactic Database (NED) which is operated by the Jet Propulsion Laboratory, California Institute of Technology, under contract with the National Aeronautics and Space Administration.


\begin{thebibliography}{99}

\bibitem[\protect\citeauthoryear{Abazajian et al.}{2009}]{2009ApJS..182..543A} Abazajian K.~N., et al., 2009, ApJS, 182, 543 
\bibitem[\protect\citeauthoryear{Abdo et al.}{2009}]{2009ApJ...707L.142A} Abdo A.~A., et al., 2009, ApJ, 707, L142 
\bibitem[\protect\citeauthoryear{Aoki, Kawaguchi, \& Ohta}{2005}]{2005ApJ...618..601A} Aoki K., Kawaguchi T., Ohta K., 2005, ApJ, 618, 601 
\bibitem[\protect\citeauthoryear{Baskin \& Laor}{2004}]{2004MNRAS.350L..31B} Baskin A., Laor A., 2004, MNRAS, 350, L31 
\bibitem[\protect\citeauthoryear{Becker, White, \& Helfand}{1994}]{1994ASPC...61..165B} Becker R.~H., White R.~L., Helfand D.~J., 1994, ASPC, 61, 165 
\bibitem[\protect\citeauthoryear{Berton et al.}{2015a}]{2015A&A...578A..28B} Berton M., et al., 2015$a$, A\&A, 578, A28 
\bibitem[\protect\citeauthoryear{Berton et al.}{2015}]{2015arXiv150605800B} Berton M., Foschini L., Ciroi S., Cracco V., La Mura G., Di Mille F., Rafanelli P., 2015, arXiv, arXiv:1506.05800 
\bibitem[\protect\citeauthoryear{Bian, Yuan, \& Zhao}{2005}]{2005MNRAS.364..187B} Bian W., Yuan Q., Zhao Y., 2005, MNRAS, 364, 187 
\bibitem[\protect\citeauthoryear{Boller, Brandt, \& Fink}{1996}]{1996A&A...305...53B} Boller T., Brandt W.~N., Fink H., 1996, A\&A, 305, 53 
\bibitem[\protect\citeauthoryear{Boroson \& Green}{1992}]{1992ApJS...80..109B} Boroson T.~A., Green R.~F., 1992, ApJS, 80, 109 
\bibitem[\protect\citeauthoryear{Boroson}{2002}]{2002ApJ...565...78B} Boroson T.~A., 2002, ApJ, 565, 78 
\bibitem[\protect\citeauthoryear{Boroson}{2005}]{2005AJ....130..381B} Boroson T., 2005, AJ, 130, 381
\bibitem[\protect\citeauthoryear{Botte et al.}{2005}]{2005MNRAS.356..789B} Botte V., Ciroi S., di Mille F., Rafanelli P., Romano A., 2005, MNRAS, 356, 789 
\bibitem[\protect\citeauthoryear{Cardelli, Clayton, \& Mathis}{1989}]{1989ApJ...345..245C} Cardelli J.~A., Clayton G.~C., Mathis J.~S., 1989, ApJ, 345, 245 
\bibitem[\protect\citeauthoryear{Collin \& Joly}{2000}]{2000NewAR..44..531C} Collin S., Joly M., 2000, NewAR, 44, 531
 \bibitem[\protect\citeauthoryear{Condon et al.}{1998}]{1998AJ....115.1693C} Condon J.~J., Cotton W.~D., Greisen E.~W., Yin Q.~F., Perley R.~A., Taylor G.~B., Broderick J.~J., 1998, AJ, 115, 1693
\bibitem[\protect\citeauthoryear{Contini \& Viegas}{2001}]{2001ApJS..132..211C} Contini M., Viegas S.~M., 2001, ApJS, 132, 211 
\bibitem[\protect\citeauthoryear{Cracco et al.}{2011}]{2011MNRAS.418.2630C} Cracco V., et al., 2011, MNRAS, 418, 2630 
\bibitem[\protect\citeauthoryear{Croom et al.}{2002}]{2002MNRAS.337..275C} Croom S.~M., et al., 2002, MNRAS, 337, 275
\bibitem[\protect\citeauthoryear{Decarli et al.}{2008}]{2008MNRAS.386L..15D} Decarli R., Dotti M., Fontana M., Haardt F., 2008, MNRAS, 386, L15 
\bibitem[\protect\citeauthoryear{Dietrich, Crenshaw, \& Kraemer}{2005}]{2005ApJ...623..700D} Dietrich M., Crenshaw D.~M., Kraemer S.~B., 2005, ApJ, 623, 700 
\bibitem[\protect\citeauthoryear{Dong et al.}{2008}]{2008MNRAS.383..581D} Dong X., Wang T., Wang J., Yuan W., Zhou H., Dai H., Zhang K., 2008, MNRAS, 383, 581 
\bibitem[\protect\citeauthoryear{Dong et al.}{2009}]{2009ApJ...703L...1D} Dong X.-B., Wang T.-G., Wang J.-G., Fan X., Wang H., Zhou H., Yuan W., 2009, ApJ, 703, L1 
\bibitem[\protect\citeauthoryear{Dong et al.}{2011}]{2011ApJ...736...86D} Dong X.-B., Wang J.-G., Ho L.~C., Wang T.-G., Fan X., Wang H., Zhou H., Yuan W., 2011, ApJ, 736, 86 
\bibitem[\protect\citeauthoryear{Dong et al.}{2012}]{2012ApJ...755..167D} Dong X.-B., Ho L.~C., Yuan W., Wang T.-G., Fan X., Zhou H., Jiang N., 2012, ApJ, 755, 167
\bibitem[\protect\citeauthoryear{Ebrero et al.}{2011}]{2011A&A...535A..62E} Ebrero J., Costantini E., Kaastra J.~S., de Marco B., Dadina M., 2011, A\&A, 535, A62 
\bibitem[\protect\citeauthoryear{Ferland \& Netzer}{1983}]{1983ApJ...264..105F} Ferland G.~J., Netzer H., 1983, ApJ, 264, 105
\bibitem[\protect\citeauthoryear{Ferland et al.}{2013}]{2013RMxAA..49..137F} Ferland G.~J., et al., 2013, RMxAA, 49, 137 
\bibitem[\protect\citeauthoryear{Foschini}{2011}]{2011nlsg.confE..24F} Foschini L., 2011, in: Foschini, L., Colpi, M., Gallo, L., et al. (eds), Narrow-Line Seyfert 1 Galaxies and Their Place in the Universe, Proceedings of Science, Trieste, vol. NLS1, id 24. 
\bibitem[\protect\citeauthoryear{Foschini et al.}{2015}]{2015A&A...575A..13F} Foschini L., et al., 2015, A\&A, 575, A13 
\bibitem[\protect\citeauthoryear{Gallo et al.}{2004}]{2004MNRAS.347..269G} Gallo L.~C., Boller T., Tanaka Y., Fabian A.~C., Brandt W.~N., Welsh W.~F., Anabuki N., Haba Y., 2004, MNRAS, 347, 269 
\bibitem[\protect\citeauthoryear{Gaskell}{1982}]{1982ApJ...263...79G} Gaskell C.~M., 1982, ApJ, 263, 79 
\bibitem[\protect\citeauthoryear{Gaskell}{1985}]{1985ApJ...291..112G} Gaskell C.~M., 1985, ApJ, 291, 112 
\bibitem[\protect\citeauthoryear{Gaskell \& Goosmann}{2013}]{2013ApJ...769...30G} Gaskell C.~M., Goosmann R.~W., 2013, ApJ, 769, 30 
\bibitem[\protect\citeauthoryear{Goad, Korista, \& Ruff}{2012}]{2012MNRAS.426.3086G} Goad M.~R., Korista K.~T., Ruff A.~J., 2012, MNRAS, 426, 3086 
\bibitem[\protect\citeauthoryear{Gon{\c c}alves, V{\'e}ron, \& V{\'e}ron-Cetty}{1999}]{1999A&A...341..662G} Gon{\c c}alves A.~C., V{\'e}ron P., V{\'e}ron-Cetty M.-P., 1999, A\&A, 341, 662 
\bibitem[\protect\citeauthoryear{Goodrich}{1989}]{1989ApJ...342..224G} Goodrich R.~W., 1989, ApJ, 342, 224 
\bibitem[\protect\citeauthoryear{Grandi}{1980}]{1980ApJ...238...10G} Grandi S.~A., 1980, ApJ, 238, 10 
\bibitem[\protect\citeauthoryear{Greene \& Ho}{2005}]{2005ApJ...627..721G} Greene J.~E., Ho L.~C., 2005, ApJ, 627, 721 
\bibitem[\protect\citeauthoryear{Greene et al.}{2010}]{2010ApJ...723..409G} Greene J.~E., et al., 2010, ApJ, 723, 409 
\bibitem[\protect\citeauthoryear{Grevesse \& Anders}{1991}]{1991sia..book.1227G} Grevesse N., Anders E., 1991, sia..book, 1227 
\bibitem[\protect\citeauthoryear{Grupe}{2004}]{2004AJ....127.1799G} Grupe D., 2004, AJ, 127, 1799 
\bibitem[\protect\citeauthoryear{Grupe \& Mathur}{2004}]{2004ApJ...606L..41G} Grupe D., Mathur S., 2004, ApJ, 606, L41 
\bibitem[\protect\citeauthoryear{Heckman et al.}{1981}]{1981ApJ...247..403H} Heckman T.~M., Miley G.~K., van Breugel W.~J.~M., Butcher H.~R., 1981, ApJ, 247, 403 
\bibitem[\protect\citeauthoryear{Holt, Tadhunter, \& Morganti}{2008}]{2008MNRAS.387..639H} Holt J., Tadhunter C.~N., Morganti R., 2008, MNRAS, 387, 639 
\bibitem[\protect\citeauthoryear{Hu et al.}{2008}]{2008ApJ...687...78H} Hu C., Wang J.-M., Ho L.~C., Chen Y.-M., Zhang H.-T., Bian W.-H., Xue S.-J., 2008, ApJ, 687, 78 
\bibitem[\protect\citeauthoryear{Hu et al.}{2015}]{2015ApJ...804..138H} Hu C., et al., 2015, ApJ, 804, 138 
\bibitem[\protect\citeauthoryear{Kaspi et al.}{2000}]{2000ApJ...533..631K} Kaspi S., Smith P.~S., Netzer H., Maoz D., Jannuzi B.~T., Giveon U., 2000, ApJ, 533, 631 
\bibitem[\protect\citeauthoryear{Kellermann et al.}{1989}]{1989AJ.....98.1195K} Kellermann K.~I., Sramek R., Schmidt M., Shaffer D.~B., Green R., 1989, AJ, 98, 1195 
\bibitem[\protect\citeauthoryear{Kollatschny \& Zetzl}{2011}]{2011Natur.470..366K} Kollatschny W., Zetzl M., 2011, Natur, 470, 366 
\bibitem[\protect\citeauthoryear{Kollatschny \& Zetzl}{2013}]{2013A&A...549A.100K} Kollatschny W., Zetzl M., 2013, A\&A, 549, AA100
\bibitem[\protect\citeauthoryear{Komossa \& Mathur}{2001}]{2001A&A...374..914K} Komossa S., Mathur S., 2001, A\&A, 374, 914 
\bibitem[\protect\citeauthoryear{Komossa \& Xu}{2007}]{2007ApJ...667L..33K} Komossa S., Xu D., 2007, ApJ, 667, L33 
\bibitem[\protect\citeauthoryear{Komossa et al.}{2008}]{2008ApJ...680..926K} Komossa S., Xu D., Zhou H., 
Storchi-Bergmann T., Binette L., 2008, ApJ, 680, 926 
\bibitem[\protect\citeauthoryear{Kova{\v c}evi{\'c}, Popovi{\'c}, \& Dimitrijevi{\'c}}{2010}]{2010ApJS..189...15K} Kova{\v c}evi{\'c} J., Popovi{\'c} L.~{\v C}., Dimitrijevi{\'c} M.~S., 2010, ApJS, 189, 15
\bibitem[\protect\citeauthoryear{Kova{\v c}evi{\'c}-Doj{\v c}inovi{\'c} 
\& Popovi{\'c}}{2015}]{2015ApJS..221...35K} Kova{\v c}evi{\'c}-Doj{\v c}inovi{\'c} J., Popovi{\'c} L.~{\v C}., 2015, ApJS, 221, 35 
\bibitem[\protect\citeauthoryear{Kraemer \& Crenshaw}{2000}]{2000ApJ...532..256K} Kraemer S.~B., Crenshaw D.~M., 2000, ApJ, 532, 256 
\bibitem[\protect\citeauthoryear{Kuraszkiewicz et al.}{2000}]{2000ApJ...542..631K} Kuraszkiewicz J., Wilkes B.~J., Brandt W.~N., Vestergaard M., 2000, ApJ, 542, 631 
\bibitem[\protect\citeauthoryear{Landt et al.}{2008}]{2008ApJS..174..282L} Landt H., Bentz M.~C., Ward M.~J., Elvis M., Peterson B.~M., Korista K.~T., Karovska M., 2008, ApJS, 174, 282 
\bibitem[\protect\citeauthoryear{Laor et al.}{1994}]{1994ApJ...435..611L} Laor A., Fiore F., Elvis M., Wilkes B.~J., McDowell J.~C., 1994, ApJ, 435, 611 
\bibitem[\protect\citeauthoryear{Laor}{2000}]{2000NewAR..44..503L} Laor A., 2000, NewAR, 44, 503 
\bibitem[\protect\citeauthoryear{Leighly}{1999}]{1999ApJS..125..317L} Leighly K.~M., 1999, ApJS, 125, 317 
\bibitem[\protect\citeauthoryear{Ludwig et al.}{2012}]{2012ApJ...756...51L} Ludwig R.~R., Greene J.~E., Barth A.~J., Ho L.~C., 2012, ApJ, 756, 51 
\bibitem[\protect\citeauthoryear{Lutz et al.}{2000}]{2000ApJ...530..733L} Lutz D., et al., 2000, ApJ, 530, 733 
\bibitem[\protect\citeauthoryear{Marconi et al.}{1996}]{1996A&A...315..335M} Marconi A., van der Werf P.~P., Moorwood A.~F.~M., Oliva E., 1996, A\&A, 315, 335 
\bibitem[\protect\citeauthoryear{Marinello et al.}{2016}]{2016arXiv160205159M} Marinello A.~O.~M., Rodriguez-Ardila A., Garcia-Rissmann A., Sigut T.~A.~A., Pradhan A.~K., 2016, arXiv, arXiv:1602.05159 
\bibitem[\protect\citeauthoryear{Markwardt}{2009}]{2009ASPC..411..251M} Markwardt C.~B., 2009, ASPC, 411, 251 
\bibitem[\protect\citeauthoryear{Mart{\'{\i}}nez-Aldama et al.}{2015}]{2015ApJS..217....3M} Mart{\'{\i}}nez-Aldama M.~L., Dultzin D., Marziani P., Sulentic J.~W., Bressan A., Chen Y., Stirpe G.~M., 2015, ApJS, 217, 3 
\bibitem[\protect\citeauthoryear{Marziani et al.}{2001}]{2001ApJ...558..553M} Marziani P., Sulentic J.~W., Zwitter T., Dultzin-Hacyan D., Calvani M., 2001, ApJ, 558, 553 
\bibitem[\protect\citeauthoryear{Marziani et al.}{2003a}]{2003MNRAS.345.1133M} Marziani P., Zamanov R.~K., Sulentic J.~W., Calvani M., 2003$a$, MNRAS, 345, 1133 
\bibitem[\protect\citeauthoryear{Marziani et al.}{2003b}]{2003ApJS..145..199M} Marziani P., Sulentic J.~W., Zamanov R., Calvani M., Dultzin-Hacyan D., Bachev R., Zwitter T., 2003$b$, ApJS, 145, 199 
\bibitem[\protect\citeauthoryear{Marziani et al.}{2013}]{2013AstRv...8d...4M} Marziani P., Martnez-Aldama M.~L., Dultzin D., Sulentic J.~W., 2013, AstRv, 8, 4
\bibitem[\protect\citeauthoryear{Marziani et al.}{2015}]{2015arXiv151107138M} Marziani P., Sulentic J.~W., Stirpe G.~M., Dultzin D., Del Olmo A., Mart{\'{\i}}nez-Carballo M.~A., 2015, arXiv, arXiv:1511.07138 
\bibitem[\protect\citeauthoryear{Mas-Hesse et al.}{1994}]{1994A&A...283L...9M} Mas-Hesse J.~M., Rodriguez-Pascual P.~M., de Cordoba L.~S.~F., Boller T., 1994, A\&A, 283, L9 
\bibitem[\protect\citeauthoryear{Mathur}{2000}]{2000MNRAS.314L..17M} Mathur S., 2000, MNRAS, 314, L17 
\bibitem[\protect\citeauthoryear{Matsuoka et al.}{2007}]{2007ApJ...663..781M} Matsuoka Y., Oyabu S., Tsuzuki Y., Kawara K., 2007, ApJ, 663, 781 
\bibitem[\protect\citeauthoryear{Moran, Halpern, \& Helfand}{1996}]{1996ApJS..106..341M} Moran E.~C., Halpern J.~P., Helfand D.~J., 1996, ApJS, 106, 341 
\bibitem[\protect\citeauthoryear{Morris \& Ward}{1985}]{1985MNRAS.215P..57M} Morris S.~L., Ward M.~J., 1985, MNRAS, 215, 57P 
\bibitem[\protect\citeauthoryear{Mullaney \& Ward}{2008}]{2008MNRAS.385...53M} Mullaney J.~R., Ward M.~J., 2008, MNRAS, 385, 53 
\bibitem[\protect\citeauthoryear{Mullaney et al.}{2013}]{2013MNRAS.433..622M} Mullaney J.~R., Alexander D.~M., Fine S., Goulding A.~D., Harrison C.~M., Hickox R.~C., 2013, MNRAS, 433, 622 
\bibitem[\protect\citeauthoryear{Nagar et al.}{2002}]{2002A&A...391L..21N} Nagar N.~M., Oliva E., Marconi A., Maiolino R., 2002, A\&A, 391, L21
\bibitem[\protect\citeauthoryear{Nelson \& Whittle}{1996}]{1996ApJ...465...96N} Nelson C.~H., Whittle M., 1996, ApJ, 465, 96 
\bibitem[\protect\citeauthoryear{Netzer et al.}{2004}]{2004ApJ...614..558N} Netzer H., Shemmer O., Maiolino R., Oliva E., Croom S., Corbett E., di Fabrizio L., 2004, ApJ, 614, 558 
\bibitem[\protect\citeauthoryear{Netzer \& Trakhtenbrot}{2007}]{2007ApJ...654..754N} Netzer H., Trakhtenbrot B., 2007, ApJ, 654, 754 
\bibitem[\protect\citeauthoryear{Oke \& Shields}{1976}]{1976ApJ...207..713O} Oke J.~B., Shields G.~A., 1976, ApJ, 207, 713 
\bibitem[\protect\citeauthoryear{Osterbrock \& Pogge}{1985}]{1985ApJ...297..166O} Osterbrock D.~E., Pogge R.~W., 1985, ApJ, 297, 166
\bibitem[\protect\citeauthoryear{Ozaki}{2009}]{2009PASJ...61..259O} Ozaki S., 2009, PASJ, 61, 259 
\bibitem[\protect\citeauthoryear{Peng et al.}{2014}]{2014RAA....14..913P} Peng Z.-X., Chen Y.-M., Gu Q.-S., Zhang K., 2014, RAA, 14, 913 
\bibitem[\protect\citeauthoryear{Penston et al.}{1990}]{1990A&A...236...53P} Penston M.~V., et al., 1990, A\&A, 236, 53 
\bibitem[\protect\citeauthoryear{Persson}{1988}]{1988ApJ...330..751P} Persson S.~E., 1988, ApJ, 330, 75 
\bibitem[\protect\citeauthoryear{Peterson, McHardy, \& Wilkes}{2000}]{2000NewAR..44..491P} Peterson B.~M., McHardy I.~M., Wilkes B.~J., 2000, NewAR, 44, 491 
\bibitem[\protect\citeauthoryear{Phillips}{1976}]{1976ApJ...208...37P} Phillips M.~M., 1976, ApJ, 208, 37 
\bibitem[\protect\citeauthoryear{Pogge}{2011}]{2011nlsg.confE...2P} Pogge R.~W., 2011, nlsg.conf, 2 
\bibitem[\protect\citeauthoryear{Puchnarewicz et al.}{1995}]{1995MNRAS.276.1281P} Puchnarewicz E.~M., Branduardi-Raymont G., Mason K.~O., Sekiguchi K., 1995, MNRAS, 276, 1281 
\bibitem[\protect\citeauthoryear{Radovich, Hasinger, \& Rafanelli}{1998}]{1998AN....319..325R} Radovich M., Hasinger G., Rafanelli P., 1998, AN, 319, 325 
\bibitem[\protect\citeauthoryear{Rodr{\'{\i}}guez-Ardila, Pastoriza, \& Donzelli}{2000}]{2000ApJS..126...63R} Rodr{\'{\i}}guez-Ardila A., Pastoriza M.~G., Donzelli C.~J., 2000, ApJS, 126, 63 
\bibitem[\protect\citeauthoryear{Rodr{\'{\i}}guez-Ardila et al.}{2002}]{2002ApJ...572...94R} Rodr{\'{\i}}guez-Ardila A., Viegas S.~M., Pastoriza M.~G., Prato L., Donzelli C.~J., 2002, ApJ, 572, 94 
\bibitem[\protect\citeauthoryear{Shapovalova et al.}{2012}]{2012ApJS..202...10S} Shapovalova A.~I., et al., 2012, ApJS, 202, 10 
\bibitem[\protect\citeauthoryear{Shen et al.}{2011}]{2011ApJS..194...45S} Shen Y., et al., 2011, ApJS, 194, 45 
\bibitem[\protect\citeauthoryear{Shen \& Ho}{2014}]{2014Natur.513..210S} Shen Y., Ho L.~C., 2014, Natur, 513, 210 
\bibitem[\protect\citeauthoryear{Shields, Ludwig, \& Salviander}{2010}]{2010ApJ...721.1835S} Shields G.~A., Ludwig R.~R., Salviander S., 2010, ApJ, 721, 1835 
\bibitem[\protect\citeauthoryear{Spoon \& Holt}{2009}]{2009ApJ...702L..42S} Spoon H.~W.~W., Holt J., 2009, ApJ, 702, L42 
\bibitem[\protect\citeauthoryear{Sulentic et al.}{2000}]{2000ApJ...536L...5S} Sulentic J.~W., Zwitter T., Marziani P., Dultzin-Hacyan D., 2000$b$, ApJ, 536, L5 
\bibitem[\protect\citeauthoryear{Sulentic et al.}{2002}]{2002ApJ...566L..71S} Sulentic J.~W., Marziani P., Zamanov R., Bachev R., Calvani M., Dultzin-Hacyan D., 2002, ApJ, 566, L71 
\bibitem[\protect\citeauthoryear{Tremaine et al.}{2002}]{2002ApJ...574..740T} Tremaine S., et al., 2002, ApJ, 574, 740 
\bibitem[\protect\citeauthoryear{Vaona et al.}{2012}]{2012MNRAS.427.1266V} Vaona L., Ciroi S., Di Mille F., Cracco V., La Mura G., Rafanelli P., 2012, 
MNRAS, 427, 1266 
\bibitem[\protect\citeauthoryear{Veilleux}{1991}]{1991ApJS...75..383V} Veilleux S., 1991, ApJS, 75, 383 
\bibitem[\protect\citeauthoryear{V{\'e}ron-Cetty, V{\'e}ron, \& Gon{\c c}alves}{2001}]{2001A&A...372..730V} V{\'e}ron-Cetty M.-P., V{\'e}ron P., Gon{\c c}alves A.~C., 2001, A\&A, 372, 730
\bibitem[\protect\citeauthoryear{V{\'e}ron-Cetty, Joly, \& V{\'e}ron}{2004}]{2004A&A...417..515V} V{\'e}ron-Cetty M.-P., Joly M., V{\'e}ron P., 2004, A\&A, 417, 515 
\bibitem[\protect\citeauthoryear{V{\'e}ron-Cetty et al.}{2006}]{2006A&A...451..851V} V{\'e}ron-Cetty M.-P., Joly M., V{\'e}ron P., Boroson T., Lipari S., Ogle P., 2006, A\&A, 451, 851 
\bibitem[\protect\citeauthoryear{Vrtilek \& Carleton}{1985}]{1985ApJ...294..106V} Vrtilek J.~M., Carleton N.~P., 1985, ApJ, 294, 106 
\bibitem[\protect\citeauthoryear{Wandel, Peterson, \& Malkan}{1999}]{1999ApJ...526..579W} Wandel A., Peterson B.~M., Malkan M.~A., 1999, ApJ, 526, 579 
\bibitem[\protect\citeauthoryear{Wang, Brinkmann, \& Bergeron}{1996}]{1996A&A...309...81W} Wang T., Brinkmann W., Bergeron J., 1996, A\&A, 309, 81 
\bibitem[\protect\citeauthoryear{Wang \& Lu}{2001}]{2001A&A...377...52W} Wang T., Lu Y., 2001, A\&A, 377, 52
\bibitem[\protect\citeauthoryear{Wang, Dai, \& Zhou}{2008}]{2008ApJ...674..668W} Wang T., Dai H., Zhou H., 2008, ApJ, 674, 668 
\bibitem[\protect\citeauthoryear{Wang, Mao, \& Wei}{2011}]{2011ApJ...741...50W} Wang J., Mao Y.~F., Wei J.~Y., 2011, ApJ, 741, 50 
\bibitem[\protect\citeauthoryear{Whittle}{1985}]{1985MNRAS.213....1W} Whittle M., 1985, MNRAS, 213, 1 
\bibitem[\protect\citeauthoryear{Woo et al.}{2015}]{2015ApJ...801...38W} Woo J.-H., Yoon Y., Park S., Park D., Kim S.~C., 2015, ApJ, 801, 38 
\bibitem[\protect\citeauthoryear{Woo et al.}{2016}]{2016ApJ...817..108W} 
Woo J.-H., Bae H.-J., Son D., Karouzos M., 2016, ApJ, 817, 108 
\bibitem[\protect\citeauthoryear{Xiao et al.}{2011}]{2011ApJ...739...28X} Xiao T., Barth A.~J., Greene J.~E., Ho L.~C., Bentz M.~C., Ludwig R.~R., Jiang Y., 2011, ApJ, 739, 28 
\bibitem[\protect\citeauthoryear{Xu et al.}{2007}]{2007ApJ...670...60X} Xu D., Komossa S., Zhou H., Wang T., Wei J., 2007, ApJ, 670, 60 
\bibitem[\protect\citeauthoryear{Xu et al.}{2012}]{2012AJ....143...83X} Xu D., Komossa S., Zhou H., Lu H., Li C., Grupe D., Wang J., Yuan W., 2012, AJ, 143, 83
\bibitem[\protect\citeauthoryear{Zakamska et al.}{2015}]{2015arXiv151202642Z} Zakamska N.~L., et al., 2015, arXiv, arXiv:1512.02642 
\bibitem[\protect\citeauthoryear{Zamanov et al.}{2002}]{2002ApJ...576L...9Z} Zamanov R., Marziani P., Sulentic J.~W., Calvani M., Dultzin-Hacyan D., Bachev R., 2002, ApJ, 576, L9 
\bibitem[\protect\citeauthoryear{Zhang et al.}{2011}]{2011ApJ...737...71Z} Zhang K., Dong X.-B., Wang T.-G., Gaskell C.~M., 2011, ApJ, 737, 71 
\bibitem[\protect\citeauthoryear{Zhou et al.}{2006}]{2006ApJS..166..128Z} Zhou H., Wang T., Yuan W., Lu H., Dong X., Wang J., Lu Y., 2006, ApJS, 166, 128 
\end{thebibliography}
\end{document}